\documentstyle[lscape]{ar}
\begin{document}

\input psfig.sty
\def\nhi{\noindent \hangindent=0.3cm}
\def\be{\begin{equation}}
\def\ee{\end{equation}}
\def\msun{{~M}_{\odot}}
\def\msunyr{M_{\odot} \ {\rm yr^{-1}}}
\def\medd{{\dot M_{Edd}}}
\def\mdot{{\dot M}}
\def\ergs{{\rm\,erg\,s^{-1}}}
\def\ergscc{\rm \  \ erg \ cm^{-3} \ s^{-1}}
\def\gs{\rm \,g\,s^{-1}}

\jname{Annu. Rev. Astron. Astrophys.}
\jyear{2013}
\jvol{51}

\def\sles{\lower2pt\hbox{$\buildrel {\scriptstyle <}
   \over {\scriptstyle\sim}$}}

\def\sgreat{\lower2pt\hbox{$\buildrel {\scriptstyle >}
   \over {\scriptstyle\sim}$}}

\title{Hot Accretion Flows Around Black Holes}

\markboth{Feng Yuan and Ramesh Narayan}{Hot Accretion Flow}

\author{Feng Yuan
\affiliation{Shanghai Astronomical Observatory, Chinese Academy of Sciences, 80 Nandan Road, Shanghai 200030, China; email: fyuan@shao.ac.cn}
Ramesh Narayan
\affiliation{Harvard-Smithsonian Center for Astrophysics, 60 Garden Street, Cambridge, MA 02138, USA; email: rnarayan@cfa.harvard.edu}}

\begin{keywords}
accretion disks, active galactic nuclei, active galactic feedback, black holes, black hole X-ray binaries, jet, outflow
\end{keywords}

\begin{abstract}
Black hole accretion flows can be divided into two broad classes: cold
and hot. Cold accretion flows, which consist of cool optically thick
gas, are found at relatively high mass accretion rates. Prominent
examples are the standard thin disk, which occurs at a fraction of the
Eddington mass accretion rate, and the slim disk at super-Eddington
rates. These accretion flows are responsible for luminous systems such
as active galactic nuclei radiating at or close to the Eddington
luminosity and black hole X-ray binaries in the soft state. Hot
accretion flows, the topic of this review, are virially hot and
optically thin.  They occur at lower mass accretion rates, and are
described by models such as the advection-dominated accretion flow and
luminous hot accretion flow. Because of energy advection, the
radiative efficiency of these flows is in general lower than that of a
standard thin accretion disk. Moreover, the efficiency decreases with
decreasing mass accretion rate. Detailed modeling of hot accretion
flows is hampered by theoretical uncertainties on the heating of
electrons, equilibration of electron and ion temperatures, and
relative roles of thermal and non-thermal particles. Observations show
that hot accretion flows are associated with jets. In addition,
theoretical arguments suggest that hot flows should produce strong
winds. This link between the hot mode of accretion and outflows of
various kinds is currently being explored via hydrodynamic and
magnetohydrodynamic computer simulations. Hot accretion flows are
believed to be present in low-luminosity active galactic nuclei and in
black hole X-ray binaries in the hard and quiescent states. The
prototype is Sgr A*, the ultra-low-luminosity supermassive black hole
at our Galactic center. The jet, wind and radiation from a
supermassive black hole with a hot accretion flow can interact with
the external interstellar medium and modify the evolution of the host
galaxy. Details of this ``maintenance-mode feedback'' could, in
principle, be worked out through theoretical studies and numerical
simulations of hot accretion flows.

\end{abstract}

\maketitle

\section{Introduction}

Black hole accretion is a fundamental physical process in the
universe, and is the primary power source behind active galactic
nuclei (AGNs), black hole X-ray binaries (BHBs) and possibly gamma-ray
bursts. The first genuine model of an accretion disk --- by which we
mean a rotating flow with viscous transport of angular momentum --- is
the celebrated thin disk model developed in the early 1970s (Shakura
\& Sunyaev 1973, Novikov \& Thorne 1973, Lynden-Bell \& Pringle 1974;
see reviews by Pringle 1981, Frank et al.~2002, Kato et al.~2008,
Abramowicz \& Fragile 2013, Blaes 2013). Depending on the mass
of the central black hole, the gas temperature in this model lies in
the range $10^4-10^7$\,K, which is quite cold relative to the virial
temperature. The disk is geometrically thin, while the gas is
optically thick and radiates thermal blackbody-like radiation. Many
accreting black hole sources have been successfully modeled as thin
disks, e.g., luminous AGNs (see reviews above; but also Koratkar \&
Blaes 1999) and BHBs in the thermal state (Remillard \& McClintock
2006, McClintock et al.~2013).

The thin disk model applies whenever the disk luminosity $L$ is
somewhat below the Eddington luminosity $L_{\rm Edd}$, or
equivalently, when the mass accretion rate $\dot{M}$ is below the
Eddington rate,\footnote{While everyone agrees on the definition of
  the Eddington luminosity, viz., $L_{\rm Edd} = 4\pi GMc/\kappa_{\rm
    es}$, where $\kappa_{\rm es}$ is the electron scattering opacity,
  usually taken to be $0.4\,{\rm cm^2g^{-1}}$, many definitions are
  used for $\dot{M}_{\rm Edd}$.  In this article we use a definition
  which assumes that the accretion disk has a nominal radiative
  efficiency of $10\%$, hence $L_{\rm Edd}=0.1\dot{M}_{\rm Edd}c^2$.
  Some authors use $L_{\rm Edd}=\dot{M}_{\rm Edd}c^2$, others use
  $L_{\rm Edd}=\dot{M}_{\rm Edd}c^2/12$ or $\dot{M}_{\rm Edd}c^2/16$,
  and yet others use $L_{\rm Edd}=\eta(a_*) \dot{M}_{\rm Edd}c^2$
  where $\eta(a_*)$ is the relativistic radiative efficiency of a thin
  disk around a black hole with dimensionless spin parameter $a_*
  \equiv a/M$.}  $\dot{M}_{\rm Edd}\equiv 10L_{\rm Edd}/c^2 =
1.39\times 10^{18} (M/M_\odot)\,{\rm g\,s^{-1}}$, where $M$ is the
mass of the black hole. When $\dot{M}$ approaches or exceeds
$\dot{M}_{\rm Edd}$, the accreting gas becomes optically too thick to
radiate all the dissipated energy locally (a key requirement of the
thin disk model). Radiation is then trapped and advected inward with
the accretion flow. Consequently, the radiative efficiency becomes
lower, and $L$ becomes progressively smaller than $0.1\dot{M}c^2$. The
disk solution that describes such a system is called the slim disk, or
equivalently, optically thick advection-dominated accretion flow (Katz
1977, Begelman 1979, Begelman \& Meier 1982, Abramowicz et
al.~1988, Chen \& Taam 1993, Ohsuga et al.~2005).  The slim disk model
has been applied to narrow-line Seyfert galaxies (Mineshige et
al.~2000), objects like SS433 (Fabrika 2004) and ultraluminous X-ray
sources (Watarai et al.~2001).

The thin disk and slim disk both belong to the class of cold accretion
flows. Both consist of optically thick gas. The first hot
accretion flow model was described by Shapiro et al.~(1976; hereafter
SLE). In contrast to the thin disk and slim disk, the temperature of
the gas in the SLE solution is much higher, approaching virial,
  and the gas is optically thin. A key innovation of the SLE model is
the introduction of a two-temperature accreting plasma, where the ions
are much hotter than the electrons.  The main success of the SLE
solution, indeed its motivation, is that, for the first time, it was
able to explain the hard X-ray emission seen in some black hole
sources.
Unfortunately, soon after the SLE model was introduced, it was
realized that it is thermally unstable, so the model as originally
developed is unlikely to be realized in nature.

The important role of advection in hot accretion flows was first
emphasized by Ichimaru (1977) who pointed out that, in certain
regimes, the viscously dissipated accretion energy can go into heating
the accretion flow rather than being radiated away. This is the most
important feature of the general class of advection-dominated
accretion flows (ADAFs), one of the hot accretion solutions we will
discuss in this review. Ichimaru further argued that, because of the
inclusion of advection, his hot accretion solution should be thermally
stable. Similar ideas were described independently by Rees et
al.~(1982) in their two-temperature ``ion torus'' model, though they
did not emphasize the relation between their model and those of SLE
and Ichimaru, nor did they discuss stability. Sadly, these pioneering
studies were not followed up for many years.

With the re-discovery of the ADAF solution in the mid-1990s (Narayan
\& Yi 1994, 1995a, 1995b; Abramowicz et al.~1995; Chen et al.~1995),
and the subsequent detailed study of its properties, hot accretion
flow models finally became established in the accretion
literature.\footnote{Note that ``advection-dominated accretion'' is
  not synonymous with ``hot accretion''. For instance, the slim disk
  is advection-dominated, although for a very different reason (long
  radiative diffusion time) compared to a hot ADAF (long cooling
  time). In this article, we classify accretion solutions as hot or
  cold, and focus our attention on the hot solutions. While our
  classification is somewhat arbitrary, at least in certain respects,
  e.g., Fig.~\ref{fig:solutionsummary}, the two solution branches are
  clearly distinct.}  The dynamical and radiative properties of the
ADAF solution have been studied in significant detail, and the model
has been applied to various black hole systems, including the
supermassive black hole in our Galactic center, Sagittarius A* (Sgr
A*), low-luminosity AGNs, and BHBs in the hard and quiescent states
This article reviews our current understanding of hot accretion
flows. The reader is encouraged to read other reviews (e.g., Narayan
et al.~1998b, Lasota 1999, Quataert 2001, Narayan \& McClintock 2008,
Ho 2008, Abramowicz \& Fragile 2013, Blaes 2013), which discuss
certain aspects of ADAFs in greater detail than we can here.

Before concluding this introduction, we briefly explain some
terminology. The most popular and widely-used name for hot accretion
flows is ``advection-dominated accretion flow'' (ADAF). Two variants
of the ADAF are advection-dominated inflow-outflow solution (ADIOS)
and convection-dominated accretion flow (CDAF), which emphasize the
roles of two distinct physical phenomena in hot accretion flows:
outflows and convection (\S\ref{outflow}). Hot accretion flows are
usually radiatively inefficient. Perhaps to emphasize that the low
efficiency is not just because of advection, but may also be the
result of other effects such as outflows and convection, some authors
use the name ``radiatively inefficient accretion flow''
(RIAF). However, as we will see in \S\ref{efficiency}, the radiative
efficiency of a hot accretion flow increases with increasing mass
accretion rate. In fact, the efficiency can even be comparable to that
of the standard thin disk. This is especially the case in the
``luminous hot accretion flow'' (LHAF, \S\ref{solutionenergetics}), an
extension of the ADAF to accretion rates above the original range of
validity of the ADAF solution. Since the common feature of all these
accretion solutions is that the gas is very hot, we use the generic
name of ``Hot Accretion Flow.'' Nevertheless, because of its
popularity, we sometimes also use the term ADAF.

\section{One-dimensional dynamics and radiation}
\label{onedimension}

\subsection{One-dimensional equations and self-similar solution}
\label{self-similar}

Consider a steady axisymmetric accretion flow, and focus for
now only on the dynamics. Conservation of mass, radial momentum,
angular momentum and energy are described by the following
height-integrated differential equations (e.g., Abramowicz et
al.~1988, Narayan \& Yi 1994, Narayan et al.~1998b):
\begin{eqnarray}
{d \over dR}(\rho R H v)
&=& 0,
\label{masscons} \\
v {dv \over dR} - \Omega^2 R &=& -
\Omega_K^2R - {1 \over \rho} \, {d \over dR } (\rho c_s^2),
\label{radialeq} \\
v \, {d(\Omega R^2) \over dR} &=& {1 \over
\rho R H} \, {d \over dR} \left( \nu \rho R^3 H {d\Omega \over dR}
\right), \label{angulareq} \\
\rho v \left(\frac{de}{dR}-\frac{p}{\rho^2}\frac{d\rho}{dR}\right)&=&
\rho \nu R^2 \left( {d\Omega \over dR} \right)^2 - q^- ,
\label{energyeq}
\end{eqnarray}
where $\rho$ is the mid-plane density of the gas, $R$ is the radius,
$H\approx c_s/\Omega_K$ is the vertical scale height, $v$ is the
radial velocity, $\Omega$ is the angular velocity, $\Omega_K$ is the
Keplerian angular velocity, $c_s \equiv \sqrt{p/\rho}$ is the
isothermal sound speed, $p$ is the pressure, $e$ is the
specific internal energy, and $q^-$ is the radiative cooling rate per
unit volume. The kinematic viscosity coefficient $\nu$ may be
parameterized via the Shakura \& Sunyaev (1973) prescription, \be \nu
\equiv \alpha c_s H = \alpha {c_s^2 \over \Omega_K}, \ee where the
dimensionless parameter $\alpha$ is generally assumed to be a constant.

Equations (\ref{masscons})--(\ref{energyeq}) are quite general in the
sense that they encompass all accretion models, including the thin
disk, the slim disk and the ADAF (\S\ref{solutionsummary}).  Note,
however, that equation (\ref{masscons}) implies the mass accretion
rate $\dot{M} = 4\pi \rho RH|v|$ is independent of radius. While this
may be a reasonable approximation for a thin disk (though even these
systems can have non-constant $\dot{M}$ if there are radiatively- or
magnetically-driven winds), numerical simulations of hot accretion
flows indicate that outflows are almost inevitable (\S\ref{outflow}),
causing the mass accretion rate to decrease with decreasing
radius. Therefore, assuming a power-law variation for simplicity, it
is useful to generalize equation (\ref{masscons}) to (Blandford \&
Begelman 1999)
\begin{equation}
\dot{M}(R)=4\pi \rho RH|v| =
\dot{M}_{\rm BH}\left(\frac{R}{R_S}\right)^s, \quad R_S \leq R \leq
R_{\rm out}, \label{MdotBH}
\end{equation}
where $R_S=2GM/c^2$ is the Schwarzschild radius of the black hole,
$\dot{M}_{\rm BH}$ is the mass accretion rate at this radius and
$R_{\rm out}$ is the outer radius of the accretion flow.
The index $s$ is a measure of the strength of the outflow; $s$ cannot
exceed 1 for energetic reasons (Blandford \& Begelman 1999), while
$s=0$ corresponds to a constant mass accretion rate (no
outflow). Equations (\ref{radialeq})--(\ref{energyeq}) should also be
modified when there is mass outflow (e.g., Poutanen et al. 2007, Xie
\& Yuan 2008). However, the main effect of an outflow is probably
through the density profile. Hence, simply replacing equation
(\ref{masscons}) by equation (\ref{MdotBH}) and retaining equations
(\ref{radialeq})--(\ref{energyeq}) as written is a reasonable first
approximation. Another caveat is that the power-law variation of
$\dot{M}$ with $R$ is not likely to continue all the way down to $R_S$
as written above, but probably ceases at some inner radius $R_{\rm
  in}$ of order ten (or even tens of) $R_S$ (\S\ref{outflow} and
Fig. \ref{Fig:inflowrate}).

Equation (\ref{energyeq}) needs more discussion. The two terms on the
left hand side represent the rate of change of the internal energy per
unit volume and the work done by compression; we will call the latter
$q^c$. Together, the two terms represent energy advection, which we
write compactly as $q^{\rm adv}$. More precisely, $q^{\rm adv}$
corresponds to $\rho v Tds/dR$, where $T$ and $s$ are the temperature
and specific entropy of the gas; $q^{\rm adv}$ is thus the radial rate
of advection of entropy. The first term on the right hand side of
equation (\ref{energyeq}) is the heating rate per unit volume, or more
precisely the rate at which entropy is added to the gas via viscous
dissipation. Calling this term $q^+$, equation (\ref{energyeq}) takes
the simple form
\begin{equation}
\rho v \frac{de}{dR}-q^c\equiv q^{\rm adv} = q^+-q^- \equiv f q^+, \label{energycompact}
\end{equation}
where the parameter $f\equiv q^{\rm adv}/q^+$ measures the relative
importance of advection. Out of the total heat energy $q^+$ released
by viscous dissipation per unit volume per unit time, a fraction $f$
is advected and the rest $(1-f)$ is radiated. The standard thin disk
and SLE models assume $q^+ = q^-$ and thus correspond to $f=0$, i.e.,
vanishing energy advection. The slim disk and various hot accretion
flows have non-zero $f$. Quite often, e.g., when $\dot{M}_{\rm BH} \gg
\dot{M}_{\rm Edd}$ (slim disk) or $\dot{M}_{\rm BH} \ll \dot{M}_{\rm
  Edd}$ (hot accretion flow), one finds $q^+\gg q^-$, $f\to1$. These
accretion flows are then strongly advection-dominated.

Assuming a Newtonian gravitational potential and taking the advection
parameter $f$ to be independent of radius, Narayan \& Yi (1994, 1995b)
showed that equations (\ref{masscons})--(\ref{energyeq}) have a
self-similar solution.\footnote{The same solution was obtained earlier
  by Spruit et al.~(1987) in a different context.}  Their solution
corresponds to a constant mass accretion rate without outflows
($s=0$).  Including mass outflow via equation (\ref{MdotBH}), and
making the reasonable assumption that the only important change is in
the density profile (Xie \& Yuan 2008), the Narayan \& Yi self-similar
solution becomes approximately (Yuan et al.~2012b)
\begin{eqnarray}
v &\approx &
-1.1\times 10^{10}\alpha r^{-1/2}~{\rm
  cm~s^{-1}},\label{radialvelocity} \\
\Omega &\approx&
  2.9\times 10^4m^{-1}r^{-3/2}~{\rm s^{-1}},\label{angularvelocity} \\
c_s^2 &\approx&
1.4\times 10^{20}r^{-1}~{\rm cm}^2~{\rm
  s}^{-2},\label{soundspeedeq} \\
n_e &\approx& 6.3\times 10^{19}\alpha^{-1}m^{-1}\,\dot{m}_{\rm
  BH}\,r^{-3/2+s}~{\rm cm}^{-3},\label{density} \\
B &\approx& 6.5\times 10^8 (1+\beta)^{-1/2}
\alpha^{-1/2}m^{-1/2}\,\dot{m}_{\rm BH}^{1/2}\,
  r^{-5/4+s/2}~{\rm G}, \label{magfield} \\
p &\approx& 1.7\times
10^{16}\alpha^{-1}m^{-1}\,\dot{m}_{\rm BH}\,
  r^{-5/2+s}~{\rm g~cm^{-1}~s^{-2}},\label{pressure}
\end{eqnarray}
where the black hole mass $M$, the mass accretion rate $\dot{M}$, and
the radius $R$, have been scaled to solar, Eddington, and
Schwarzschild units, respectively,
\begin{equation}
m\equiv \frac{M}{M_\odot},
\quad \dot{m} \equiv \frac{\dot{M}}{\dot{M}_{\rm Edd}},
\quad r\equiv \frac{R}{R_S}.
\label{scaled}
\end{equation}
Correspondingly, $\dot{m}_{\rm BH} = \dot{M}_{\rm BH}/\dot{M}_{\rm
  Edd}$, where $\dot{M}_{\rm BH}$ is defined in
equation~(\ref{MdotBH}).
The parameter $\beta$ is a measure of the strength of the magnetic
field:\footnote{This is the standard definition of $\beta$ as used in
  plasma physics. However, following Narayan \& Yi (1995b), much of
  the ADAF literature uses a different $\beta_{\rm ADAF} \equiv p_{\rm
    gas}/(p_{\rm gas}+p_{\rm mag})$, which is confusingly also called
  $\beta$. The two $\beta$'s are related by $\beta_{\rm ADAF} =
  \beta/(\beta+1)$.} \be \beta\equiv \frac{p_{\rm gas}}{p_{\rm
    mag}},\label{betadefinition}\ee where $p_{\rm gas}$ is the gas
pressure and $p_{\rm mag}\equiv B^2/8\pi$ is the magnetic
pressure. Numerical magnetohydrodynamic (MHD) simulations usually give
$\beta\simeq 10$ (\S\ref{MHDsimulation}).

The advection parameter $f$ is generally a function of radius $r$ and,
more importantly, the mass accretion rate $\dot{m}$. When $\dot{m}$ is
very much smaller than unity (say $\sles\ 10^{-4}$), $f$ is nearly equal to unity and the flow is
well-described as a true ADAF. As $\dot{m}$ becomes larger, radiation
plays an increasingly important role and $f$ becomes smaller, even
negative in some regimes (LHAF). \S\ref{solutionenergetics} discusses
the energetics of various kinds of hot accretion flows.

Apart from being convenient for estimating gas properties in hot
accretion flows, the self-similar solution reveals several distinct
features of these solutions, which distinguish hot flows from the
standard (cool) thin disk.
\begin{itemize}
\item The temperature of a hot accretion flow is almost virial, \be
  T\simeq G M m_p/6kR\sim (10^{12}/r)\,{\rm K},\label{temperature}\ee
  which is much larger than the temperature of a thin disk.  Because
  of the near-virial temperature, the accretion flow is geometrically
  quite thick, $H/R\sim 0.5$.  Nevertheless, the height-integrated
  equations used in the 1D analysis appear to be reasonably accurate
  (Narayan \& Yi 1995a).
\item The radial velocity is much larger than in a thin disk. This is
  because accretion theory predicts $v\sim \alpha c_s H/R$ (e.g.,
  eq.~\ref{angulareq}), and both $c_s$ and $H/R$ are much larger in a
  hot accretion flow.
\item The angular velocity is sub-Keplerian. This is because the
  pressure is much larger than in a thin disk (higher temperature) and
  so gravity is partially balanced by the radial pressure gradient
  (right hand side of eq.~\ref{radialeq}).
\item The large radial velocity and the low mass accretion rate
  generally cause the optical depth of a hot accretion flow to be less
  than unity. Therefore, the emitted radiation is almost never
  blackbody, but is dominated by processes like synchrotron,
  bremsstrahlung and inverse Compton scattering. In addition, as we
  discuss in \S\ref{radiation}, the radiative efficiency,
\begin{equation}
\epsilon \equiv \frac{L}{\dot{M}_{\rm BH}c^2},
\label{eff}
\end{equation}
where $L$ is the luminosity of the accretion flow, is much lower than
the fiducial 10\% efficiency of a standard thin accretion disk,
especially when $\dot{m}$ is small.
\item In the low radiative efficiency limit, since the gas is heated
  but hardly cools, the entropy increases with decreasing radius.  Hot
  accretion flows are therefore potentially unstable to
  convection.\footnote{Rotation can stabilize a system against
    convection even if the entropy gradient is unstable. The role of
    magnetic fields is less clear (\S\ref{outflow}).}
\item Finally, the self-similar solution implies that the Bernoulli
  parameter $Be$ of the flow is positive, which suggests that hot
  accretion flows should have strong outflows and jets (Narayan \& Yi
  1994, 1995a; Blandford \& Begelman 1999). Global solutions
  (\S\ref{globalsolution}) indicate that $Be$ may be either positive
  or negative, depending on outer boundary conditions (Nakamura 1998,
  Yuan 1999).
\end{itemize}

While much work on hot accretion flows has focused on the time-steady
self-similar solution described above, Ogilvie (1999) has derived a
beautiful similarity solution which describes the radiatively
inefficient evolution of an initially narrow ring of viscous orbiting
fluid. This solution confirms several of the features discussed
above. In addition, it avoids an annoying singularity that is present
in the time-steady self-similar solution when the gas adiabatic index
approaches 5/3 (Narayan \& Yi 1994; Quataert \& Narayan 1999a;
Blandford \& Begelman 1999).

\subsection{Two-temperature flow: Thermal properties}
\label{thermal}

\subsubsection{The two-temperature scenario}
\label{2T}

In the discussion so far, we focused on the dynamics. When dealing
with the thermodynamics of a hot accretion flow, it is customary to
follow the pioneering work of SLE and to allow the ions and electrons
to have different temperatures. For such two-temperature plasmas, the
energy equation (\ref{energyeq}) or (\ref{energycompact}) is replaced
by two coupled equations (e.g., Nakamura et al.~1997, Quataert \&
Narayan 1999b): \be q^{\rm adv,i}\equiv \rho
v\left(\frac{de_i}{dR}-\frac{p_i}{\rho^2}\frac{d\rho}{dR}\right)
\equiv \rho v\frac{de_i}{dR}-q^{i,c}=(1-\delta) q^+ -q^{\rm
  ie},\label{ionseq}\ee \be q^{\rm adv,e}\equiv \rho
v\left(\frac{de_e}{dR}-\frac{p_e}{\rho^2}\frac{d\rho}{dR}\right)\equiv
\rho v\frac{de_e}{dR}-q^{e,c}= \delta q^++q^{\rm
  ie}-q^-.\label{electroneq}\ee Here $e_i\equiv
kT_i/[(\gamma_i-1)\mu_i m_{p}]$ and $e_e\equiv kT_e/[(\gamma_e-1)\mu_e
  m_{p}]$ are the internal energies of ions and electrons per unit
mass of the gas. Similarly, $\gamma_i$, $\gamma_e$ are the respective
adiabatic indices; $p_i$, $p_e$ are the respective pressures;
$q^{i,c}$, $q^{e,c}$ are the respective compression work done per unit
volume.  The quantity $q^{\rm ie}$ is the rate of transfer of thermal
energy from ions to electrons via Coulomb collisions.  The parameter
$\delta$ denotes the fraction of the viscously dissipated energy that
directly heats electrons; the remainder $(1-\delta)$ goes into the
ions. There have been attempts to estimate this important parameter
from first principles (\S\ref{heating}), but $\delta$ is often treated
as a free parameter. The above energy equations are further modified
when the contribution of the magnetic field is included (Quataert \&
Narayan 1999a), but we ignore this complication here.

It is important to note that the two-temperature nature of the gas in
a hot accretion flow is not simply an assumption but rather a generic
consequence of the physics of these solutions. First, electrons
radiate much more efficiently than ions (which is why we include a
cooling term $q^-$ only in eq.~\ref{electroneq}), and thus have a
tendency to be cooler. Second, the primary channel whereby ions cool
is by transfering their energy to the electrons. Coupling via Coulomb
collisions is inefficient at the low densities found in hot flows,
thus Coulomb equilibration of temperatures is suppressed. Third, we
see from the energy equation that gravitational energy is transformed
into thermal energy of the gas via two comparably important channels:
viscous heating ($q^+$) and compressional heating ($q^{i,c}$,
$q^{e,c}$).  As we discuss in \S\ref{heating},
 viscous heating probably deposits comparable amounts of energy in
  the ions and electrons, with electrons perhaps receiving a somewhat
  smaller share ($\delta\sim0.1-0.5$).  As for compressional heating,
under adiabatic conditions this causes the temperature to scale as $T
\propto \rho^{\gamma-1}$. Since the ions remain non-relativistic
throughout the accretion flow (even at $T_i\sim10^{12}$\,K), they have
$\gamma_i\sim5/3$.  However, in the inner regions of the accretion
flow, the electrons become relativistic, $kT_e > m_e c^2$, and
$\gamma_e\to 4/3$.  Therefore, while ions heat up by compression as
$T_i\sim\rho^{2/3}$, electrons heat up only as $T_e \sim
\rho^{1/3}$. This drives the gas to a two-temperature state at radii
$r\ \sles\ 10^3$.

Despite all these arguments, the gas would still be single-temperature
if there were efficient modes of energy transfer (over and above
Coulomb collisions) from ions to electrons.  Only one mechanism has
been discussed in the literature (Begelman \& Chiueh 1988), and it is
unclear how important this particular mechanism is in situations of
interest (Narayan \& Yi 1995b). Of course, plasmas are complicated and
there may well be some as-yet unidentified mechanism that succeeds in
maintaining the gas at a single temperature.  On the other hand, the
two-temperature nature of hot accretion flows seems to be supported by
observations (Yuan et al.~2006). Furthermore, the plasma in the solar
wind is found to be both two-temperature and anisotropic (Marsch
2012), and the plasma behind shocks in supernova remnants is also
two-temperature (Rakowski 2005). So it is certainly not the case that
nature abhors a two-temperature plasma.

\subsubsection{Heating and acceleration of electrons and ions}
\label{heating}

Early work on the two-temperature ADAF model assumed that most of the
turbulent viscous energy goes into the ions (Ichimaru 1977, Rees et
al.~1982, Narayan \& Yi 1995b), and that only a small fraction $\delta
< 10^{-2}$ goes into the electrons.  However, the existence of neither
a two-temperature plasma (\S\ref{2T}) nor a radiatively inefficient
flow (\S\ref{solutionenergetics}) requires such a small value of
$\delta$. What is essential is that $\dot{m}$ needs to be low.

A few attempts have been made to estimate $\delta$ from microphysics,
by considering magnetic reconnection (Bisnovatyi-Kogan \& Lovelace
1997; Quataert \& Gruzinov 1999; Ding et al.~2010; Hoshino 2012,
  2013), or MHD turbulence (Quataert 1998, Quataert \& Gruzinov 1999,
Blackman 1999, Medvedev 2000, Lehe et al.~2009), or dissipation of
pressure anisotropy in a collisionless plasma (Sharma et
al.~2007a). There is no consensus at the moment, but the work so far
generally suggests that $\delta \gg 10^{-2}$.

By modeling astrophysical observations of hot accretion flows, weak
constraints have been obtained on the value of $\delta$.  In the case
of Sgr A*, where we have perhaps the most detailed observations of a
hot accretion flow, Yuan et al.~(2003; see \S\ref{sgra} for details)
estimated $\delta\approx0.5$. However, from modeling black hole
sources at higher luminosities, it appears that $\delta\sim 0.1$ (Yu
et al.~2011, Liu \& Wu 2013).  The best we can say at the moment is
that $\delta$ probably lies in the range $0.1-0.5$.\footnote{This
  revision in the value of $\delta$ means that our understanding of
  hot accretion flows has evolved significantly since the early work
  reviewed in Narayan et al.~(1998b), which was based entirely on
  models with $\delta<0.01$. One consequence is that the radiative
  efficiency of a hot accretion flow is not as low as previously
  imagined, even when $\dot{m}$ is small
  (\S\ref{solutionenergetics}).}

Is the energy distribution of the hot electrons thermal or
non-thermal?  Obviously, this depends on the details of energy
dissipation, particle acceleration and thermalization.  Processes like
magnetic reconnection, weak shocks and turbulent dissipation are all
likely to accelerate a fraction of the ions and electrons into a
non-thermal power-law distribution (e.g., Ding et al. 2010, Hoshino
2013). How rapidly are the distributions then thermalized?

Mahadevan \& Quataert (1997) showed that Coulomb collisions are far
too inefficient to thermalize the ions, so ions retain whatever energy
distribution they acquire through viscous dissipation and
heating. Coulomb coupling between ions and electrons is also
inefficient (though less so), which is why hot accretion flows develop
a two-temperature structure in the first place (\S\ref{2T}). On the
other hand, electrons can exchange energy quite efficiently through
Coulomb collisions, as well as by the emission and absorption of
synchrotron photons. Thus, for accretion rates $\dot{m} > 10^{-3}$,
the electrons are expected to have a more or less thermal distribution
throughout the accretion flow. However, very high-energy electrons are
not easily thermalized and could, in principle, retain a power-law
distribution even at these high accretion rates. The electron energy
distribution may thus be Maxwellian for the bulk of the electrons, but
power-law for a small population of electrons at higher energies.

At lower accretion rates, thermalization is less efficient and the
electron distribution function is expected to retain a stronger memory
of the heating/acceleration process. So a hybrid thermal-nonthermal
energy distribution should form readily. Observationally, non-thermal
electrons are needed to explain the quiescent low-frequency radio
emission in Sgr A* (Mahadevan 1998, \"Ozel et al.~2000, Yuan et
al.~2003; see \S\ref{sgra}) and other low-luminosity AGNs (Liu \& Wu
2013), as well as the X-ray emission in flares in Sgr A* (Yuan et
al.~2004).

\subsection{Global solutions}
\label{globalsolution}

The great virtue of the self-similar solution presented in
\S\ref{self-similar} is that it is analytic and provides a transparent
way of understanding the key properties of an ADAF. However, since the
self-similar solution is scale-free, it cannot describe the flow near
the inner or outer boundary. Especially for calculating the radiation
spectrum one requires a global solution, since most of the radiation
comes from the region close to the inner boundary where the
self-similar solution is invalid.

A global solution refers to a numerical solution obtained by solving
directly the differential equations of the problem, e.g., equations
(\ref{masscons})--(\ref{energyeq}). Usually, an integrated version of
the angular momentum equation (\ref{angulareq}) is used, \be
\frac{d\Omega}{dR}=\frac{v\Omega_K(\Omega R^2-j)}{\alpha R^2c_s^2},\ee
where the integration constant $j$ is the angular momentum per unit
mass accreted by the central mass. This constant is an eigenvalue of
the problem and is obtained as part of the numerical solution.  If the
model under consideration includes mass loss in a wind, then (at the
simplest level) equation (\ref{masscons}) is simply replaced by
equation (\ref{MdotBH}) with the chosen value of $s$ (which is assumed
to be independent of radius\footnote{ Numerical simulations suggest
  that mass loss begins only at radii greater than ten or tens of
  $R_S$ (\S\ref{outflow}), so it is an oversimplification to assume a
  constant $s$ all the way down to $R_S$
  (eq.~\ref{MdotBH}). Presumably, the error introduced is not large,
  though this has not been checked.}). If one wishes to study the
thermodynamics of the two-temperature gas consistently, equation
(\ref{energyeq}) is replaced by equations (\ref{ionseq}) and
(\ref{electroneq}). If one wishes to go beyond the assumption of a
constant value of $f$, then the radiative cooling term $q^-$ is kept
in the energy equation, with contributions from relevant radiative
processes (\S\ref{radiation}). In many studies, a pseudo-Newtonian
gravitational potential
(see Paczy\'nski \& Wiita 1980) is adopted to mimic the effective
potential of a Schwarzschild black hole.

Mathematically, obtaining a global solution involves a three-point
boundary value problem. Since the radial velocity of the accreting gas
at large radius is highly subsonic, whereas the gas falls into the
black hole horizon at the speed of light, there has to be an
intermediate sonic radius $R_{\rm sonic}$ where the radial velocity
equals the sound speed.  The global solution must satisfy two boundary
conditions at this radius, one of which is $v=c_s$.  In addition,
since the black hole cannot support a shear stress, the viscous torque
must be zero at the horizon. This boundary condition is not always
applied at the horizon; sometimes it is transferred to the sonic
radius. Finally, at the outer edge of the solution ($R=R_{\rm out}$),
the flow should match the properties of the gas flowing in from the
outside.

The above boundary value problem is usually solved by one of two
numerical methods (see Press et al.~1992, 2002 for details):
relaxation (Narayan et al.~1997c, Chen et al.~1997, Esin et al.~1997),
or shooting (Nakamura et al.~1996, 1997; Manmoto et al.~1997; Yuan
1999, 2001; Yuan et al.~2003). The main parameters are: black hole
mass $M$, mass accretion rate $\dot{M}_{\rm BH}$, viscosity parameter
$\alpha$, magnetization parameter $\beta$, wind parameter $s$, and
electron heating parameter $\delta$. Among these, $M$ is usually known
through observations, $\delta$ was discussed in \S\ref{heating}, and
rough values of $\alpha$, $\beta$ and $s$ may be obtained from
numerical simulations, though $s$ in particular is somewhat
  uncertain (\S\ref{outflow}); $\dot{M}_{\rm BH}$ is a free parameter
which is either allowed to range over many values if one is doing a
parameter study (e.g., Fig.~\ref{fig:spectrum}) or is fitted to
observations such as the luminosity and spectrum of a source. The
global solution then gives the radial distributions of $v$, $\Omega$,
$c_s$, $\rho$, $T_i$, $T_e$ and $B$, together with the eigenvalue $j$
and the sonic radius $R_{\rm sonic}$. Away from the boundaries, global
solutions generally agree well with the self-similar solution (Narayan
et al.~1997c, Chen et al.~1997), confirming the validity and value of
the latter.

The relativistic global problem, where the Newtonian equations
discussed here are replaced by their general relativistic versions
corresponding to the Kerr metric, has been solved by several authors
(Abramowicz et al.~1996, Peitz \& Appl 1997, Gammie \& Popham 1998,
Popham \& Gammie 1998, Manmoto 2000).  Solutions of the relativistic
equations are similar to those of the Newtonian problem for radii
$R\ \sgreat\ 10R_S$, but differ significantly at smaller radii.  In
addition, the black hole's spin has a substantial effect at small
radii, and this can impact the observed spectrum (Jaroszynski \&
Kurpiewski 1997).

\subsection{Radiation processes, spectrum, and radiative efficiency}
\label{radiation}

Since gas close to the black hole in a hot accretion flow has a very
high temperature and is moreover optically thin and magnetized, the
relevant radiation processes are synchrotron emission and
bremsstrahlung, modified by Comptonization. The radiative cooling
rate, the shape of the spectrum, the different components in the
spectrum, and how all these scale with parameters, are described in
various papers (e.g., Narayan \& Yi 1995b; Narayan 1996; Mahadevan
1997; Esin et al.~1997; Nakamura et al.~1997; Manmoto et al.~1997;
Narayan et al.~1998b; Quataert \& Narayan 1999b; Yuan et al.~2003).

\begin{figure}
\hskip 0.0truein
\psfig{figure=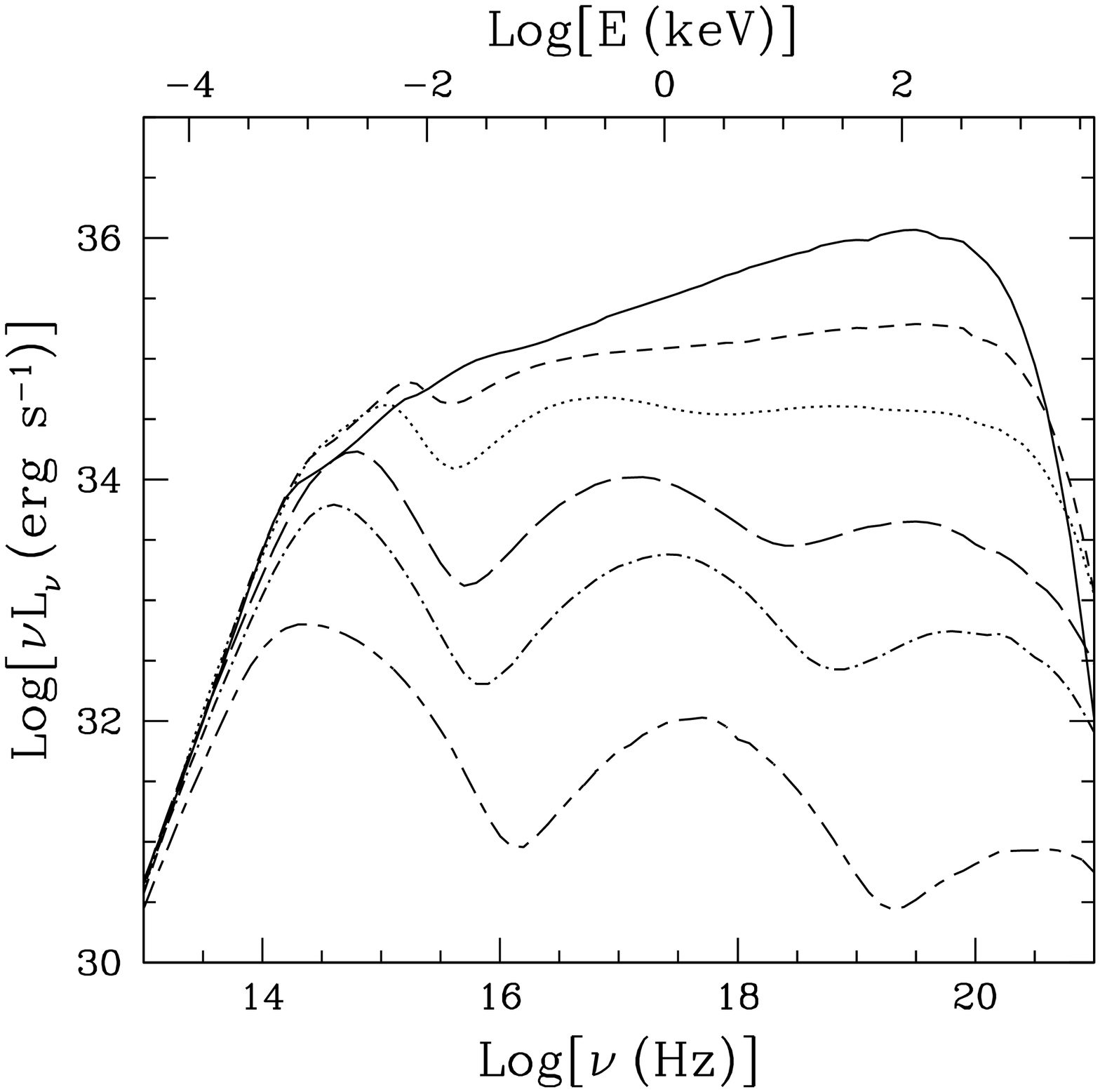,width=6.5cm,angle=0}
\hskip 0.0truein
\psfig{figure=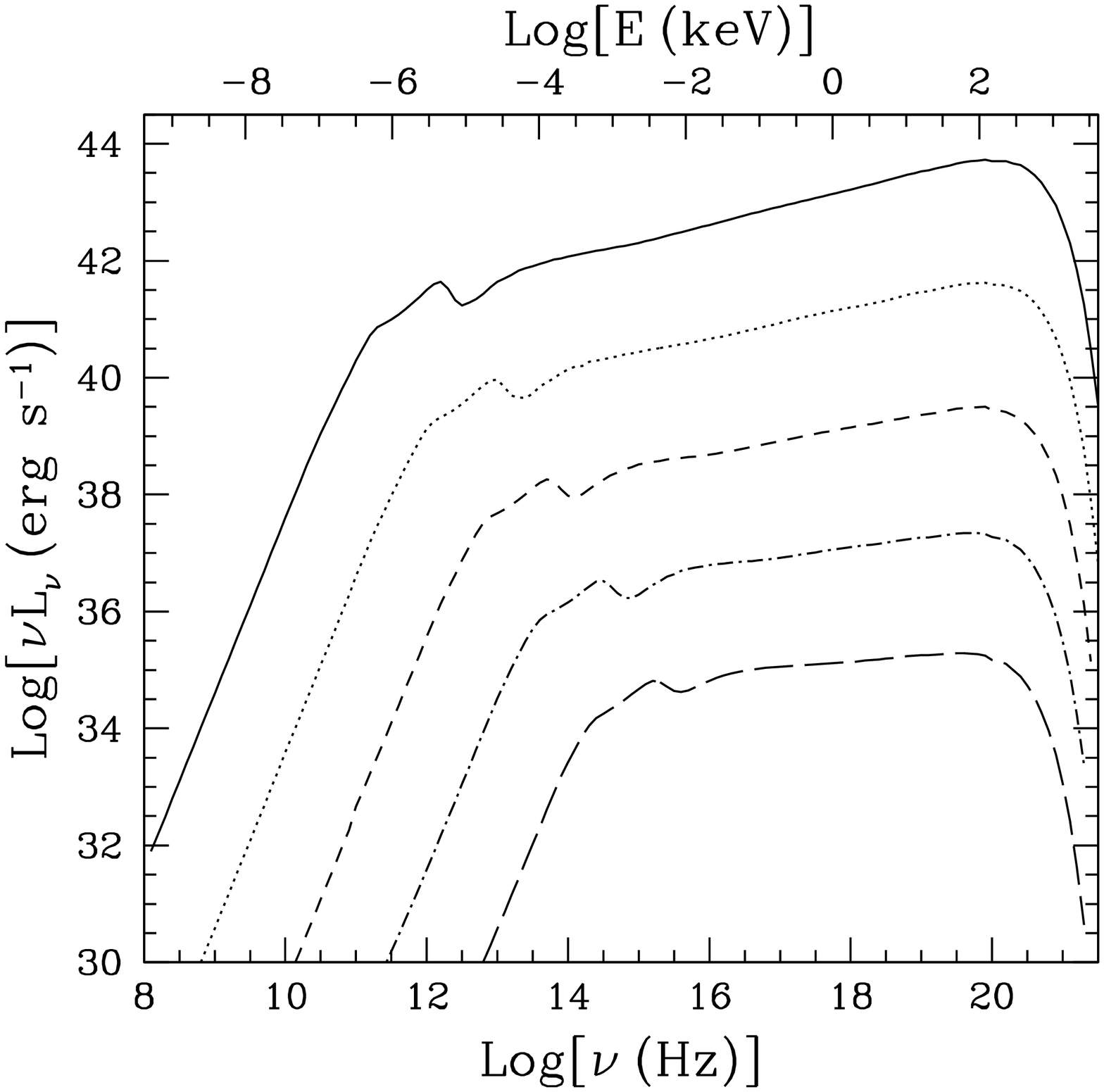,width=6.5cm,angle=0}
\caption{Model spectra of hot accretion flows for the following
  parameters: viscosity parameter $\alpha=0.1$, magnetization
  parameter $\beta=9$, electron heating parameter $\delta=0.5$, wind
  parameter $s=0.4$ (\S\S\ref{self-similar},\,\ref{thermal}).  {\em
    Left:} Spectra corresponding to a $10M_\odot$ black hole accreting
  with a mass accretion rate, from bottom to top, $\dot{m}_{\rm BH} =
  8\times 10^{-6}$, $5\times 10^{-5}$, $1.6\times 10^{-4}$, $8\times
  10^{-4}$, $2.4\times 10^{-3}$ and $5\times 10^{-3}$, respectively.
  {\em Right:} Spectra corresponding to $\dot{m}_{\rm BH}=2.4\times
  10^{-3}$ for black hole masses, from bottom to top, $M/M_\odot =10$,
  $10^3$, $10^5$, $10^7$ and $10^9$, respectively. The model spectra
  shown here are for hot thermal accretion flows. When there is a cool
  outer disk beyond a transition radius, the spectrum has an
  additional thermal blackbody-like component (see
  Fig.~\ref{Fig:1118}). If there is a jet with non-thermal electrons,
  or if the hot flow itself has non-thermal particles, there is
  enhanced emission at radio and infrared wavelengths, and the
  prominent inverse Compton bumps shown here at low mass accretion
  rates are smoothed out to some degree (Figs.~\ref{Fig:sgra},
  \ref{Fig:1118}). (Adapted from Narayan 1996, but with modern
  parameters.)}
\label{fig:spectrum}
\end{figure}

Figure \ref{fig:spectrum} shows model spectra for hot accretion flows
with different mass accretion rates $\dot{m}$ (left) and black hole
masses $m$ (right). The results can be understood as follows (based on
Mahadevan 1997). At photon energies below and up to the first peak in
the spectrum, the radiation is primarily due to synchrotron emission
from the thermal electrons. The emission is highly self-absorbed and
is very sensitive to the electron temperature ($\nu L_{\nu}\propto
T_e^7$). The emission at the peak comes from gas near the black hole,
while the radiation at lower frequencies comes from larger radii. The
peak frequency scales roughly as $\nu_{\rm peak} \propto
m^{-1/2}\dot{m}^{1/2}$. Synchrotron photons are Compton-upscattered by
the hot electrons and produce hard radiation extending up to about the
electron temperature: $kT_e\ \sgreat\ 100$\,keV for typical
two-temperature models. The importance of this Compton component
depends on $\dot{m}$. At high values of $\dot{m}$, it dominates the
spectrum, becoming even stronger than the primary synchrotron peak. As
$\dot{m}$ decreases, the Compton component is softer and becomes
weaker (bolometrically) than the synchrotron component. At a
sufficiently low $\dot{m}$, Comptonization is so weak that the X-ray
spectrum is dominated by bremsstrahlung emission, which again cuts off
at $h\nu \sim kT_e$.

The above discussion pertains to a pure hot accretion flow. If the hot
flow is surrounded by a standard thin disk at larger radii
(\S\ref{SSDADAFgeometry}), there will be an additional multicolor
blackbody component in the spectrum from the thermal disk.  Also, the
Compton component will be modified because, in addition to synchrotron
photons, there is a second source of soft photons from the outer
disk. The importance of the latter depends on where the transition
radius $R_{\rm tr}$ between the hot flow and the thin disk is located.

In addition to thermal radiation from hot electrons, proton-proton
collisions in a hot accretion flow can create pions, whose decay will
give gamma-rays (Mahadevan et al.~1997). The same collisions will also
produce a population of relativistic nonthermal electrons whose
synchrotron radiation might explain the excess radio emission observed
in Sgr A* (Mahadevan 1998, 1999),\footnote{This process is less
  important in current models, which use higher values of $\delta$
  than in the past and thus have lower mass accretion rates.} although
other processes can also produce such nonthermal electrons
(\S\ref{heating}).  Interestingly, although the electrons in a hot
accretion flow reach relativistic temperatures, pair processes are
generally unimportant (Bj\"ornsson et al.~1996; Kusunose \& Mineshige
1996; Esin 1999; Mo\'scibrodzka et al.~2011), since the low opacity
and low radiation energy density mean that there are very few
pair-producing interactions in the medium.

At radii $\ \sgreat\ 10^4R_S$, the gas in a hot accretion flow is cool
enough that heavier atomic species, especially iron-peak elements, are
able to retain one or two electrons. As a result, the X-ray emission
from these regions is expected to show emission lines on top of the
inverse Compton and bremsstrahlung continuum (Narayan \& Raymond
1999). The utility of these lines lies in their ability to constrain
the run of gas density with radius and to thereby provide an
observational estimate of the outflow parameter $s$ (Perna et
al.~2000, Xu et al.~2006, Wang et al.~2013).

Figure \ref{efficiency} shows the radiative efficiency of a hot
accretion flow as a function of the mass accretion rate for various
values of the electron heating parameter $\delta$.  Mass loss has been
included via eq.~(\ref{MdotBH}) with $s=0.4$.  As can be seen,
the efficiency depends strongly on the assumed value of
$\delta$. Also, for a given $\delta$, the efficiency increases steeply
with increasing mass accretion rate. Indeed, near the upper end, the
efficiency of a hot accretion flow approaches the efficiency
$\epsilon_{\rm SSD}\approx10\%$ of a standard Shakura-Sunyaev
disk. Xie \& Yuan (2012) give piecewise power-law fitting formulae for
the dependence of the radiative efficiency on $\dot{m}$ and $\delta$.

\subsection{Energetics: eADAF, ADAF, LHAF, and beyond}
\label{solutionenergetics}

We now consider the energy equation of a hot accretion flow and
discuss the role of the various terms that appear in it: viscous
heating, compressional heating, energy advection, Coulomb energy
transfer, radiative cooling. For simplicity, we begin with the simple
energy equation (\ref{energycompact}), which corresponds to a
single-temperature flow.

When $\dot{M}_{\rm BH}$ is very low, the gas density $\rho$ is also
low, and the radiative cooling rate $q^-$ (which decreases rapidly
with decreasing $\rho$) becomes negligibly small
(\S\ref{radiation}). The viscous heating rate is then balanced
primarily by energy advection rather than cooling.  Hence we have \be
q^+\approx q^{\rm adv}\gg q^-, \quad f\approx 1.\ee That is, most of
the viscous heat energy is stored in the flow and advected into the
black hole rather than being radiated away. This is the classic regime
of an ADAF. In the terminology used in this field, advection plays a
``cooling'' role.

\begin{figure}	
\hspace{1cm}
\psfig{figure=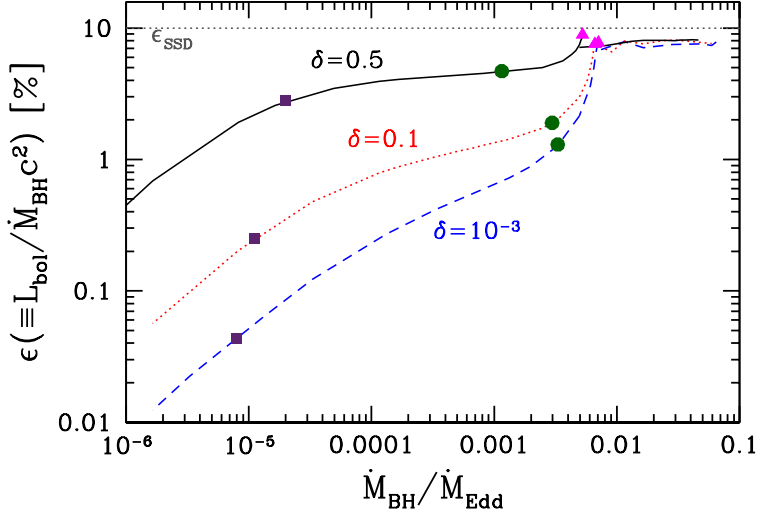,width=10.cm}
\caption{Radiative efficiency (eq.~\ref{eff}) of a hot accretion flow
  as a function of the mass accretion rate at the black hole
  $\dot{M}_{\rm BH}$ for three values of the electron heating
  parameter $\delta$.  Model parameters: $\alpha=0.1$, $\beta=9$,
  $s=0.4$. The nominal radiative efficiency of a standard thin disk,
  $\epsilon_{\rm SSD}= 10\%$, is indicated by the horizontal dotted
  line at the top. When $\delta$ is large, the efficiency of a hot
  accretion flow is within a factor of a few of $\epsilon_{\rm SSD}$
  for a wide range of $\dot{M}_{\rm BH}$ down to
  $\sim10^{-5}\dot{M}_{\rm Edd}$. In contrast, when $\delta$ is small,
  the efficiency drops precipitously for $\dot{M}_{\rm
    BH}\ \sles\ 10^{-2}\dot{M}_{\rm Edd}$. Squares, filled circles and
  triangles indicate $\dot{M}_{\rm eADAF}$, $\dot{M}_{\rm crit,ADAF}$
  and $\dot{M}_{\rm crit,LHAF}$, respectively, for each value of
  $\delta$ (\S\ref{solutionenergetics} defines these quantities).  The
  horizontal extensions of the curves above $\sim 7\times
  10^{-3}\dot{M}_{\rm Edd}$ show approximate radiative efficiencies
  assuming a two-phase accretion flow.  (Adapted from Xie \& Yuan
  2012.)}
\label{efficiency}
\vspace{-0.0cm}
\end{figure}

With increasing $\dot{M}_{\rm BH}$, the radiative cooling $q^-$
increases faster than $q^{\rm adv}$, and thus advective cooling
becomes progressively less dominant. At a critical accretion rate
$\dot{M}_{\rm crit,ADAF}$, the condition, \be q^+ =q^-, \quad
f\approx0,\label{criticalADAF0}\ee is satisfied. An ADAF is allowed
only for $\dot{M}_{\rm BH} \le \dot{M}_{\rm crit,ADAF}$.

What happens when $\dot{M}_{\rm BH}>\dot{M}_{\rm crit,ADAF}$? Clearly
we will have $q^+<q^-$, i.e., radiative cooling will be stronger than
the rate of heating by viscosity ($f<0$). Yuan (2001) showed that hot
accretion flows are still permitted in this regime up to a second
critical accretion rate $\dot{M}_{\rm crit,LHAF}$, which is determined
by the condition \be q^c+q^+=q^-. \label{criticalLHAF0}\ee Solutions
over the range $\dot{M}_{\rm crit,ADAF}<\dot{M}_{\rm BH}<\dot{M}_{\rm
  crit,LHAF}$ are called luminous hot accretion flows (LHAFs) --- they
are hot, but unlike ADAFs, they are radiatively efficient and
luminous. The gas in these solutions remains hot despite the strong
cooling because of the action of compressional heating $q^{c}$. Even
though the entropy of the gas decreases with decreasing radius, the
quantity $\rho v (de/dR)=q^+ + q^{c}-q^-$ is still positive. Thus the
gas temperature continues to increase inward, and the flow remains hot
(provided it starts out hot at a large radius). Over the entire LHAF
branch, we have \be q^{c}+q^+ > q^- > q^+. \ee Thus, $q^{\rm
  adv}=q^+-q^-<0$ and $f<0$, so energy advection plays a ``heating''
role.  In other words, the extra energy to heat the gas is supplied,
not by viscous dissipation, but by the entropy already stored in the
gas at large radius. Because of the high radiative efficiency and
relatively large $\dot{M}_{\rm BH}$, LHAFs are expected to be much
more luminous than ADAFs.

Consider now the more realistic case of a two-temperature hot
accretion flow, where eq.~(\ref{energycompact}) is replaced by
eqs. (\ref{ionseq}) and (\ref{electroneq}). In the early literature on
ADAFs, this case was treated in an approximate fashion by considering
only the energy equation (\ref{ionseq}) of the ions.  The neglect of
the electron energy equation (\ref{electroneq}) is valid whenever
$\delta$ is small, as was the case in these early studies which
assumed $\delta \approx 0-0.01$.  In this limit, almost all of the
viscous heat goes into the ions. Moreover, the critical bottleneck
that prevents gas from radiating is the rate of transfer of energy
from ions to electrons, $q^{\rm ie}$; whenever $q^{\rm ie}$ is
substantial (as happens at larger values of $\dot{M}_{\rm BH}$), the electrons
have no trouble radiating whatever energy they receive from the ions,
i.e., $q^{\rm ie}\approx q^-$. Thus, the approximation is
self-consistent, though it does require very small values of $\delta$.
Numerically, in this regime it is found that $\dot{M}_{\rm
  crit,ADAF}\approx 0.4\alpha^2 \dot{M}_{\rm Edd}$ and $\dot{M}_{\rm
  crit,LHAF}\approx \alpha^2 \dot{M}_{\rm Edd}$ (Narayan 1996, Esin et
al.~1997, Yuan 2001)\footnote{Another way to define $\dot{M}_{\rm
    crit,ADAF}$ is to require the timescale for ion-electron
  equilibration via Coulomb collisions to be equal to the accretion
  timescale (Narayan et al.~1998b). The two definitions are physically
  slightly different but give approximately the same result.}.

As discussed in \S\ref{heating}, the current consensus is that hot
accretion flows have a larger value of $\delta \sim 0.1-0.5$.  Viscous
heating of electrons is then no longer negligible, nor is the Coulomb
energy transfer rate $q^{\rm ie}$ the sole bottleneck. Consequently,
it is now necessary to consider both the ion and electron energy
equations (see Nakamura et al.~1997; Mahadevan \& Quataert 1997; Yuan
2001 for early works on electron advection). As before, the two
critical mass accretion rates, $\dot{M}_{\rm crit,ADAF}$ and
$\dot{M}_{\rm crit,LHAF}$, may still be defined by the conditions
given in equations (\ref{criticalADAF0}) and
(\ref{criticalLHAF0}).\footnote{Xie \& Yuan (2012) adopt
  $(1-\delta)q^+ = q^{\rm ie}$ to define $\dot{M}_{\rm
    crit,ADAF}$. But the definition (\ref{criticalLHAF0}) is more
  physical. There is little difference in the numerical results.}
However, a third critical accretion rate, $\dot{M}_{\rm eADAF}$,
appears, which is explained below. By computing global models for
$\delta$ in the range $0.1-0.5$, Xie \& Yuan (2012) obtain the
following rough estimates for the three critical accretion rates
(measured at the black hole),\footnote{Two caveats should be
  mentioned. First, all quantities such as $q^{\rm adv}$, $q^c$,
  $q^+$, $q^-$ are functions of radius. So we should, in principle,
  define a ``local'' critical accretion rate as a function of radius
  (e.g., see the review of Narayan et al.~1998b). Here we adopt a
  simpler and more ``global'' definition where we check if the
  relevant condition is satisfied at any radius within the range of
  interest. For example, we call a solution an LHAF whenever the
  condition $q^+<q^-$ is satisfied at any radius. Second, the results
  quoted here are from Xie \& Yuan (2012), who assume $s=0.4$ and
  $\beta=9$. The results are likely to change
  for other choices of the parameters.}
\begin{eqnarray}
\dot{M}_{\rm eADAF} &\approx& 0.001\,\alpha^2\dot{M}_{\rm
  Edd}, \label{eADAF} \\
\dot{M}_{\rm crit,ADAF} &\approx& (0.1-0.3)\,\alpha^2\dot{M}_{\rm
  Edd}, \label{criticalADAF} \\
\dot{M}_{\rm crit,LHAF} &\approx& (0.06-0.08)\,\alpha \dot{M}_{\rm
  Edd}. \label{criticalLHAF}
\end{eqnarray}
In Figure \ref{efficiency}, these critical rates are indicated by
squares, filled circles, and triangles, respectively.  Note that
$\dot{M}_{\rm crit,ADAF}$ is smaller than the value ($0.4\alpha^2
\dot{M}_{\rm Edd}$) mentioned above for $\delta\approx 0-0.01$. This
is because electrons now receive more energy directly by viscous
heating and hence radiate more efficiently.

The three critical mass accretion rates listed above separate
different regimes of hot accretion as follows:

\begin{itemize}
\item $\dot{M}_{\rm BH}<\dot{M}_{\rm eADAF}$: Here both ions and
  electrons are radiatively inefficient. In particular, the electrons
  are unable to radiate either the viscous heat they acquire directly
  ($\delta q^+$) or the small amount of energy they receive from ions
  via Coulomb collisions ($q^{\rm ie}$). Systems in this regime are
  truly radiatively inefficient since even the electrons are
  advection-dominated; we call this regime ``electron ADAF'' or
  eADAF. These systems correspond to the dimmest black hole accretion
  sources known, e.g., Sgr A* at the Galactic Center and quiescent
  BHBs.

\item $\dot{M}_{\rm eADAF} < \dot{M}_{\rm BH} < \dot{M}_{\rm
  crit,ADAF}$: Here electrons radiate efficiently their own viscous
  energy ($\delta q^+$) as well as any energy they receive from the
  ions ($q^{\rm ie}$). However, Coulomb collisions are inefficient. So
  the ions transfer only a small fraction of their energy to the
  electrons, and therefore remain advection-dominated. Systems in this
  regime are expected to be fairly radiatively efficient, with an
  efficiency of the order of a percent or more (depending on the value
  of $\delta$). They are thus substantially brighter than classic ADAF
  models (see Fig.~\ref{efficiency}). However, the flows are hot and
  geometrically thick, and are still ADAFs in the sense that
  $q^+>q^-$.

\item $\dot{M}_{\rm crit,ADAF} <\dot{M}_{\rm BH} < \dot{M}_{\rm
  crit,LHAF}$: Here we have an LHAF with $q^+<q^-$. All the energy
  terms in the ion energy equation are roughly comparable in
  magnitude.  The entropy decreases as the gas flows in, but the
  gas remains hot because of compressional heating.  The
  radiative efficiency increases rapidly with increasing $\dot{M}_{\rm
    BH}$, { as shown in Fig.~\ref{efficiency}. This plot further
    shows that, when $\delta\ll1$, the LHAF branch is restricted to
    quite a narrow range of $\dot{M}_{\rm BH}$. However, for currently
    accepted values of $\delta\sim0.1-0.5$, the LHAF solution extends
    over a factor of several in $\dot{M}_{\rm BH}$.}

\item $\dot{M}_{\rm BH}>\dot{M}_{\rm crit,LHAF}$: In this regime, the
  one-dimensional global equations have no hot accretion flow
  solution.  Radiative cooling is too strong, and even compressional
  heating is insufficient to keep the gas hot.  In the traditional
  view, the accretion flow transitions to a standard thin accretion
  disk. However, there are large uncertainties, and Yuan (2003)
  speculates that the gas may transition to a two-phase medium with
  cold dense clumps embedded in hot gas. Alternatively, Oda et
  al.~(2010) propose that a magnetically dominated accretion flow may
  form (see \S\ref{mad}).

\end{itemize}

\subsection{Stability and relationship to other accretion solutions}
\label{solutionsummary}

Hot optically thin gas generally tends to be thermally unstable.
Therefore, all hot accretion flows are potentially unstable.  What
saves them is the fact that the accretion time scale is shorter than
the instability growth time.

Narayan \& Yi (1995b) and Abramowicz et al.~(1995) showed that ADAFs
are stable to long wavelength perturbations. For small scale
perturbations, however, the results are somewhat subtle. Wu \& Li
(1996) and Wu (1997) showed that ADAFs are stable under most reasonable
conditions, while Kato et al.~(1996, 1997) showed that ADAFs are
potentially unstable at short wavelenghts. Using a time-dependent
analysis, Manmoto et al.~(1996) showed that small scale density
perturbations in a one-temperature ADAF grow as the gas flows in, but
not sufficiently quickly to affect the global viability of the
solution. All these results were derived assuming that advection
dominates. Hence they do not apply to the SLE or LHAF solutions.

\begin{figure}
\hskip -0.01cm
\vspace{0.01cm}
\hskip 1.truein
\psfig{figure=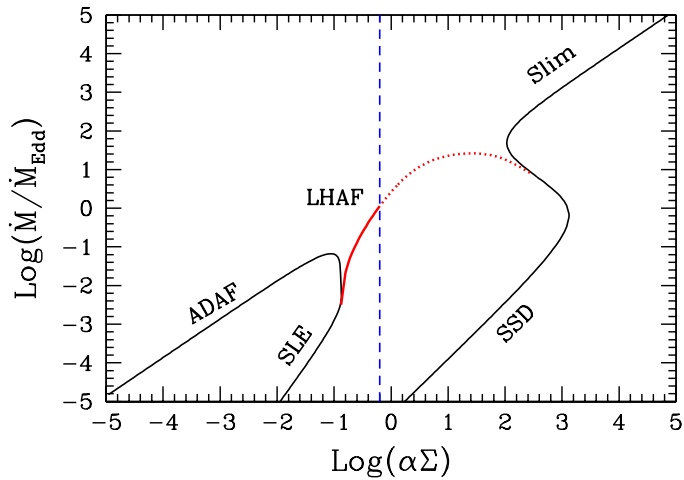,width=9cm,angle=0}
\caption{Thermal equilibrium curves of various accretion solutions for
  the following parameter values: $M=10M_\odot$, $\alpha=0.1$, $r=5$.
  The accretion rate is normalized to $\dot{M}_{\rm Edd}$ and the
  surface density is in units of ${\rm g\,cm^{-2}}$. The black solid
  lines correspond to the classic solution branches, viz., the hot
  branch consisting of ADAF and SLE, and the cold branch consisting of
  Slim disk and SSD. The blue vertical dashed line separates optically
  thin solutions on the left from optically thick solutions on the
  right. The red line corresponds to the LHAF solution. While the LHAF
  branch appears to go all the way across from the hot to the cold
  branch, global models indicate that this solution is self-consistent
  only to the left of the vertical line (shown by the solid red
  segment).  (Adapted from Yuan 2003; see Abramowicz et al.~1995, Chen
  et al.~1995, for similar plots without the LHAF solution branch.)}
\label{fig:solutionsummary}
\end{figure}

In the case of the SLE solution, Piran (1978) showed that the
  model is thermally unstable.  Yuan (2003) studied the thermal
stability of LHAFs and concluded that these flows are thermally
unstable. However, if the accretion rate is below $\dot{M}_{\rm
  crit,LHAF}$, the growth timescale of the instability remains longer
than the accretion timescale and the solution can survive. Above this
accretion rate, however, the instability will grow quickly.  It is
possible that the instability will not destroy the solution but will
lead instead to a two-phase medium in which cold dense blobs are
intermixed with hot gas (\S\ref{solutionenergetics}). Dynamically, the
hot phase would behave like an LHAF with radiative cooling stronger
than viscous heating. Xie \& Yuan (2012) estimated the luminosity of
such a two-phase accretion flow
and found that, very approximately, the radiative efficiency is
expected to be around 10\%, as indicated by the horizontal extensions
in Fig.~\ref{efficiency}.

The above discussion deals with thermal stability. What about viscous
stability? The most convenient way to investigate this is by plotting
the locus of accretion solutions in the two-dimensional plane of
accretion rate $\dot{M}$ and surface density $\Sigma\equiv 2\rho H$
(e.g., Frank et al.~2002). If the solution track has a
positive slope, the solution is viscously stable, and vice versa.  By
including all solutions (both hot and cold) in such a diagram, one can
appreciate the relationship between the various solutions.

Figure \ref{fig:solutionsummary} shows an example of such a plot,
taken from Yuan (2003; see also Abramowicz et al.~1995, Chen et
al.~1995). The various solution tracks shown have been obtained by
solving simple equations such as (\ref{masscons})--(\ref{energyeq}),
or more complex versions of these that include radiative transfer,
two-temperature plasma, etc.  Usually, approximations are needed,
e.g., assuming $\Omega=\Omega_K$ and $q^{\rm adv}=\xi (\dot{M}
c_s^2/2\pi R^2 H)$ with a constant $\xi$ (see Abramowicz et al.~1995).
Then, for a given set of parameters, $\alpha$, $M$, $R$, $\xi$, one
can solve for $\dot{M}$ as a function of $\Sigma$.

The black solid lines in Fig.~\ref{fig:solutionsummary} show all the
standard solution branches: the ADAF and SLE solutions belong to the
sequence of hot solutions on the left, and the Shakura-Sunyaev disk
(SSD) and slim disk solutions belong to the sequence of cold solutions
on the right. All four of these solution branches are viscously stable
since each track has a positive slope. At a given $\Sigma$, if there
are multiple solutions, the uppermost (highest $\dot{M}$) solution is
thermally stable to long wavelength perturbations, the next one below
is unstable, and the next is stable. Therefore, the ADAF, slim disk
and SSD solutions are thermally stable. However, the SLE solution on
the left is thermally unstable, as is the segment between the SSD and
slim disk on the right.\footnote{Recently, there has been considerable
  interest in the thermal stability of this solution branch, which
  corresponds to a radiation pressure dominated thin disk. Using
  numerical radiation MHD simulations, Hirose et al.~(2009) concluded
  that a thin disk in this regime is thermally stable. However, later
  work by Jiang et al.~(2013) showed that the disk is, in fact,
  thermally unstable. The reason for the discrepancy is discussed by
  the latter authors. Viscous stability of this solution branch is yet
  to be investigated via numerical simulations.}  The latter solution
is also viscously unstable (Lightman \& Eardley 1974) since this
branch has a negative slope. In terms of the advection parameter $f$,
we have $f\approx 0$ for SSD and SLE, and $f\approx 1$ for ADAF and
slim disk. The red line in Fig.~\ref{fig:solutionsummary}
  corresponds to the LHAF solution. Here $\xi$ is negative, advection
  plays a heating role (\S\ref{solutionenergetics}) and $f<0$.

The results presented in Fig.~\ref{fig:solutionsummary} are
approximate since they are based on a single-temperature model
(however, see Fig.~1c in Chen et al.~1995 for equivalent results for a
two-temperature model) and are, moreover, based on a local rather than
a global analysis.  Nevertheless, these plots are believed to be
qualitatively correct and therefore raise an important question. For
certain ranges of the mass accretion rate, both hot and cold solutions
are available, and both are thermally stable.  Which solution does
nature pick?  Narayan \& Yi (1995b) discussed a number of options, of
which the following two deserve mention.

One option is that, if the accreting gas is hot at the outer feeding
radius where mass first enters the accretion disk, and if a hot
accretion solution is permitted at that radius, then the gas will
start off in the hot mode of accretion and will remain in the hot
accretion state all the way down to the black hole. On the other hand, if the gas starts out on the
cool SSD branch on the outside, then it will remain in that branch
down to the black hole unless the disk enters the Lightman \& Eardley
(1974) viscous instability zone (where the gas would become a hot
accretion flow, the only stable solution remaining).

The second, more revolutionary, option is that the accretion flow will
switch to the hot accretion branch whenever the latter solution is
allowed, i.e., whenever $\dot{M}<\dot{M}_{\rm crit,LHAF}(R)$
corresponding to the local radius $R$. In other words, accretion
occurs via the SSD solution only if it is the sole stable solution
available.  This so-called ``strong ADAF principle'' appears to be
generally consistent with observations (\S\ref{SSDADAFgeometry}).

\section{Numerical simulations}
\label{simulation}

The one-dimensional solutions considered so far are easy to calculate
and often capture the important physics. However, hot accretion flows
are geometrically thick, so one cannot be sure that the vertically
integrated equations from which 1D solutions are derived are valid. In
particular, height-integration eliminates multidimensional structures
such as outflows.

Analytical two-dimensional solutions have been obtained by a number of
authors over the years (e.g., Begelman \& Meier 1982, Narayan \& Yi
1995b, Xu \& Chen 1997, Blandford \& Begelman 2004, Xue \& Wang 2005,
Tanaka \& Menou 2006, Jiao \& Wu 2011, Begelman 2012). However, these
models make simplifying assumptions such as self-similarity, and
therefore have limited applicability.  If we wish to understand the
multidimensional structure of hot accretion flows, numerical
simulations are the only way.

\subsection{Hydrodynamic simulations}
\label{hydrosimulation}
Although it is known that angular momentum transport in hot accretion
flows is via magnetohydrodynamic (MHD) turbulence driven by the
magnetorotational instability (\S\ref{shearingbox}), early computer
simulations were carried out without magnetic fields, using 2D
hydrodynamic (HD) codes and an $\alpha$-like prescription for the
viscous stress (Igumenshchev et al.~1996, Igumenshchev \& Abramowicz
1999, Stone et al.~1999, Igumenshchev \& Abramowicz 2000; Igumenshchev
et al.~2000, De Villiers \& Hawley 2002; Fragile \& Anninos
2005). There were some differences in the adopted form of the shear
stress, e.g., Stone et al.~(1999) assumed that only the azimuthal
component of the shear stress tensor is present, while Igumenshchev \&
Abramowicz (1999) included also poloidal components.

HD simulations of hot accretion flows reveal rich and complicated
time-dependent structures. In particular, there are convective motions
(Igumenshchev \& Abramowicz 1999, Stone et al.~1999, Igumenshchev et
al.~2000), confirming an early prediction of Narayan \& Yi (1994;
\S\ref{self-similar}). The level of convective turbulence depends on
details; for example, convection becomes weaker if a larger value of
$\alpha$ is used or if poloidal components of the shear stress are
included (Igumenshchev \& Abramowicz 1999, Stone et al.~1999, Yuan \&
Bu 2010).  The radial dynamic range of simulations is usually fairly
limited (even more so for the 3D MHD simulations discussed below), but
Yuan et al.~(2012b) recently achieved an unprecedented four decades of
dynamic range in a HD simulation using a ``two-zone'' approach.

The time-averaged ``steady-state'' flow in HD simulations is usually
well described by a radial power law distribution of various
quantities, with the power law indices depending on the specific form
of the adopted shear stress.  For the usual Shakura \& Sunayev
$\alpha$-prescription, the radial scalings are consistent with the
self-similar solution (eqs.~\ref{radialvelocity}--\ref{pressure}, see
Stone et al.~1999). On the other hand, the initial conditions used in
the simulations appear to affect some results such as the streamline
structure and the Bernoulli parameter (Yuan et al.~2012a).

\subsection{Magnetohydrodynamic simulations}
\label{MHDsimulation}

\subsubsection{Magnetorotational instability (MRI)}
\label{shearingbox}

It is now widely accepted that the mechanism of angular momentum
transport in ionized accretion flows is the magnetorotational
instability (MRI; Balbus \& Hawley 1991, 1998). This instability takes
a seed magnetic field in the accreting gas and amplifies it
exponentially, until the system becomes nonlinear and develops MHD
turbulence.  The Maxwell and Reynolds stresses in the turbulent state
transport angular momentum outward, causing gas to accrete inward.

While the basic MRI is a linear instability and can be understood
analytically, the nonlinear turbulent state relevant for disk
accretion can be studied only with numerical simulations. A number of
codes have been used, notably, ZEUS (Stone \& Norman 1992a, 1992b),
HARM (Gammie et al.~2003), the GRMHD code of De Villiers \& Hawley
(2003a), COSMOS++ (Anninos et al.~2005), and ATHENA (Stone et
al.~2008). Many studies have been done in the limit of a local
shearing box, which permits high spatial resolution (e.g., Hawley \&
Balbus 1991; Hawley et al.~1995, 1996; Brandenburg et al.~1995;
Matsumoto \& Tajima 1995; Stone et al.~1996). These show that MHD
turbulence is inevitable so long as the gas and the magnetic field are
well-coupled, and that the Maxwell stress dominates over the Reynolds
stress by a factor of several.

When data from different published 3D shearing box simulations are
combined, a tight correlation is seen between the parameter $\beta$
(eq.~\ref{betadefinition}) and the viscosity parameter $\alpha$, viz.,
$\alpha \beta\sim 0.5$ (Blackman et al.~2008, Guan et al.~2009,
Sorathia et al.~2012, see also Hawley \& Balbus 1996).  However, the
individual values of $\beta$ and $\alpha$ vary substantially from one
numerical experiment to the next; for example, Hawley et al. (2011) obtained
$\beta$ values in the range $10-200 $, corresponding to $\alpha\sim
0.01-0.003$, and other authors have found even larger variations. The
value of $\alpha$ is thus not constrained. It seems to depend on the
magnitude of the net initial magnetic field (Hawley et al.~1995, 1996;
Pessah et al.~2007), a dependence that is confirmed also in localized
regions of global simulations (Sorathia et al.~2010). Some shearing
box simulations even find $\alpha$ values larger than unity (e.g., Bai
\& Stone 2013).  Numerical resolution also plays a role. Generally,
better resolution gives a larger $\alpha$ (up to some saturation
value). However, when the net magnetic flux is zero, increasing the
resolution actually causes $\alpha$ to decrease (Fromang \& Papaloizou
2007, Fromang et al.~2007), with $\alpha$ going to zero in the limit
of infinite resolution.

Interestingly, the uncertainty in the value of $\alpha$ is eliminated
if shearing box simulations include vertical stratification to mimic
the effect of vertical disk gravity (Davis et al.~2010, Bai \& Stone 2013). Perhaps
because of this, global disk simulations, which automatically include
vertical gravity, show less variations in the effective value of
$\alpha$. These simulations generally evolve to steady state with
$\alpha\sim 0.05-0.2$ (Hawley \& Balbus 2002, Penna et al.~2013b).

\subsubsection{Global simulations: General results}
\label{globalsimulation}

Global MHD simulations of hot accretion flows are more realistic than
global HD simulations as they self-consistently generate shear stress
through MRI-induced MHD turbulence, whereas HD simulations must
include an ad hoc viscosity. Early pioneers in global MHD simulations
include Matsumoto \& Shibata (1997), Armitage (1998), Hawley (2000),
Machida et al.~(2000), Stone \& Pringle (2001), Hawley \& Krolik
(2001), and Igumenshchev \& Narayan (2002).

Compared to shearing box simulations, global simulations enable the
MRI to sample much larger radial and azimuthal wavelengths.  The
largest radial dynamic ranges are achieved in 2D\footnote{With current
  computer resources, 3D simulations can reach inflow equilibrium over
  at best only about 2 orders of magnitude in radius (e.g., Pang et
  al.~2011, McKinney et al.~2012, Narayan et al.~2012b, Sadowski et
  al.~2013a), a substantial part of which is dominated by either inner
  or outer boundary conditions. 2D simulations can achieve a factor of
  several larger dynamic range.}. However, 2D simulations do not treat
the MRI accurately because of Cowling's antidynamo theorem (Cowling
1933), which limits the growth of the poloidal magnetic field and
causes turbulence to die away.
Thus, no steady accretion is possible in 2D and one has to carefully
select a period of time after the disk has become turbulent but before
the turbulence dies out. There has been no systematic study of how
well the properties of this intermediate period in 2D simulations
agree with those of 3D simulations with sustained
turbulence. Qualitatively, it appears that the differences are not
large.

Hawley (2000) compared the results of local shearing box and global
simulations using two different initial configurations of the magnetic
field: toroidal and vertical. In terms of growth of the MRI and
transition to MHD turbulence, he found global and shearing box
simulations to behave similarly.  The magnetic shear stress
$T_{R\phi}$ is directly proportional to the magnetic pressure,
$2\langle B_RB_{\phi}\rangle \approx (0.4-0.5)\langle B^2\rangle$;
this is equivalent to $\alpha\beta\approx 0.5$ mentioned earlier.
Depending on the value of $\beta$, the resulting $\alpha\approx
0.05-0.2$ in the interior of the disk (e.g., Penna et al.~2013b). The
Maxwell (magnetic) stress is always larger than the Reynolds stress by
a factor of several. Also, the toroidal component of the field is
significantly larger than the radial component, which is itself
somewhat larger than the vertical component. While most global
simulations start with a weak magnetic field (initial $\beta\sim100$),
Machida et al.~(2000) used a strong initial toroidal field with
$\beta=1$. There was no MRI in their simulation, but they found the
Parker instability, which led to the formation of a magnetized corona.

Global 3D MHD simulations have been run by various groups, and the
results are fairly similar.  Early work assumed Newtonian dynamics and
modeled the black hole at the center via a pseudo-Newtonian potential
(Armitage 1998; Hawley 2000; Machida et al.~2000, 2004; Hawley \&
Krolik 2001; De Villiers \& Hawley 2003b; Igumenshchev et al.~2003).
General relativistic MHD (GRMHD) codes were later developed (Koide et
al.~1999, Gammie et al.~2003, De Villiers \& Hawley 2003a, Fragile et
al.~2007). Much of the recent work in this field is based on the
latter codes, which provide a more realistic description of phenomena
close to the black hole. Nevertheless, pseudo-Newtonian simulations
are still useful for studying large scale properties of the accretion
flow.

\begin{landscape}
\begin{figure*}
\vskip -2.5cm
\centerline{\psfig{figure=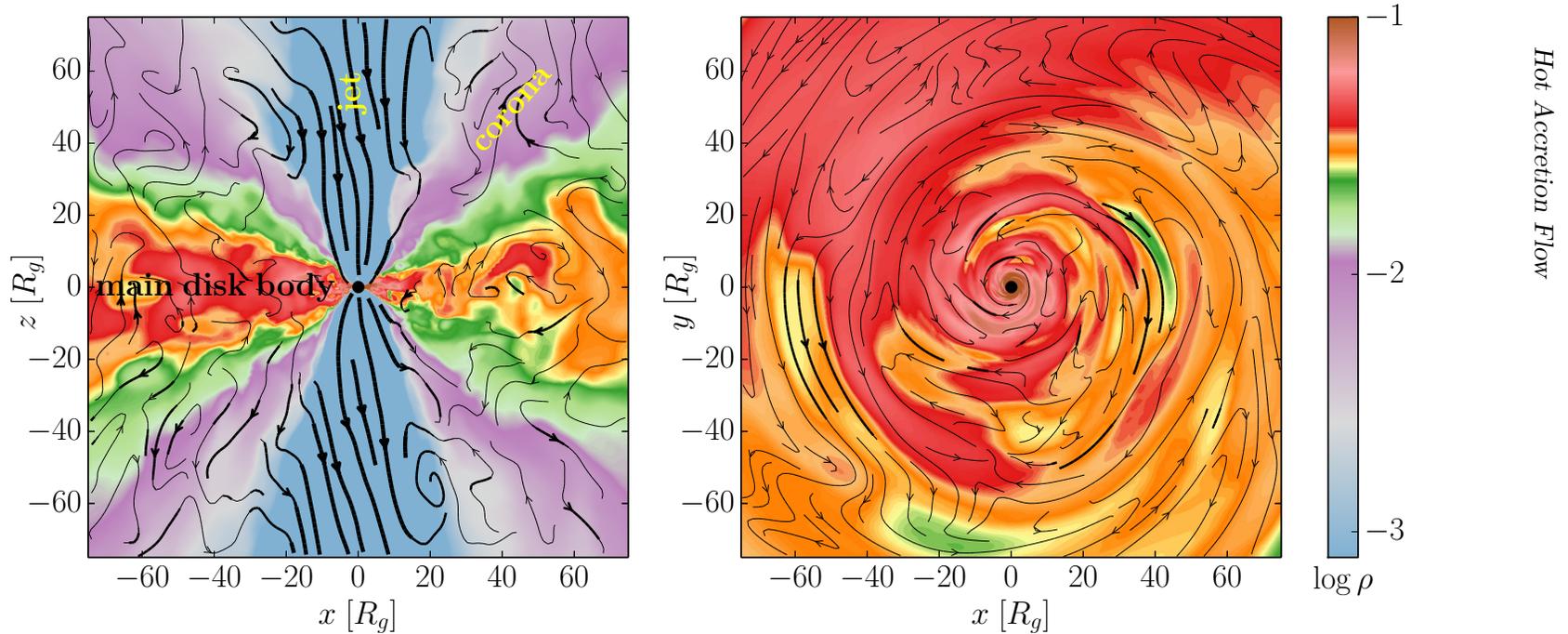,angle=0,width=20cm}}
\vskip -0 cm
\caption{An instantaneous poloidal slice at a fixed azimuthal angle
  ({\it left}) and a slice through the equatorial plane ({\it right})
  from a 3D GRMHD simulation of a hot accretion flow around a spinning
  black hole ($a_*\equiv a/M=0.7$). The black hole is at (0,0) and
  lengths are in units of $R_g=GM/c^2$. Three regions of the flow are
  identified: main disk body, corona, jet. Color indicates density,
  with fluctuations evident due to turbulence. Lines trace the
  magnetic field in the two image planes, with the out-of-plane
  component ignored. Arrows show the direction of the magnetic field
  and line thickness indicates magnetic energy density relative to
  other energy densities: the thickest lines correspond to regions
  with  comoving $B^2/4\pi > \rho c^2$ (found primarily in the region of the
  jet), intermediate thickness lines indicate regions with (see
  eq.~\ref{betadefinition}) $\beta<4$ (mostly in the corona), and thin
  lines correspond to regions with the weakest magnetization,
  $\beta>4$ (primarily in the main disk body). (Figure courtesy of
  A. Tchekhovskoy, based on data from Sadowski et al. 2013a.)}
\vskip -0.90cm
\label{diskstructure}
\end{figure*}
\end{landscape}

Representative results from 3D GRMHD simulations can be found in the
series of early papers by De Villiers and collaborators (De Villiers
et al.~2003, 2005; Hirose et al.~2004; Krolik et al.~2005) and Gammie
and collaborators (Gammie et al.~2004, McKinney \& Gammie 2004,
McKinney 2006). The simulations are initialized with a rotating torus
in hydrostatic equilibrium and embedded with a weak poloidal magnetic
field.  Accretion occurs self-consistently as a result of MHD
turbulence generated by the MRI, and the accretion flow separates into
three qualitatively different regions (Fig.~\ref{diskstructure}): disk
body, corona, axial funnel.

The disk body is turbulent and dense and has a roughly constant value
of $H/R$, consistent with 1D self-similar and global solutions
(\S\ref{onedimension}).  The magnetic field within the disk is
sub-equipartition ($\beta\sim 10-100$), and both the magnetic field
and velocity streamlines are chaotic, as expected for
turbulence. Inside the innermost stable circular orbit (ISCO) is the
plunging region. Here the flow spirals in rapidly toward the black
hole horizon and the motion is almost laminar. While the ISCO is
roughly where the turbulent gas in the disk transitions to laminar
inflow, there is no other specific signature in the flow dynamics
associated with the ISCO. This is in contrast to the case of thin
disks, where the flow changes dramatically across the ISCO (Reynolds
\& Fabian 2008, Shafee et al.~2008, Penna et al.~2010, but see Noble
et al.~2010).

Above and below the main disk is the corona, where the gas density is
much lower. The magnetic field here is more regular than in the disk
body, and tends to be toroidal.  The magnetic and gas pressure are
roughly comparable ($\beta \sim 1$).
The value of $\beta$ decreases with increasing distance away from the
midplane, with $\beta\sim 0.1$ above about two density scale heights
(De Villiers et al.~2003, 2005).  The corona is the launchpad for the
disk wind (\S\ref{outflow}).

The axial funnel is a magnetically dominated region in which the gas
is very tenuous.  This is the location of the jet
(\S\ref{jetformation}). The magnetic field is predominantly radial
close to the black hole, where the jet extracts rotational energy from
the black hole spin (\S\ref{jet1}). Far from the black hole the field
becomes mostly toroidal and carries the jet power in the form of a
Poynting flux. The boundary of the funnel (the funnel wall)
corresponds to the centrifugal barrier associated with material
originating from the innermost region of the disk. The jet here is
less relativistic and is powered at least partly by the rotation of
the accretion flow (\S\ref{jet2}). Hence it is less sensitive to the
black hole spin.  Overall, there is a smooth variation of properties,
going from highly magnetically dominated conditions at the axis to
progressively larger gas content with increasing distance from the
axis. Past the funnel wall, the jet merges with the corona where the
disk wind is launched (\S\ref{outflow})

The effect of different initial magnetic field geometries has been
investigated by a number of authors (e.g., Hawley \& Krolik 2002,
Igumenshchev et al.~2003 with a pseudo-Newtonian potential; and
Beckwith et al.~2008, McKinney \& Blandford 2009 with GRMHD). Models
with a purely toroidal initial field evolve much more slowly than
those with poloidal initial field. This is because the former have
neither an initial vertical field, which is needed for the linear MRI,
nor a radial field, which is needed for field amplification via shear.
Inflow begins only after the MRI has produced turbulence of sufficient
amplitude (Hawley \& Krolik 2002), which happens much later when the
initial field is toroidal. Generally, once saturation has been
reached, the disk properties do not depend much on the initial field
topology. On the other hand, jet properties are very sensitive
(\S\ref{jetformation}).

In the case of geometrically thin accretion disks, it is believed that
Lense-Thirring precession causes a tilted disk to align with the spin
axis of the black hole out to a fairly large radius (Bardeen \&
Petterson 1975, Scheuer \& Feiler 1996, Lodato \& Pringle 2006).
Fragile and collaborators have carried out a number of numerical
simulations of tilted hot accretion flows (Fragile et al.~2007, 2009;
Fragile 2009; Dexter \& Fragile 2011, 2013). They find that the disk
does not align with the black hole, in agreement with theoretical
predictions for a geometrically thick disk (Papaloizou \& Lin
1995). Instead the disk precesses as a whole out to some
radius. Alignment does happen when accretion occurs in the MAD regime
(\S\ref{mad}) and might have observational implications for
relativistic jets (McKinney et al.~2013).

The precession of a tilted disk will lead to time-dependence in the
observed radiation from a hot accretion flow, and could potentially
explain some low frequency quasi-periodic oscillations (QPOs) seen in
BHBs (Ingram et al.~2009, Ingram \& Done 2011, Veledina et
al.~2013). It is unlikely that high frequency QPOs can be explained in
a similar fashion (Dexter \& Fragile 2011). A number of authors (e.g.,
Bursa et al.~2004, Blaes et al.~2006, Abramowicz et al.~2006) have
explored oscillation modes of tori in this connection.  Recently, high
frequency QPOs have been reported in numerical GRMHD simulations of
geometrically thick hot accretion flows (Dolence et al.~2012) {
  and magnetic arrested disks (\S\ref{mad}, McKinney et al. 2012,
  Shcherbakov \& McKinney 2013).}

A few authors have carried out simulations of magnetized spherical
accretion (Igumenshchev \& Narayan 2002, Igumenshchev 2006) as well as
accretion of low-angular momentum gas (Proga \& Begelman 2003,
Mo\'scibrodzka et al.~2007, Janiuk et al.~2009). These topics are beyond
the scope of this review.

We conclude with some general remarks on numerical accuracy. Since
energy advection plays a key role in hot accretion flows, it is
important to ensure that numerical codes conserve energy accurately.
Early numerical simulations were based on codes that evolve the
internal energy of the gas. Such codes do not conserve total energy
and can introduce an effective numerical cooling that is hard to
quantify.  An alternative Godunov-based approach has come to the fore
in recent years. This enforces strict mass, momentum and total energy
conservation, as exemplified by the pioneering relativistic MHD code
of Komissarov (1999, 2001), the GRMHD codes HARM (Gammie et al.~2003) and
COSMOS++ (Anninos et al.~2005), and the Newtonian MHD code ATHENA
(Stone et al.~2008). Direct comparison of simulations using both
techniques shows that accretion flows simulated with non-energy
conserving codes tend to be geometrically thinner than they should be.

The effect of numerical resolution on global simulation results has
been investigated recently by a number of authors (Sorathia et
al.~2010; Hawley et al.~2011, 2013; Shiokawa et al.~2012). These
studies achieved numerical convergence in terms of shell-averaged
quantities, azimuthal correlation length of fluid variables, and
synthetic spectra. It appears that most previous global simulations
might have been somewhat under-resolved.

\subsubsection{Magnetically arrested disk}
\label{mad}

Magnetically dominated hot accretion flows have recently come to the
fore, thanks to the advent of numerical MHD simulations. One version
of these models, called a ``magnetically arrested disk'' (MAD, Narayan
et al.~2003, Igumenshchev et al.~2003, Igumenshchev 2008) or a
``magnetically choked accretion flow'' (MCAF, McKinney et al.~2012),
is based on an idea originally proposed by Bisnovatyi-Kogan \&
Ruzmaikin (1974), in which a strong vertical bipolar magnetic field is
pushed into the central black hole by the thermal and ram pressure of
the accreting gas. A significant amount of magnetic flux threads the
horizon. As a result, the field outside the black hole becomes so
strong that it disrupts the axisymmetric accretion flow, forcing the
gas to move inward via streams and blobs through an interchange
instability. This behavior was first noted by Igumenshchev et
al.~(2003) in 3D Newtonian MHD simulations and was later confirmed in
3D GRMHD simulations (Tchekhovskoy et al.~2011, 2012; McKinney et
al.~2012). Current interest in MAD accretion is driven by the
discovery that it leads to very powerful relativistic jets
(\S\ref{jetformation}).

All magnetized accretion flows cause a certain amount of magnetic flux
to thread the black hole. The MAD state is special in that the flux
threading the hole is at its maximum saturation value for the given
mass accretion rate $\dot{M}_{\rm BH}$.  This saturation flux is
approximately\footnote{$\Phi_{\rm MAD}$ has a weak dependence on
  the black hole spin as well as the disk thickness (see Tchekhovskoy
  et al.~2013), but we ignore this complication for clarity.}
(Tchekhovskoy et al.~2011, 2012)
\begin{equation}
\Phi_{\rm MAD} \approx 50\, \dot{M}_{\rm BH}^{1/2} R_g c^{1/2}
= 1.5 \times10^{21}\,
m^{3/2} \dot{m}_{\rm BH}^{1/2}~{\rm G\,cm^2},
\label{PhiMAD}
\end{equation}
where $m$ and $\dot{m}$ are defined in eq.~(\ref{scaled}) and
$R_g=GM/c^2 = R_S/2$ (half the Schwarzschild radius) is the
gravitational radius of the black hole. The corresponding field
strength at the horizon is roughly\footnote{Simulation results are
  often given in Heaviside-Lorentz units, whereas numerical estimates
  in this article are in gaussian units. The two differ by a factor of
  $\sqrt{4\pi}$. For instance, the magnetic pressure is $B^2/8\pi$ in
  gaussian units but $B^2/2$ in Heaviside-Lorentz units. Note
    also that the magnetic field strength is frame-dependent. For
    instance, when evaluating the magnetic pressure, especially for
    computing the value of $\beta$ (eq.~\ref{betadefinition},
    Fig.~\ref{diskstructure}), one must consider the field strength in
    the comoving fluid frame, i.e., $B^2/4\pi \to b^2$ in the notation
    of Komissarov (1999) and Gammie et al.~(2003). On the other hand,
    $B_{\rm MAD}$ and $\Phi_{\rm MAD}$ in eq.~(\ref{BMAD}) are
    evaluated in the stationary coordinate frame or ``lab frame''.}
(compare eq.~\ref{magfield})
\begin{equation}
B_{\rm MAD} \approx \frac{\Phi_{\rm MAD}}{2\pi R_g^2} = 10^{10}\,
m^{-1/2} \dot{m}_{\rm BH}^{1/2}~{\rm G}.
\label{BMAD}
\end{equation}
Systems that have not reached the MAD limit have been referred to as
SANE (``standard and normal evolution'', Narayan et al.~2012b). They
span a one-parameter family of models extending from $\Phi=0$ up to a
magnetic flux just below $\Phi_{\rm MAD}$.  Structural differences are
evident between MAD and SANE models (Narayan et al.~2012b, Sadowski et
al.~2013a), most notably in the jet.

It should be noted that, unlike small-scale fields, a large-scale
vertical magnetic field cannot be dissipated locally (since the plasma
has negligible resistivity), nor can it be absorbed by the central
black hole (e.g., even when field lines thread the horizon, the
external magnetic flux is unaffected, see Igumenshchev et
al.~2003). But how does vertical field reach the center in the first
place? Presumably the field is advected in from whatever external mass
reservoir feeds the accretion flow. Such advection happens quite
efficiently in numerical simulations, especially in the case of
geometrically thick hot accretion flows. However, most simulations are
limited to radii relatively close to the black hole and it is not
clear that the same physics will necessarily operate at larger
radii. If outward diffusion of the magnetic field via reconnection is
inefficient (Spruit \& Uzdensky 2005, Bisnovatyi-Kogan \& Lovelace
2007, Rothstein \& Lovelace 2008, Cao 2011), as seems especially
likely for geometrically thick accretion flows (Livio et al.~1999;
Guilet \& Ogilvie 2012, 2013), magnetic field should be readily
advected in from large radii. Magnetic field can also be brought in
efficiently via the corona (Beckwith et al.~2009). At least in the
case of supermassive black holes accreting from an external medium,
plenty of magnetic flux is available in the mass reservoir (e.g.,
Narayan et al.~2003).  Therefore, all supermassive black holes with
hot accretion flows should quickly approach the MAD limit, provided
flux is advected efficiently. The situation is less clear in the
  case of BHBs, since the supply of net magnetic flux depends on the
  properties of the companion star and the details of the mass
  transfer.

Two additional magnetically dominated accretion models have been
discussed. Meier (2005, see also Fragile \& Meier 2009) proposed that
the inner regions ($R\ \sles\ 50R_S$) of a hot accretion flow may be
converted into a magnetically-dominated magnetosphere-like phase in
which a strong, well-ordered field is present rather than the weak,
turbulent field usually seen in a hot flow. A different possibility is
the ``magnetically supported accretion flow'' model proposed by Oda et
al.~(2010), stimulated by MHD simulations described in Machida et
al.~(2006). This kind of flow exists only when the accretion rate is
relatively high, well above $\dot{M}_{\rm crit,ADAF}$.
Both of these models have a magnetic field geometry dominated by
radial and toroidal fields, which is different from the vertical
poloidal field envisaged in the MAD model.

\subsection{Jets in hot accretion flows}
\label{jetformation}

It was conjectured already in early papers (Narayan \& Yi 1994, 1995a;
Blandford \& Begelman 1999) that hot accretion flows should have
strong winds and, by extension, jets.  Observational evidence for such
an association has accumulated in recent years with the recognition
that essentially all low-luminosity AGNs are radio loud (Nagar et
al.~2000, Falcke et al.~2000, Ho 2002) and the parallel discovery that
virtually every BHB in the hard state has radio emission (Corbel et
al.~2000; Fender 2001, 2006; Fender et al.~2004). Since all these
systems are believed to have hot accretion flows (\S\ref{llagn}), it
seems likely that there is a direct causal connection between hot
flows and radio-emitting jets. In contrast, jets are much weaker, and
often not seen at all, in systems that have cool geometrically thin
disks. While there is no definitive explanation for this dichotomy, it
is likely that three effects play a role: (i) geometrically thick
disks more easily advect magnetic field to the black hole compared to
thin disks (Livio et al.~1999, Guilet \& Ogilvie 2012); (ii) the
Bernoulli parameter of the gas in hot accretion flows is larger, hence
enhancing winds in these systems (Narayan \& Yi 1994, 1995a; Blandford
\& Begelman 1999); (iii) strong winds help to collimate and stabilize
the jet (Appl \& Camenzind 1992, 1993; Beskin \& Malyshkin 2000).

Although many models of jets have been proposed over the years, the
current consensus is that jets arise from a combination of magnetic
fields and rotation. Especially influential in this field are the
Blandford-Znajek model (BZ, Blandford \& Znajek 1977; see also Ruffini
\& Wilson 1975, Lovelace 1976, MacDonald \& Thorne 1982; Phinney 1983;
Thorne et al.~1986; Punsly \& Coroniti 1989; Komissarov \& McKinney
2007, Tchekhovskoy et al.~2011; Penna et al.~2013a) and the
Blandford-Payne model (BP, Blandford \& Payne 1982; see also Pudritz \& Norman 1983;
Heyvaerts \& Norman 1989; Li
et al.~1992; Contopoulos \& Lovelace 1994;
Ostriker 1997; Vlahakis \& Konigl 2003). The primary distinction
between the two models is the energy source of the jet. In the BZ
model, the source is the rotational energy of the black hole, while in
the BP model, it is the rotational energy of the accretion
flow. Numerical simulations suggest that truly relativistic jets are
produced primarily by the BZ mechanism, while quasi-relativistic jets
and non-relativistic winds may be driven by a combination of the BP
and other mechanisms. In the following discussion we use the term ``BZ
jet'' for the truly relativistic BZ-powered outflow, and refer to the
quasi-relativistic outflow from the inner region of the accretion flow
as the ``disk jet''.

\subsubsection{Relativistic BZ jet}
\label{jet1}

In the BZ model a large-scale poloidal magnetic field passes through
the ergosphere of the rotating black hole and threads the horizon (cf
\S\ref{mad}). Frame dragging by the rotating hole leads to the
creation of toroidal field and hence a Poynting flux. The key to the
BZ process, which goes back to the influential work of Penrose (1969)
and subsequently Ruffini \& Wilson (1975), is the fact that, within
the ergosphere, it is possible to have a negative inward
electromagnetic energy flux as measured at infinity. This negative
energy flux enters the horizon, thereby reducing the mass-energy and
angular momentum of the hole. Correspondingly, there is an outgoing
Poynting-dominated jet which carries positive energy and angular
momentum. At its most basic, the outflowing power in the BZ model is
given by (Blandford \& Znajek 1977; see Tchekhovskoy et al.~2010 for
more accurate approximations)
\begin{equation}
P_{\rm BZ} = \frac{\kappa}{4\pi c} \Phi^2 \Omega_{\rm H}^2,
\label{BZpower}
\end{equation}
where $\Phi$ is the magnetic flux threading the horizon, $\Omega_{\rm
  H}=a_*c/2R_{\rm H}$ is the angular velocity of the horizon,
$a_*\equiv a/M$ is the dimensionless spin parameter of the black hole,
and $R_{\rm H} = R_g (1+\sqrt{1-a_*^2})$ is the radius of the horizon.
The numerical coefficient $\kappa$ depends weakly on the magnetic
field geometry and is approximately $\approx 0.05$.  The above formula
highlights the fact that the BZ mechanism requires two key
ingredients: an ordered magnetic flux at the horizon ($\Phi$), and
rotation of the black hole ($\Omega_{\rm H}$).

Many MHD simulations of hot accretion flows have been performed to
study jet formation (e.g., Kudoh et al.~1998; Koide et al.~1999, 2000;
Kuwabara et al.~2000; Hawley \& Balbus 2002; Koide 2003; McKinney \&
Gammie 2004; Kato et al.~2004a, 2004b; De Villiers et al.~2005;
McKinney 2005, 2006; Hawley \& Krolik 2006; Komissarov et al.~2007;
McKinney \& Blandford 2009; Tchekhovskoy et al.~2011; McKinney et
al.~2012, Sadowski et al.~2013a).  The simulations are typically
initialized with a gas torus threaded with a weak magnetic field.  It
is found that a large-scale magnetic field forms self-consistently at
the black hole horizon, as required by the BZ model, even though such
a field is not present in the initial state (e.g., Igumenshchev et
al.~2003; De Villiers et al.~2003, 2005; McKinney 2006; Tchekhovskoy
et al.~2011), and that the magnetic flux is trapped within a funnel,
causing the outgoing power to be collimated in a relativistic
jet. However, a powerful jet forms only if the initial field in the
simulation has a favorable poloidal configuration. A dipolar field is
ideal.~If a quadrupolar initial field is adopted, the field in the
funnel is much weaker, and if a toroidal field is adopted, no funnel
field at all develops (Igumenshchev et al.~2003, De Villers et
al.~2005, Beckwith et al.~2008, McKinney \& Blandford 2009). In the
latter case there is no BZ jet.

The jet power measured in simulations shows good agreement with the
predictions of the BZ model (eq.~\ref{BZpower}), modulo modest changes
in the coefficient due to the presence of the surrounding thick disk.
A rough estimate of the BZ jet power is obtained by combining
equations (\ref{PhiMAD}) and (\ref{BZpower})\footnote{This formula
  slightly underestimates the jet power for slow spins and
  overestimates the power for rapid spins. A better approximation is
  $P_{\rm jet} \approx 0.65 a_*^2 (1+0.85 a_*^2) (\Phi/\Phi_{\rm
    MAD})^2 \dot{M}_{\rm BH}c^2$ (A.~Tchekhovskoy, private
  communication).},
\begin{equation}
P_{\rm jet} \approx 2.5
\left(\frac{a_*}{1+\sqrt{1-a_*^2}}\right)^2
\left(\frac{\Phi}{\Phi_{\rm MAD}}\right)^2
\dot{M}_{\rm BH}c^2,
\label{Pjet}
\end{equation}
where $\Phi_{\rm MAD}$ is the limiting magnetic flux defined in
equation~(\ref{PhiMAD}) and $\Phi$ is the actual flux threading the
black hole horizon.  As is evident from this relation, the most
favorable situation is when the magnetic flux has reached the MAD
limit ($\Phi\to\Phi_{\rm MAD}$) and the black hole spin is maximum
($a_*\to 1$). In this limit, the BZ jet power can exceed the total
accretion energy budget of $\dot{M}_{\rm BH}c^2$ (Tchekhovskoy et
al.~2011, 2012; Tchekhovskoy \& McKinney 2012).  Although at first
sight this might appear to violate energy conservation, there is no
inconsistency since most of the jet energy comes directly from the
spinning black hole via a generalization of the Penrose (1969)
process.

Many studies have been published over the years, providing estimates
of the jet power as a function of black hole spin (e.g., Hawley \&
Balbus 2002; De Villers et al.~2005; McKinney 2005, 2006; Hawley \&
Krolik 2006).  While the numerical values given do not always agree,
the results are consistent with the above BZ-derived relation,
once the dependence on magnetic flux is taken into account.  Modest
differences are seen between prograde and retrograde disks, with
prograde producing somewhat stronger jets (Tchekhovskoy \& McKinney
2012).  In addition, the physics near the horizon in the simulations
matches very closely the physics of the BZ mechanism (Penna et
al.~2013a) as described in the membrane paradigm (Thorne et
al.~1986). It also satisfies all the requirements to be viewed as a
form of generalized Penrose process (Lasota et al.~2014).

Another quantity of interest is the asymptotic Lorentz factor of the
jet, $\gamma_{\rm jet}$. Unfortunately, the value of $\gamma_{\rm
  jet}$ depends on how much mass is loaded on magnetic field lines,
the physics of which is very poorly understood. The current best guess
is that mass-loading occurs via pair creation through breakdown of a
vacuum gap (Beskin et al.~1992; Hirotani \& Okamoto 1998).  However,
in simulations, mass-loading is treated in an entirely ad hoc fashion
by applying a minimum ``floor'' value for the gas density. The
resulting jet Lorentz factor tends to be large on the axis and to
decrease outward, but the values obtained do not mean much without a
physical model of mass-loading.  Mass loss in the jet $\dot{M}_{\rm
  jet}$ and jet Lorentz factor $\gamma_{\rm jet}$ are related by
$P_{\rm jet} \approx \gamma_{\rm jet}\dot{M}_{\rm jet}c^2$ (including
the rest mass energy of the ejected gas). Whereas $P_{\rm jet}$ can be
calculated with reasonable confidence from simulations, neither
$\dot{M}_{\rm jet}$ nor $\gamma_{\rm jet}$ can be estimated reliably.

\subsubsection{Quasi-relativistic disk jet}
\label{jet2}

Surrounding the relativistic BZ jet discussed in the previous
subsection is a quasi-relativistic disk jet. The key distinctions are:
(i) the disk jet is matter-dominated, not Poynting flux
dominated,\footnote{The recent detection of Doppler-shifted X-ray
  emission lines in the candidate black hole binary, 4U1630--47 (Diaz
  Trigo et al. 2013), suggests that at the time of the observations
  this system might have had a baryon-loaded jet. What was observed
  was conceivably a disk-jet, and perhaps also an episodic jet
  (\S\ref{jet3}).} and (ii) the disk jet is powered by the inner
regions of the accretion disk, not directly by the black hole. There
is no unambiguous way to demarcate the two regions, and different
authors have used different prescriptions to identify the boundary
(e.g., Hawley \& Krolik 2006, Tchekhovskoy et al.~2011). The outflow
power of the disk jet is typically $< 0.1\dot{M}_{\rm BH}c^2$, and it
varies only modestly with the parameters $\Phi$ and $a_*$, in contrast
to the steep dependence in the case of the BZ jet (eq.~\ref{Pjet}).
As a result, when $\Phi \ll \Phi_{\rm MAD}$ (extreme SANE limit), the
power in the disk jet can be larger than that in the BZ jet even
though the black hole may be spinning rapidly (De Villiers et
al.~2005, Sadowski et al.~2013a). Observationally, the disk jet will
produce radio emission and will behave in many respects like a
relativistic jet. However, its Lorentz factor is usually modest.

The quasi-relativistic disk jet receives energy from the disk via
magnetic fields anchored in the accretion flow. In the BP model, if
the field lines are angled outward sufficiently with respect to the
disk rotation axis, there is a net outward centrifugal force on matter
threading the field. As gas is accelerated outward along the rotating
field lines, its angular momentum increases, thereby causing further
acceleration. The relevance of the BP mechanism to quasi-relativistic
outflows in GRMHD simulations has not been explored systematically.

An alternative magnetic tower mechanism has been proposed in which a
strong toroidal magnetic field is produced by the differential
rotation of the accretion flow, and the resulting magnetic pressure
gradient causes gas to be accelerated away from the disk surface
(Lynden-Bell 2003). Structures analogous to a magnetic tower have been
seen in some MHD simulations (e.g., Shibata \& Uchida 1985, 1986; Kato
et al.~2004b). In addition, there is a suggestion (Hawley \& Krolik
2006) that acceleration of the disk jet is caused, not by centrifugal
force (BP), but by the gradient of magnetic and gas pressure (magnetic
tower).

\subsubsection{Episodic jet}
\label{jet3}

Observations show that BHBs have two distinct kinds of jets (Fender \&
Belloni 2004): steady jets and episodic (or ballistic) jets. Episodic
jets are most obviously observed in BHBs during the transition from
the hard to the soft state (McClintock \& Remillard 2006, Remillard \&
McClintock 2006), often at luminosities close to Eddington. However,
there is a hint that these jets can occur also at lower luminosities
(Cyg X-1, Fender et al.~2006; Sgr A*, Yusef-Zadeh et al.~2006,
  \S\ref{flare}; other examples are reviewed in Yuan et al.~2009a).
The most distinctive difference between the two jets is
that the episodic jet is transient and is in the form of discrete,
isolated blobs, while the steady jet behaves like a continuous
outflow. Other differences have been noted in the polarization,
spectrum, and power (Fender \& Belloni 2004).

It is unclear whether the models reviewed in \S\ref{jet1} and
\S\ref{jet2} are applicable to episodic jets since the simulations
discussed there generally give quasi-steady jets.  Numerical MHD
simulations of accretion disks in other contexts do produce episodic
ejections of magnetized blobs from the disk surface (e.g., Romanova et
al.~1998, Kudoh et al.~2002, Kato et al.~2004a, Dyda et al.~2013),
although the underlying physics has not been identified
clearly\footnote{Interestingly, similar episodic ejection have also
  been found in MHD simulations of accretion disks around young
  stellar objects, and are possibly better understood there (e.g., Hayashi
  et al.~1996, Goodson et al.~1999, Goodson \& Winglee 1999).}.

By analogy with coronal mass ejections in the Sun, which is another
example of episodic mass ejection, Yuan et al.~(2009a) proposed an MHD
model for the formation of episodic jets.\footnote{Massi \& Poletto
  (2011) discuss other interesting similarities between coronal mass
  ejections and AGN jets.} In this model, a flux rope is first formed
in the corona due to the twisting of magnetic loops emerging from the
disk body via the Parker instability. The flux rope is initially in
force equilibrium between magnetic tension and magnetic pressure.
However, with further twisting of the field lines, a threshold energy
is reached. The flux rope jumps upward, causing a reconnection event
to occur. This results in a substantial enhancement of the magnetic
pressure force and weakening of the tension force, causing the flux
rope to be ejected. This model is similar to the magnetic tower model
(Lynden-Bell 2003) discussed earlier, except that here it is
time-dependent and involves a flux rope.

\subsection{Disk Wind from hot accretion flows}
\label{outflow}

Outside the quasi-relativistic disk jet and above the main disk body
lies the bulk of the mass outflow from the disk (Yuan et al.~2012a,
Narayan et al.~2012b, Sadowski et al.~2013a). We call this the disk
wind. In contrast to the BZ jet and the disk jet, the disk wind is
non-relativistic and moves slowly. However, it occupies a much larger
solid angle. The mass loss rate is also quite high, although the rate
of outflow of energy is small compared to the power in the BZ jet or
the disk jet (Yuan et al.~2012a, Sadowski et al.~2013a). As in the case
of the boundary between the BZ jet and the disk jet, here again there
is no unabiguous way to identify the boundary between the disk jet and
the disk wind.

The likelihood that ADAFs will have strong winds was pointed out by
Narayan \& Yi (1994, 1995a), but these authors were unable to come up
with a quantitative prediction for the amount of mass loss in the
wind.  Blandford \& Begelman (1999) described a family of self-similar
solutions with a wide range of assumed outflow efficiencies, again
emphasizing the inability of analytical models to say anything
definite about the magnitude of mass and energy loss in winds. This
uncertainty has been a serious bottleneck in the development of
  one-dimensional models of hot accretion flows, and is an important
motivation for doing numerical simulations. From a practical
standpoint, it is essential to understand the nature of disk winds in
hot accretion flows because (i) mass loss can strongly affect the
dynamics of the accreting gas, and (ii) disk winds canx be a powerful
contributor to AGN feedback (\S\ref{agnfeedback}).

\begin{figure}
\vspace{-0.6cm}
\hskip 0.0truein
\psfig{figure=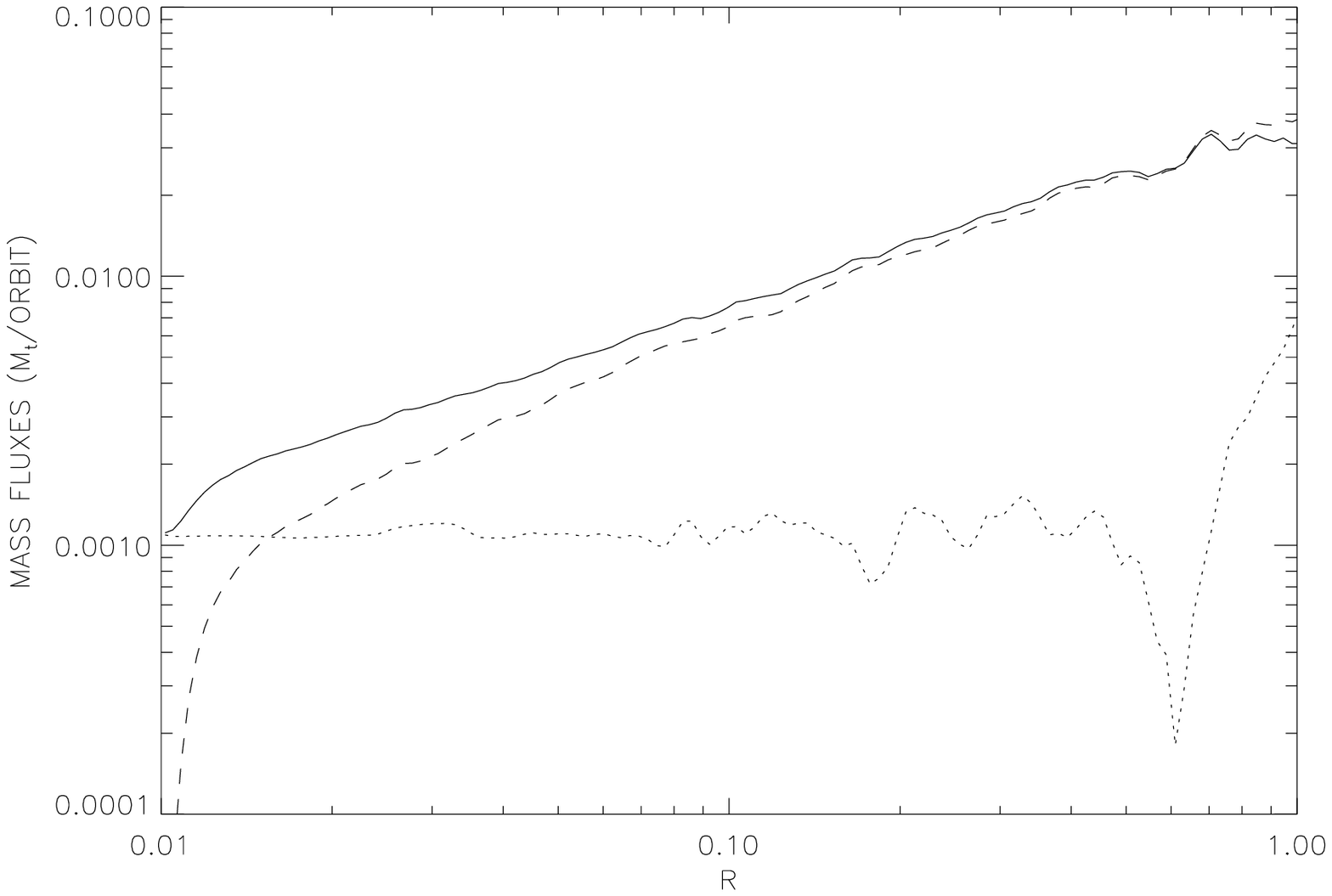,width=6.cm,angle=0}
\psfig{figure=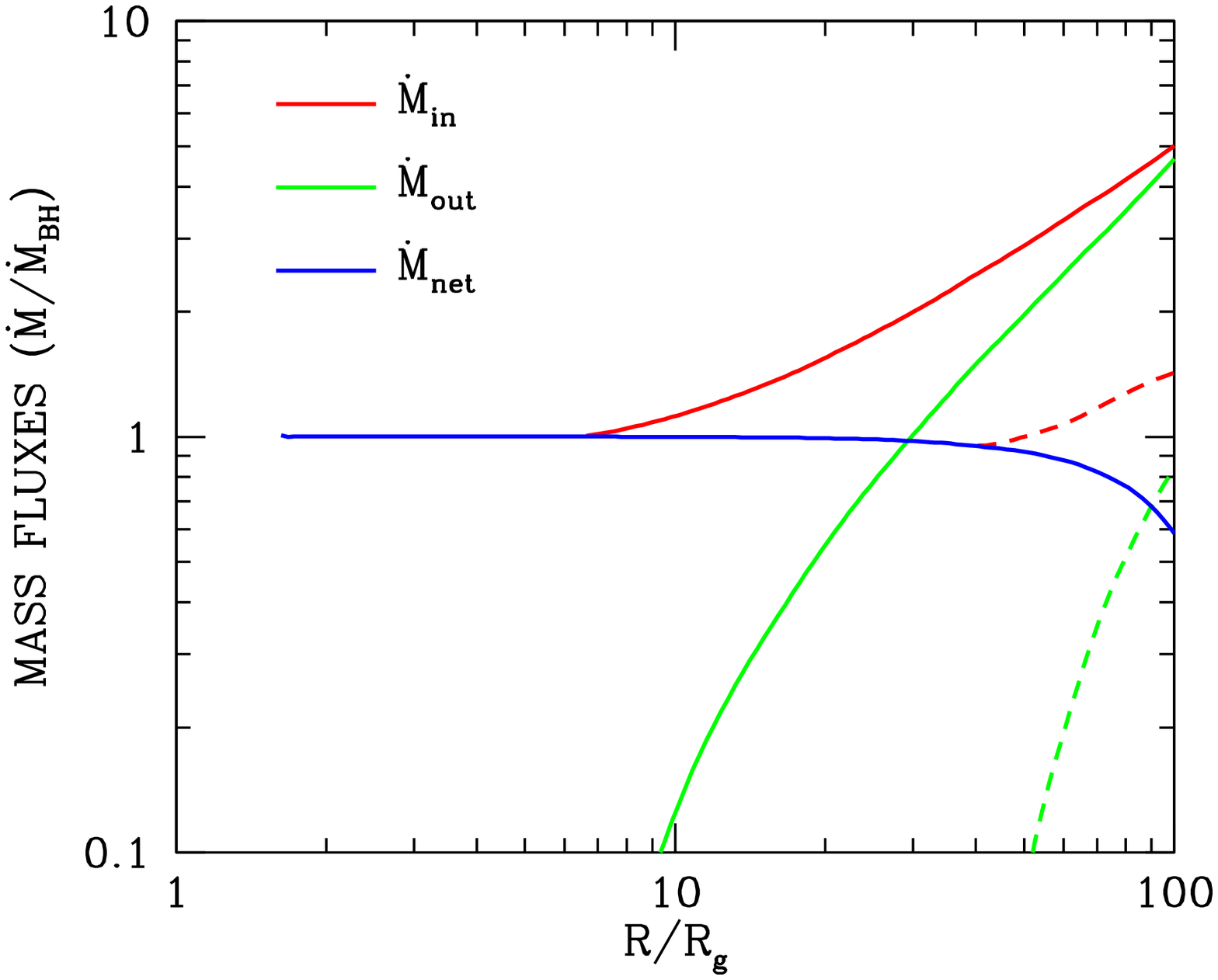,width=7.2cm,height=5.8cm,angle=0}
\caption{Radial profiles of mass inflow rate $\dot{M}_{\rm in}$, mass
  outflow rate $\dot{M}_{\rm out}$ and net mass accretion rate
  $\dot{M}_{\rm net}$ (eqs.~\ref{inflowrate}, \ref{outflowrate},
  \ref{netrate}). {\em Left:} Results from a 2D Newtonian HD
  simulation of a hot accretion flow (Stone et al.~1999). Solid,
  dashed, and dotted lines correspond to $\dot{M}_{\rm in}$,
  $\dot{M}_{\rm out}$ and $\dot{M}_{\rm net}$, respectively. Note the
  rapid increase of $\dot{M}_{\rm out}$ with increasing radius. {\em
    Right:} Solid lines show equivalent results from a 3D GRMHD
  simulation of a hot accretion flow around a non-spinning black hole
  ($a_*=0$, Narayan et al.~2012b). Mass outflow becomes important only
  beyond a radius $\sim30R_g$, though the slope outside this radius is
  similar to that in the panel on the left. Dashed lines show results
  for a different kind of time-averaging, as described in the
  text. Here the estimated mass outflow rate is very much smaller (see
  also the top-left panel of Fig.~14 in Yuan et al.~2012a). The true
  mass outflow rate is likely to be in between the solid and dashed
  green lines.}
\label{Fig:inflowrate}
\end{figure}

In an important pioneering study of winds from hot accretion flows,
Stone et al.~(1999) carried out numerical 2D HD simulations and
calculated the mass inflow, outflow and net accretion rates via the
following integrals,
\begin{eqnarray}
\label{inflowrate}  \dot{M}_{\rm in}(r) &=& -2\pi r^{2}
\left\langle \int_{0}^{\pi} \rho \min(v_{r},0)
   \sin \theta d\theta \right\rangle_{t\phi}, \\
\label{outflowrate} \dot{M}_{\rm out}(r) &=& 2\pi r^{2}
\left\langle \int_{0}^{\pi} \rho \max(v_{r},0)
    \sin \theta d\theta \right\rangle_{t\phi}, \\
\label{netrate} \dot{M}_{\rm net} &=& \dot{M}_{\rm in} - \dot{M}_{\rm out},
\end{eqnarray}
where the angle brackets represent an average over time (and also
azimuthal angle $\phi$ in the case of 3D simulations).  The quantity
$\dot{M}_{\rm net}$ is the net mass accretion rate; in steady state,
it is equal to the accretion rate on the black hole $\dot{M}_{\rm
  BH}$.  Stone et al.~(1999) found, as has been confirmed in many
later simulations (see references below), that both $\dot{M}_{\rm in}$
and $\dot{M}_{\rm out}$ decrease inward, following roughly a power-law
behavior (see eq.~\ref{MdotBH}),
\begin{equation} \dot{M}_{\rm in}(r)=\dot{M}_{\rm in}(r_{\rm out})\left(\frac{r}{r_{\rm out}}\right)^s, \hspace{0.5cm} s>0.
\end{equation}
This is illustrated in the left panel in Fig.~\ref{Fig:inflowrate}.
Correspondingly, the radial profile of density becomes flatter than in
a self-similar ADAF solution: $\rho(r)\propto r^{-p}$ with $p<1.5$. These
statements appear to be true regardless of whether one considers hydro
simulations (Stone et al.~1999, Yuan \& Bu 2010, Yuan et al.~2012b, Li
et al.~2013a) or MHD simulations (Stone \& Pringle 2001, Machida et
al.~2001, Hawley et al.~2001, Hawley \& Balbus 2002, Igumenshchev et
al.~2003, Pen et al.~2003, Kato et al.~2004b, Pang et al.~2011, Yuan
et al.~2012a).

The values of $s$ and $p$ in various simulations are summarized in
Yuan et al.~(2012b): $s=0.4-0.8$, $p=0.5-1$. The variations between
different simulations seem to be due to differences in the value of
the viscosity parameter $\alpha$ (in the case of hydro simulations),
choice of Newtonian gravity versus general relativity, initial
configuration of the magnetic field (toroidal or poloidal), and the
strength of the initial field (weak or strong).  Observationally,
there is support for a value of $s\sim 0.3$, $p\sim 1$ (see the
discussion of Sgr A* in \S\ref{sgra}). The apparent discrepancy
between theory and observations may be due to low angular momentum of
the accretion flow in Sgr A* (Bu et al.~2013) or dynamical importance
of thermal conduction (\S\ref{collisionless}; Johnson \& Quataert
2007).

Competing models have been proposed to explain the radially varying
inflow and outflow rates seen in numerical simulations. In the
adiabatic inflow-outflow solution (ADIOS; Blandford \& Begelman 1999,
2004; Begelman 2012), the inward decrease of $\dot{M}_{\rm in}$ is due
to a genuine mass loss in a wind. What drives the wind is
unspecified in the model. Assuming merely that a mechanism exists for
draining energy from the interior of the accretion flow to launch a
wind, the authors construct 1D and 2D self-similar solutions. In
their models, the index $s$ is left as a free parameter, limited only
by the condition $0\leq s \leq 1$.  In the most recent version of the
ADIOS model, however, Begelman (2012) considers the inflow and outflow
zones on an equal footing and, using a conserved outward energy flux,
finds that $s$ should be close to unity. This is somewhat larger than
the range of values seen in numerical simulations.

An alternative scenario is the convection-dominated accretion flow
(CDAF) model (Narayan et al.~2000, Quataert \& Gruzinov 2000,
Abramowicz et al.~2002a, Igumenshchev 2002), which is based on the
assumption that a hot accretion flow is convectively unstable (Narayan
\& Yi 1994). In this model, inward angular momentum transport by
convection and outward transport by viscous stresses almost cancel
each other. A convective envelope solution can then be constructed
which has a conserved (outward) convective luminosity and
automatically produces a flat density profile. The gas constantly
moves in and out in turbulent convective eddies, and this motion gives
the impression that there are large fluxes of inflowing and outflowing
matter.  However, none of the outgoing gas really escapes, and the net
accretion rate is quite small.  There is unresolved discussion in the
literature on whether the CDAF model can be applied to MHD accretion
flows (Stone \& Pringle 2001, Hawley \& Balbus 2002, Narayan et
al.~2002).

The ADIOS and CDAF scenarios are very different from each other and it
would seem that numerical simulations ought to be able to discriminate
between them easily. In this context, the key question is: how strong
is the ``real'' wind in a simulated hot accretion disk? This is not
easy to answer.  Returning to equations (\ref{inflowrate}),
(\ref{outflowrate}), note that the integrals are computed at each
instant of time using instantaneous velocities, and the integrals are
then averaged over $t$ and $\phi$ to obtain $\dot{M}_{\rm in}(r)$ and
$\dot{M}_{\rm out}(r)$. This procedure gives undue importance to
turbulent motions. Especially at large radii, where a given turbulent
eddy will consist of roughly half the gas moving in and half moving
out, one is likely to overestimate both the inflow rate and outflow
rate. A parcel of gas that is moving out at a particular time will
likely soon turn around and begin to flow in.  Thus, the inflow and
outflow rates estimated via equations (\ref{inflowrate}) and
(\ref{outflowrate}) will both be overestimates of the true values.

An alternative approach is to move the $t\phi$ average inside the
integrals, i.e., to integrate $\min(\langle\rho v_r\rangle_{t\phi},
0)$ and $\max(\langle\rho v_r\rangle_{t\phi}, 0)$ (see Narayan et
al.~2012b). This eliminates contributions from the to-and-fro motion
due to turbulence, and not surprisingly produces substantially lower
estimates for the mass outflow rate (see the dashed lines in the right
panel of Fig.~\ref{Fig:inflowrate}).  However, this procedure too is
problematic --- it could underestimate the real outflow rate if some
genuine outflowing streams wander around in three-dimensional space
(Yuan et al.~2012a). Therefore, the estimated mass outflow rate
obtained by this method is a lower limit.

Yuan et al.~(2012a) present an alternative way to roughly estimate the
strength of the wind. They calculate and compare the various
properties of inflow and outflow such as angular momentum and
temperature. They find that the properties are quite different
whereas, if the inflowing and outflowing motion were dominated by
turbulence, the properties would be roughly similar\footnote{In the
  case of convection (the HD case in Yuan et al.~2012a), some
  differences are expected between inflow and outflow, but perhaps
  this will not affect the final conclusion.}.  They therefore
conclude that systematic outflow must exist and the rate of real
outflow should be a significant fraction of that indicated by
eq.~(\ref{outflowrate}). Yuan et al.~(2012a) investigate the
production mechanism of the wind. Based on the much larger angular
momentum of outflow compared to inflow, they argue that the
magneto-centrifugal force must play an important role. They also
briefly discuss the velocity of the wind (see also Li et al.~2013a).

An influential concept in theoretical discussions of outflows is the
Bernoulli parameter $Be$, which is the sum of the kinetic energy,
potential energy and enthalpy.  It measures the ratio of energy flux
to mass flux. For a steady inviscid hydrodynamic flow, $Be$ is
conserved along streamlines.  Therefore, any parcel of gas with a
positive $Be$ can escape to infinity while a parcel with negative $Be$
cannot.  One-dimensional hot accretion flow models often have $Be>0$
(Narayan \& Yi 1994, 1995a; Blandford \& Begelman 1999), which is
interpreted as a strong clue that these flows should experience heavy
mass loss in winds. However, note that $Be$ is not conserved if the
flow is either viscous or non-steady. Therefore, $Be$ is not a useful
parameter for describing gas in the interior of a turbulent disk. The
situation is somewhat better in the case of outflows, which tend to be
more laminar and quasi-steady.

In the area of numerical simulations of non-radiative accretion flows,
Igumenshchev \& Abramowicz (2000) were among the first to explore the
connection between the Bernoulli parameter and outflows. They found
that HD simulations with a large value of the viscosity parameter
$\alpha\ \sgreat\ 0.3$ have well-defined outflows with $Be>0$, whereas
simulations with a smaller $\alpha\ \sles\ 0.1$ have outflowing gas
with $Be<0$, i.e., the outward-moving gas in the latter models is
gravitationally bound and cannot escape to infinity. More recently,
Yuan et al.~(2012a) have carried out a detailed study of $Be$ in HD
and MHD simulations. They find that in the HD case the value of $Be$
of outflowing gas is always larger than that of inflowing gas.

In the case of MHD flows, the definition of $Be$ must be modified to
include the contribution of the magnetic field. The necessary
expression is well-known in the theory of relativistic hydromagnetic
winds, e.g., the ``total energy-to-mass flux ratio'' $\mu$ in Vlahakis
\& Konigl (2003; also $J$ in Lovelace et al.~1986). Sadowski et
al.~(2013a) analyzed 3D GRMHD simulations using a general relativistic
version of $\mu$ and found that gas with $\mu>0$ has an
outward-pointing velocity (outflow), while gas with $\mu<0$ has an
inward-pointing velocity (inflow). This result appears to confirm the
usefulness of the Bernoulli parameter as a diagnostic for MHD winds in
hot accretion flows. Note that the analysis was carried out using
time-averaged quantities in quasi-steady state, where the Bernoulli
parameter is expected to be particularly well conserved.

In addition to directly estimating the strength of mass outflows in
simulated hot accretion flows, the convective stability of the gas in
MHD simulations may be analyzed using the Hoiland criteria (Narayan et
al.~2012b, Yuan et al.~2012a). This reveals that gas is convectively
stable over most of the accretion flow, in contrast to HD accretion
simulations which are convectively unstable
(\S\ref{hydrosimulation}). It thus appears that the magnetic field in
MHD simulations stabilizes the gas against convection. Pen et
al.~(2003) named this state of affairs ``frustrated convection''.

On the whole, current results seem to favor the ADIOS model over the
CDAF model, but not overwhelmingly. It is likely that the truth
involves some combination of the two models. More work is required to
clarify this issue.  In particular, simulations covering a
significantly larger dynamic range in radius than currently possible
will be required before we can hope to obtain unambiguous results.

Large dynamic range 3D simulations are especially important for
estimating two critical parameters: (i) the mass-loss index $s$, and
(ii) the radius $R_{\rm in}$ inside which mass loss is unimportant. It
is vital to know the values of these parameters if we wish to
calculate quantitative global models of hot accretion flows
(\S\ref{globalsolution}) and to apply these models to real systems
(\S\ref{applications}).

\subsection{Effect of radiation}

Radiation is nearly always ignored in hydro and MHD simulations of hot
accretion flows. However, a few studies have considered optically thin
radiative cooling.  Yuan \& Bu (2010) included bremsstrahlung
radiation in the energy equation in their HD simulations and recovered
the ADAF and LHAF solutions when they varied the mass accretion
rate. Surprisingly, their simulated LHAF was convectively unstable,
whereas 1D models predict that the entropy gradient should be stable
(\S\ref{solutionenergetics}). Apparently, the 2D structure of the flow
permits an unstable entropy gradient to survive, although this
behavior is not understood.  Li et al.~(2013a) again included
bremsstrahlung cooling and showed that, by changing the mass supply
rate outside the Bondi radius, they could successfully reproduce both
a cool thin disk at high $\dot{M}_{\rm BH}$ and a hot accretion flow
at lower $\dot{M}_{\rm BH}$.

Ohsuga and collaborators (e.g., Ohsuga et al.~2009, Ohsuga \&
Mineshige 2011) have carried out simulations with full radiative
transfer and have studied a wide range of accretion rates. They
recover both the cold and hot accretion flow solutions at appropriate
values of $\dot{M}_{\rm BH}$. Radiation generally does not appear to
have a significant effect on the dynamics of their hot solutions.
However, in a recent study, Dibi et al.~(2012) include optically thin
cooling and find that, when $\dot{M}_{\rm
  BH}\ \sgreat\ 10^{-7}\dot{M}_{\rm Edd}$, radiative cooling can
significantly affect the density and temperature. Their result is
likely to be sensitive to the particular prescription they used to fix
the electron temperature in the two-temperature plasma. Nevertheless,
their study highlights the fact that, above some accretion rate,
numerical simulations need to include radiation self-consistently.

\subsection{Effect of low collisionality}
\label{collisionless}

Most studies of hot accretion flows are based on a fluid
approximation, specifically MHD. However, the density of the accreting
gas is often so low that the flow is macroscopically collisionless,
and one must carry out a kinetic analysis to determine whether MHD
simulations can capture the relevant physics.

In the case of the MRI, a kinetic treatment is in principle required
whenever the wavelength of the fastest growing mode is smaller than
the collisional mean free path. Quataert et al. (2002) found that,
while the MRI instability criterion is the same in kinetic theory as
in MHD, the growth rates of modes are different. The nonlinear
development of the kinetic MRI has been studied using numerical
simulations based on a fluid model with kinetic effects added (Sharma
et al. 2006) as well as the more precise particle-in-cell technique
(Riquelme et al. 2012).  The non-linear evolution of the axisymmetric
kinetic MRI is found to be qualitatively similar to that of the
standard MHD MRI.

The low collisionality of hot accretion flows also has an effect on
thermal conduction. For a magnetized collisionless accretion flow, the
collisional mean free path of electrons is larger than the electron
Larmor radius, and thermal conduction is the dominant mode of heat
transport. Conduction tends to be primarily along the magnetic field
lines (though cross-field diffusion is not as small as often imagined,
Narayan \& Medvedev 2001). Among other things, anisotropic conduction
modifies the convective stability criterion, as shown by Balbus
(2001). The instability in this case is usually referred to as the
magnetothermal instability (MTI). Local MHD simulations with
anisotropic electron thermal conduction have demonstrated that the MTI
amplifies the magnetic field and causes a substantial convective heat
flux (Parrish \& Stone 2007). Sharma et al. (2008) investigated the
effects of the MTI on non-rotating accretion flows and confirmed the
main results of local simulations. Bu et al. (2011) extended this
study to the case of a rotating accretion flow and found that the MTI
and MRI operate independently and can cooperatively amplify the
magnetic field.

Thermal conduction in a hot collisionless accretion flow can directly
affect the dynamics by flattening the temperature profile (e.g.,
Quataert 2004, Tanaka \& Menou 2006, Johnson \& Quataert 2007). This
mechanism has been invoked as an alternative explanation for the very
low mass accretion rate in Sgr~A* (\S\ref{sgra}; Shcherbakov \&
Baganoff 2010).

\section{Applications}
\label{applications}

\subsection{Galactic Center black hole: Sgr A*}
\label{sgra}

Sagittarius A* (hereafter Sgr A*), the compact radio source at the
center of our Galaxy, is a unique laboratory for studying black hole
accretion. Observations of O and B stars orbiting the Galactic
  Center (Sch\"odel et al.~2002; Ghez et al.~2003, 2008; Gillessen et
  al.~2009a, 2009b; Meyer et al. 2012) provide very strong evidence
  for the presence of a dark compact object of mass $(4.1\pm0.4)\times
  10^6\msun$. Measurements of the size of Sgr A* (Bower et al.~2004,
  Shen et al.~2005, Doeleman et al. 2008) leave little doubt that this
  dark object must be a supermassive black hole. Since Sgr A* is
relatively nearby, there is abundant data to constrain the nature of
the accretion flow (for details on the observations, see Genzel et
al.~2010 and references therein).

Sgr A* spends most of its time in a steady low-luminosity state,
usually referred to as the ``quiescent state'' (\S\ref{quiescent}). A
few times each day, strong variations are seen in the infrared and
X-ray bands, and sometimes also in other wavebands. These fluctuations
are referred to as ``flares'' (\S\ref{flare}). The ADAF model and its
variants explain the main features of the quiescent state. However,
despite important recent progress, the nature of the flares is still
poorly understood.

\subsubsection{Observational constraints on the quiescent state}
\label{quiescent}

The outer boundary of the accretion flow around Sgr A* is generally
assumed to be located at the Bondi (1952) radius, $R_B \sim 10^5 R_S
\approx 0.04\, {\rm pc}\approx 1^{"}$, where the thermal energy of the
external ambient gas is equal to its potential energy in the
gravitational field of the black hole.  Because of the high spatial
resolution of the {\it Chandra X-ray Observatory}, the density and
temperature of gas near $R_B$ can be measured using X-ray observations
(Baganoff et al.~2003), and we can thereby estimate the Bondi mass
accretion rate: $\dot{M}_B\sim 10^{-5} \msunyr \sim
10^{-4}\dot{M}_{\rm Edd}$.  Independently, 3D numerical simulations of
stellar winds accreting on to Sgr A* (Cuadra et al.~2008; see also
Quataert 2004, Shcherbakov \& Baganoff 2010) predict $\dot{M}_B \sim$
few$\,\times10^{-6} M_\odot {\rm yr^{-1}}$, consistent with the above
estimate. The simulations indicate that the gas has a reasonable
amount of angular momentum at the Bondi radius, corresponding to a
circularization radius $\sim 10^4R_s$. Thus, gas cannot fall directly
into the black hole, as in the classic Bondi model, but must accrete
viscously via a hot accretion flow. The measured density and
temperature of gas at the Bondi radius, and its estimated specific
angular momentum, constitute outer boundary conditions that a
successful accretion model must satisfy.

The spectral energy distribution of Sgr A* is shown in
Fig.~\ref{Fig:sgra}.  The radio spectrum has two components: below
$\sim 50$\,GHz the spectrum consists of a power-law, and above this
frequency there is a ``submillimeter bump''. The X-ray emission in the
quiescent state is spatially resolved ($\sim 1.4^{"}$; Baganoff et
al.~2003), consistent with the size of the Bondi radius. The most
outstanding feature of Sgr A* is that its bolometric luminosity is
extremely low: $L_{\rm bol}\sim 10^{36}\ergs \sim 2 \times 10^{-9}
L_{\rm Edd}$. If gas accretes on the black hole at the Bondi accretion
rate $\dot{M}_B$ estimated above, the radiative efficiency of the
accretion flow must be extremely low: $\epsilon_{\rm Sgr~A*} \sim
10^{-6}$ instead of the traditional $\epsilon_{\rm SSD} \approx
10\%$. This ultra-low efficiency is the strongest argument for
invoking an ADAF or other hot accretion flow model instead of a
standard thin accretion disk. A second argument is that the observed
spectrum does not look anything like the multi-temperature blackbody
spectrum expected for a standard thin disk. Note, however, that strong
extinction in the optical and UV could hide much of the emission from
a thin disk, so constraints come mainly from infrared observations
(Falcke \& Melia 1997).

The radio emission at submillimeter and millimeter wavelengths is
linearly polarized at a level of $2-9\%$ (Aitken et al.~2000; Bower et
al.~2003; Marrone et al., 2006, 2007), and the mean rotation measure
between 227 and 343 GHz is $-5.6 \pm 0.7 \times 10^5\, {\rm
  rad\,m^{-2}}$. The latter measurement limits the mass accretion rate
at the central black hole (Agol 2000, Quataert \& Gruzinov 2000).
Current constraints are $\dot{M}_{BH}<2\times 10^{-7}\msunyr$ if the
magnetic field is near equipartition, ordered, and largely radial, and
$\dot{M}_{BH}>2\times 10^{-9}\msunyr$ if the field is
subequipartition, disordered, or toroidal (Marrone et al.~2007). Since
these estimates of $\dot{M}_{BH}$ are very much smaller than
$\dot{M}_B$, it is inferred that most of the gas available at $R_B$
does not fall into the black hole.  Recently, emission lines from
relatively low ionization species were detected (Wang et al.~2013) in
3 million seconds of {\it Chandra} observations. The H-like Fe
K$\alpha$ line was extremely weak, indicating a flat radial density
profile $\rho(r)\propto r^{-1}$ near the Bondi radius, rather than
$r^{-3/2}$ as one expects for the classic ADAF model. The flat density
profile seems to confirm that the mass accretion rate decreases with
decreasing radius (\S\ref{outflow}), consistent with millimeter wave
polarization results. Both observations suggest that the accretion
flow has a significant outflow which causes $\dot{M}$ to decrease with
decreasing radius (\S\ref{outflow}).

\subsubsection{ADAF model for the quiescent state of Sgr A*}
\label{sgramodel}

Narayan et al.~(1995; see also Manmoto et al.~1997; Narayan et
al.~1998a) applied the ADAF model to Sgr A* and showed that the model
explains the main features of the source, viz., an ultra-low radiative
efficiency and an unusual (non thin disk) spectrum.  In these early
studies, the accretion rate was taken to be independent of radius, and
viscous dissipation was assumed to heat only ions ($\delta\ll 1$,
\S\ref{heating}). The most serious defect of this model is that it
predicts a rotation measure orders of magnitude larger than that
observed (\S\ref{quiescent}).

\begin{figure}
\vspace{-1.cm}
\centerline{\psfig{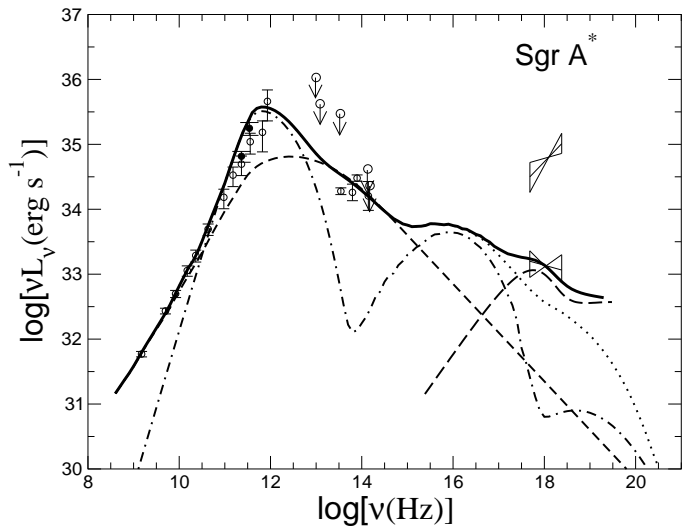}}
\caption{Spectrum corresponding to an ADAF model of the quiescent
  state of Sgr A*, the supermassive black hole at the Galactic Center
  (from Yuan et al.~2003). Circles with errorbars show measurements at
  radio and millimeter wavelengths, circles with arrows correspond to
  infrared upper limits, and the two ``bowties'' show X-ray data in
  the quiescent state (below) and during a bright flare (above).
  Additional data are shown in the infrared waveband (from Sch\"odel
  et al.~2011), which were not available when the model was originally
  developed.  The model spectrum (thick solid line) is the sum of
    three components: synchrotron emission and its Compton humps from
    thermal electrons (dot-dashed line), synchrotron emission from a
    population of non-thermal electrons (short-dashed line), and
    bremsstrahlung emission from electrons near the Bondi radius
    (long-dashed line). The dotted line shows the total synchrotron
    and inverse Compton emission.}
\label{Fig:sgra}
\end{figure}

In the years following this early work, three separate developments
led to a new paradigm. (1) Numerical simulations demonstrated that
outflows are important (\S\ref{outflow}).  (2) Electron heating was
recognized to be more efficient than previously assumed
($\delta\sim0.1-0.5$, \S\ref{heating}). (3) Faraday rotation
measurements indicated that $\dot{M}_{BH} \ll \dot{M}_B$
(\S\ref{quiescent}).  Yuan et al.~(2003, 2004) presented an updated
ADAF model of Sgr A* (sometimes called a RIAF model to distinguish it
from the ``old'' ADAF model) which allowed for an outflow and assumed
more efficient electron heating. The model spectrum is shown in
Fig.~\ref{Fig:sgra}. The submillimeter bump is produced by synchrotron
emission from thermal electrons in the ADAF (dot-dashed line in the
figure -- the additional bumps at higher frequencies are due to
inverse Compton scattering), while the low-frequency power-law radio
spectrum (short-dashed line) is produced by synchrotron radiation from
a small population of nonthermal electrons (following earlier
suggestions by Mahadevan 1998 and \"Ozel et al.~2000). The nonthermal
electrons are usually introduced into the model in an ad hoc fashion,
but are thought to be the result of electron-positron production via
proton-proton collisions (Mahadevan 1998, 1999), magnetic
reconnection, or weak shocks (e.g., Machida \& Matsumoto 2003).
Bremsstrahlung emission of thermal gas near the Bondi radius
(long-dashed line) is responsible for the X-ray emission (Quataert
2002). Sazonov et al.~(2012) recently proposed that the extended X-ray
emission may be dominated by coronal radiation from a population of
low-mass stars. However, Wang et al.~(2013) did not detect the
predicted level of Fe K$\alpha$ emission.

The net radiative efficiency of the Yuan et al.~(2003) model is very
low:\\ $L_{\rm bol}/[\dot{M}_{\rm in}(R_B)c^2] \approx 2\times
10^{-5}$.  The low efficiency is the result of two effects: (i) mass
loss in an outflow gives an effective $\epsilon_{\rm outflow} \equiv
\dot{M}_{\rm BH}/[\dot{M}_{\rm in}(R_B)] \sim 4\times 10^{-2}$, and
(ii) energy advection gives an additional ``real efficiency''
(eq.~\ref{eff}) $\epsilon \equiv L_{\rm bol}/\dot{M}_{\rm BH}c^2 \sim
4\times 10^{-4}$.  The inclusion of an outflow in the model is
consistent with independent modeling of X-ray emission lines (Wang et
al.~2013) as well as numerical MHD simulations
(\S\ref{outflow}). However, there is no direct observational evidence
for any outflowing gas.  Overall, the success of this (by no means
unique) model provides strong evidence for the presence of a hot
advection-dominated accretion flow around Sgr A*. It has also been
used to argue for the existence of an event horizon in this object
(Narayan et al.~1998a, Broderick \& Narayan 2006, Narayan \&
McClintock 2008, Broderick et al.~2009; see also Narayan et al.~1997b,
Menou et al.~1999a, Garcia et al.~2001, McClintock et al.~2004 for
related arguments in the case of quiescent BHBs, and Abramowicz et
al.~2002b for a counter-argument).

More constraints on model parameters such as the mass accretion rate,
the relative importance of the disk versus the jet
(\S\ref{sgraothermodel}), the orientation and magnitude of the BH spin
and the magnetic field, could be obtained by considering additional
millimeter-VLBI observations and polarization data
(\S\ref{futureobs}). This has been done using the Yuan et al.~(2003)
semianalytical model (Huang et al.~2007, Broderick et al.~2011) and in
numerous studies based on MHD simulations (Goldston et al.~2005; Noble
et al.~2007; Sharma et al.~2007b, 2008; Mo\'scibrodzka et al.~2009;
Chan et al.~2009; Kato et al.~2009; Dexter et al.~2010; Shcherbakov et
al.~2012; Dexter \& Fragile 2013). In most of the latter studies the
simulations do not include radiation, so the emergent spectrum is
calculated separately by post-processing the simulation output.

\subsubsection{Multiwaveband flares}
\label{flare}
Flares have been observed in Sgr A* in many wavebands, but are
strongest in X-rays (Baganoff et al.~2001) and infrared (Genzel et
al.~2003, Ghez et al.~2004, Gillessen et al.~2006), where the flux can
increase by up to a factor of 100 and 5, respectively. The variability
timescale ranges from several minutes to three hours, indicating that
the flares must be produced close to the black hole; for comparison,
the light-crossing time of the black hole is $2R_S/c \approx 30$\,s,
and the orbital period at the ISCO for a non-rotating black hole is
$2\pi/\sqrt{GM/R_{\rm ISCO}^3}\approx 2000$\,s.  Many multiwaveband
campaigns (e.g., Eckart et al.~2004, 2006; Yusef-Zadeh et al.~2006,
2009; Dodds-Eden et al.~2009; Trap et al.~2011) have provided valuable
information on flare spectra, polarization, time lags between
different wavebands, and occurrence rates. The observations suggest
that flares in different wavebands are likely physically related.

So far, most theoretical flare models are phenomenological and have
focused mainly on interpreting the observed spectrum (e.g., Markoff et
al.~2001; Yuan et al.~2004; Dodds-Eden et al.~2009, 2010). While the
IR flare is generally believed to be due to pure synchrotron emission,
there is still some debate on whether the X-ray flare is due to
synchrotron (Yuan et al.~2003, 2004; Dodds-Eden et al.~2009),
synchrotron self-Compton, or inverse Compton of external radiation
(Markoff et al.~2001; Eckart et al.~2004, 2006; Yusef-Zadeh et al.~2009).
Yusef-Zadeh et al.~(2006) found that, in the radio band, the peak
flare emission at $43$ GHz leads that at 22 GHz by $\sim20-40$
minutes. They interpret this in terms of a van der Laan (1966)-like
expanding plasma blob model. The blob is ejected from the accretion
flow and becomes optically thin as it expands. The maximum emission at
any given frequency occurs when the blob transitions from optically
thick to thin at that frequency. Thus, the peak naturally occurs later
at longer wavelengths. A similar process may be happening in GRS
1915+105 and other black hole sources (Fender \& Belloni 2004; see
references in Yuan et al.~2009a), and strongly suggests episodic jet
ejections (\S\ref{jet3}). In analogy with coronal mass ejections
associated with solar flares, the ejections in Sgr A* might be caused
by magnetic reconnection in a corona, and the same process may also be
responsible for the flares (Yuan et al.~2009a).

More multiwaveband observations will clarify many remaining
puzzles. At the same time, there is a need for theoretical models that
combine detailed gas dynamics with radiative processes.

\subsubsection{Alternative models of Sgr A*}
\label{sgraothermodel}

Two alternative models of Sgr A* have been discussed in the literature:
the jet model (Falcke \& Markoff 2000, Markoff et al.~2001, Yuan et
al.~2002a, Markoff et al.~2007), and the spherical accretion model
(Melia 1992, Melia et al.~2001).

In one version of the jet model (Falcke \& Markoff 2000, Markoff et
al.~2007), it is proposed that all the emission from radio to X-rays
is produced by the jet. In an alternate version, called the jet-ADAF
model (Yuan et al.~2002a), only the radio spectrum below $\sim 50$ GHz
is produced by the jet while the rest of the emission is assumed to
come from the ADAF. No radio jet has been convincingly detected in Sgr
A* (but see Li et al.~2013b), even though this supermassive black hole
is right in our Galaxy and has been observed in the radio with both
high sensitivity and very high spatial resolution.

The spherical accretion model (Melia 1992) is similar to the ADAF
model in that the accretion flow is assumed to be very hot, with the
temperature being nearly virial. However, in contrast to the ADAF and
other hot accretion flow models, the gas here is one-temperature. In
addition, the angular momentum of the gas is assumed to be extremely
small, with a circularization radius of only $\sim5-10 R_s$, which is
rather extreme. If gas at the Bondi radius has any reasonable angular
momentum, as seems likely (e.g., the numerical work of Cuadra et
al.~2008), accretion cannot take place spherically but must proceed
via a viscous rotating flow.

\subsubsection{Future Observations}
\label{futureobs}

As the nearest supermassive black hole, Sgr A* has always been a
favorite target for observational campaigns.  Two near-term
opportunities may provide significant new information on the nature of
the accretion flow around this black hole.

Gillessen et al.~(2012) discovered a dense cloud of gas called G2 on a
highly eccentric orbit around Sgr A*. Pericentric passage is expected
in early 2014 at a distance of $\sim 2000R_S$ from the black hole
(Phifer et al.~2013, Gillessen et al.~2013).  The interaction of G2
with the hot accretion flow could potentially be observed in radio,
infrared or X-rays (Narayan et al.~2012a, Sadowski et al.~2013b, Saitoh
et al.~2013, Yusef-Zadeh \& Wardle 2013, Crumley \& Kumar 2013,
Shcherbakov 2013). If a signal is detected, it will provide
information on the properties (density, temperature, magnetic field)
of the accretion flow in a region ($R\ \sgreat\ 10^3R_S$) where we
have hitherto had no observational constraint. In addition, gas
stripped from G2 is expected to accrete on the black hole on a viscous
time, causing the quiescent radio and millimeter emission of Sgr A* to
increase. A detection of this increase will provide a direct
measurement of the viscous time at the orbit pericenter, which
currently is poorly constrained.  A number of HD and GRMHD simulations
of the G2 encounter have already been carried out (e.g., Moscibrodzka
et al.~2012, Anninos et al.~2012, Burkert et al.~2012, Schartmann et
al.~2012, 2013, Sadowski et al.~2013b, Abarca et al.~2013).  Coupled
with future observations, such work should provide new information on
the nature of the accretion flow in Sgr A* at intermediate radii.

In recent years, ultra-high angular resolution millimeter wave
interferometry has become a reality and the first detections have been
made of event-horizon scale structure in Sgr A* (Doeleman et al.~2008,
Fish et al.~2011) and M87 (Doeleman et al.~2012). From these
observations some inferences have already been made on the spins of
the black hole and the nature of the accretion flow and jet near the
horizon (e.g., Fish et al.~2009, 2011; Broderick et al.~2009,
2011ab). Future plans are focused on commissioning the Event Horizon
Telescope (Doeleman et al.~2010), which will carry out long baseline
millimeter wave interferometry using a network of up to eight
telescopes spread all around the world. With this array, images of Sgr
A* will be measured with unprecedented sensitivity and angular
resolution, and information will be obtained both on polarization and
time variability. It is anticipated that the Event Horizon Telescope
will finally provide direct information on the physics of the hot
accretion flow in the vicinity of Sgr A*'s horizon. Questions such as
the relative importance of the disk versus the jet; the density,
temperature and optical depth (at millimeter wavelengths) of the
accreting plasma; and the direction and topology of the magnetic field
will hopefully be answered by means of direct observations. In
anticipation, numerical codes are being developed and a number of
numerical simulations have already been carried out (e.g.,
Moscibrodzka et al.~2009, Dexter et al.~2010, Shcherbakov et al.~2012,
Dexter \& Fragile 2013, Chan et al.~2013).  Much more work is
anticipated in the future.

\subsection{Other low-luminosity sources}
\label{llagn}

The hot accretion flow solution is essentially independent of black
hole mass. If $\dot{M}_{\rm BH}$ is scaled by the Eddington rate and radius is
scaled by the Schwarzschild radius, many properties of the solution,
notably the gas temperature and the Eddington-scaled luminosity, are
independent of mass (see \S\ref{self-similar}). Because of this, hot
accretion flow systems, such as low-luminosity active galactic nuclei
(LLAGNs) and black hole binaries (BHBs) in the hard and quiescent
state, show similar properties, despite the large disparity in their
black hole masses. We begin by introducing LLAGNs and BHBs, and then
describe the role of hot accretion flows in determining their
properties.

\subsubsection{Introduction}
\label{llagnintro}

{\it LLAGNs.} Although virtually every galaxy with a bulge has a
supermassive black hole in its nucleus, at any given time only a
small fraction of these black holes have luminosities close to
Eddington. The vast majority are LLAGNs with luminosities spanning the
range $L_{\rm bol}/L_{\rm Edd}\approx 10^{-9}-10^{-1}$ (Ho 2008,
2009). As in the case of Sgr A*, most of these black holes have
considerable gas available for accretion close to their Bondi radii
(Fabian \& Canizares 1988). The fact that the black holes are dim thus
suggests that they must be accreting via a radiatively inefficient
mode, i.e., a hot accretion flow (Fabian \& Rees 1995, Di Matteo et
al.~2000, Ho 2009, Russell et al.~2013b). Other distinctive features
of LLAGNs confirm this suspicion.

The ``big blue bump'', a characteristic spectral feature associated
with a thin accretion disk around a supermassive black hole, is absent
in LLAGNs (Ho 1999, 2008; Chiaberge et al.~2006; Eracleous et
al.~2010; Younes et al.~2012). In the language of $\alpha_{\rm ox}$,
defined as the two-point spectral index between 2500\,\AA\ and 2\,keV,
LLAGNs have $\alpha_{\rm ox}\ \sgreat\ -1$ while quasars and Seyferts
have $\alpha_{\rm ox}\approx -1.4$ and $-1.2$, respectively. The
optical-UV slope is also exceptionally steep (Ho 2008). These
observations strongly suggest that LLAGNs do not have a thin disk in
their inner regions; however, a disk may be present at larger radii,
as indicated by a ``big red bump'' in their spectra (Lasota et
al.~1996a, Quataert et al.~1999, Gammie et al.~1999, Yuan et al.~2002b,
Chiang \& Blaes 2003, Ptak et al.~2004, Yuan \& Narayan 2004, Nemmen
et al.~2006, Wu et al.~2007, Yu et al.~2011).\footnote{Based on the
  fact that LLAGNs lie on the low-luminosity extrapolation of the
  well-known relation between $\alpha_{ox}$ and luminosity, Maoz
  (2007) argues that LLAGNs do not differ appreciably from luminous
  AGNs, and hence that LLAGN accretion disks are similar to disks in
  luminous AGNs. However, he does not provide a physical explanation
  for the correlation, whereas Yu et al.~(2011) show that the trend is
  naturally explained in the framework of hot accretion flows.} As
further confirmation, the iron K$\alpha$ line, which is commonly
attributed to X-ray fluorescence off a cold accretion disk extending
close to the black hole, is weak or absent (e.g., Fabbiano et
al.~2003, Ptak et al.~2004, Binder et al.~2009, Younes et al.~2011,
Kawamuro et al.~2013).

{\it BHBs.} BHBs have a number of distinct spectral states (see
Zdziarski \& Gierli\'nski 2004, McClintock \& Remillard 2006, Done et
al.~2007, Zhang 2013, Poutanen \& Veledina 2014 for reviews). The most
notable among these, in order of decreasing luminosity, are the soft
or thermal state, the hard state, and the quiescent
state.\footnote{There is also a very high state or steep power-law
  state which usually, but not always, occurs at an even higher
  luminosity than the thermal state. This state is poorly understood.}
The soft state is found at luminosities down to $\sim 1.5\% L_{\rm
  Edd}$ (Kalemci et al.~2013). It is characterized by a strong
blackbody or thermal spectrum, and is well described by the standard
thin disk model.  The hard state is found at lower luminosities and
differs dramatically.  Its spectrum has only a weak blackbody
component and is dominated instead by a strong hard power-law with a
cutoff at $\sim 100$ keV. While the soft state has little time
variability, the hard state is highly variable and often exhibits
quasi-periodic oscillations (QPOs, Remmilard \& McClintock 2006). In
addition, the hard state has a continuous, steady jet, whereas the
soft state almost never has jets (Fender 2006). All of these
differences indicate that the hard state must correspond to a very
different regime of accretion compared to the thin disk.

Many studies over the years have shown that the power-law component in
the hard state must be produced by thermal Comptonization in a hot
plasma with a temperature $kT\sim 100$\,keV and optical depth
$\tau\sim 1$ (e.g., Zdziarski et al.~ 1998).  The ADAF model has the
correct density, electron temperature and optical depth needed to
reproduce the observed spectrum. Moreover, the luminosity at which a
fading thermal state source switches to the hard state ($\sim 1.5\%
L_{\rm Edd}$) is reasonably consistent with the maximum mass accretion
rate of a hot accretion flow (Fig.~\ref{efficiency}).  At luminosities
below $\sim 10^{-3}L_{\rm Edd}$, the hard state merges smoothly with
the quiescent state, which then extends down to as low as
$L\sim10^{-9}L_{\rm Edd}$. There is no clear boundary between the hard
and quiescent states, so it is likely that the same accretion physics
operates in both.

BHBs show a very interesting hysteresis effect in their transitions
between the soft and hard state. Whereas with decreasing luminosity
the soft-to-hard transition in most sources happens at roughly the
same luminosity $\sim1.5\%L_{\rm Edd}$, with increasing luminosity the
hard-to-soft transition can occur at any of a wide range of
luminosities extending up to $L\sim 30\%L_{\rm Edd}$ (Zdziarski \&
Gierli\'nski 2004, Done et al.~2007, Yu \& Yan 2009).
There is as yet no convincing physical explanation for this hysteresis
phenomenon, though Meyer-Hofmeister et al.~(2005) and Liu et
al.~(2005) show that their disk evaporation model can reproduce the
observations when Compton cooling is included.
Done et al.~(2007) have an alternative explanation, arguing that it is
simply a matter of time scales. As $\dot{M}_{\rm BH}$ increases, it takes a
viscous time for the truncated thin disk (see \S\ref{SSDADAFgeometry}
below) to move down to the ISCO, and during this time the luminosity
continues to increase and goes well above the threshold value before
the transition is completed. This plausible scenario needs to be
confirmed with detailed models.

\subsubsection{Accretion geometry: Truncated thin disk plus hot accretion flow}
\label{SSDADAFgeometry}

\begin{landscape}
\begin{figure}
\vspace{1cm}
\hskip 0.0truein
\psfig{figure=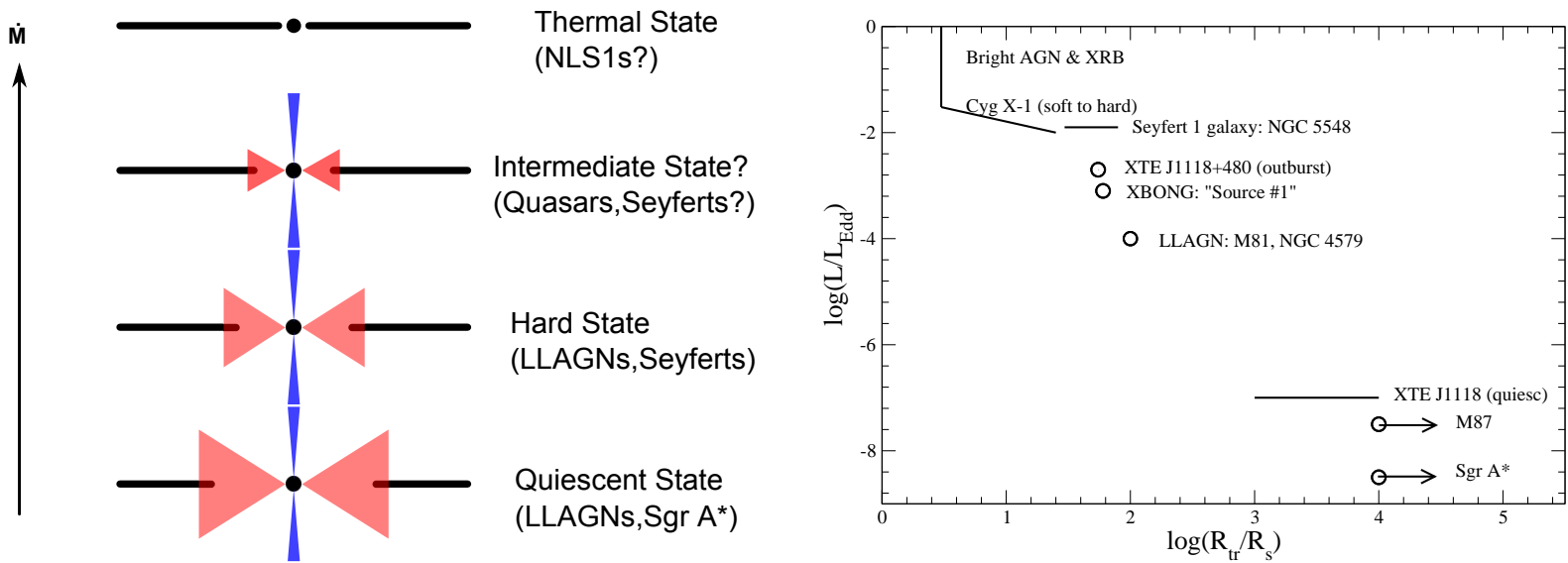,width=10cm,angle=0}
\hskip 0.3truein
\psfig{figure=fig7b.eps,width=10cm,angle=0}
\caption{
{\em Left}: Schematic diagram showing as a function of the mass
accretion rate $\dot{M}_{\rm BH}$ the configuration of the accretion
flow in different spectral states of BHBs (adapted from Esin et
al.~1997, Narayan \& McClintock 2008).  Possibly equivalent AGN
classes are indicated in parentheses.  Red triangles indicate the hot
accretion flow while thick black horizontal lines represent the
standard thin disk. The transition radius $R_{\rm tr}$ where the thin
disk is truncated becomes smaller with increasing $\dot{M}_{\rm BH}$.
In the thermal state, the disk is not truncated and its inner edge is
located at the ISCO.
{\em Right}: Plot of the Eddington-scaled accretion luminosity
$L/L_{\rm Edd}$ versus $R_{\rm tr}$ as deduced from observations. The
transition radii were estimated by modeling spectra of individual
LLAGNs and BHBs (from Yuan \& Narayan 2004).}
\label{transitionradius}
\end{figure}
\end{landscape}

Most BHBs in the hard and quiescent state, as well as many LLAGNs, are
deduced to have a two-zone accretion flow consisting of a cool thin
disk at large radii and a hot accretion flow at small
radii\footnote{In the case of Sgr A* and some elliptical galaxies such
  as M87, the flow starts out hot at the Bondi radius and remains hot
  throughout, so there is apparently no thin disk at large
  radii.}. This configuration is illustrated in the left panel of
Fig.~\ref{transitionradius}. The main parameter that controls the
transition radius $R_{\rm tr}$ between the two zones is the mass
accretion rate $\dot{M}_{\rm BH}$. When $\dot{M}_{\rm BH}$ is above
the maximum allowed for the hot accretion flow solution
(\S\ref{solutionsummary}), only the thin disk solution is available,
so the thin disk extends all the way down to the ISCO. This
corresponds to the soft state. When $\dot{M}_{\rm BH}$ becomes
smaller, the source first enters an intermediate state and then, with
decreasing $\dot{M}_{\rm BH}$, progresses to the hard state and
finally the quiescent state. In these latter states, the thin disk is
truncated at a radius $R_{\rm tr}> R_{\rm ISCO}$ and the region inside
$R_{\rm tr}$ is occupied by a hot accretion flow. The radial extent of
the hot zone increases with decreasing $\dot{M}_{\rm BH}$.  This
``truncated thin disk plus hot inner accretion flow'' configuration
was first proposed by Shapiro et al.~(1976) to explain the hard state
of Cyg X-1; however, their model was based on the unstable SLE
solution. A similar concept, but using the ADAF solution, was later
developed by Narayan (1996, see also Poutanen et al.~1997 and Esin et
al.~1997) to explain the various spectral states of BHBs.

While there is some understanding of the dynamics of the transition
from the outer thin disk to the inner hot accretion flow (e.g.,
Abramowicz et al.~1998, Manmoto et al.~2000), the physical reason why
cold gas in the outer disk is converted into hot gas on the inside is
not fully understood. It is likely that some combination of the
following models is responsible: ``evaporation'' model (Meyer \&
Meyer-Hosmeister 1994; Liu et al.~1999, 2011; Meyer et al.~2000, 2007;
R\'o\.za\'nska \& Czerny 2000; Spruit \& Deufel 2002; Mayer \& Pringle
2007; Taam et al.~2012); ``turbulent diffusion'' model (Honma 1996,
Manmoto \& Kato 2000, Manmoto et al.~2000); ``large viscosity'' model
(Gu \& Lu 2000, Lu et al.~2004). Of these, the evaporation model has
been studied most extensively, though all three models predict that
$R_{\rm tr}$ should increase with decreasing $\dot{M}_{\rm BH}$.  As the panel
on the right in Fig.~\ref{transitionradius} shows, this prediction is
in agreement with empirical data based on modeling spectra of
individual sources (Yuan \& Narayan 2004).\footnote{Not all systems
  follow the trend shown in the right panel of
  Fig.~\ref{transitionradius}.  For BHBs with short orbital periods,
  the mass transfer stream from the companion star circularizes at a
  fairly small radius $R_{\rm circ}$. When these systems go into the
  quiescent state, the continued supply of cold gas at $R_{\rm
    circ}$ will ensure that the transition radius is pinned close to
  $R_{\rm circ}$ (Menou et al.~1999b). A similar situation is possible
  even in the case of very low luminosity AGN. If the AGN is fueled
  from the external medium by cold gas clouds with low angular
  momentum, the gas clouds will first circularize and form a thin disk
  before ``evaporating'' into a hot accretion flow (Inogamov \&
  Sunyaev 2010).

The value of $R_{\rm tr}$ in the case of XTE J1118+480 shown in
Fig. ~\ref{transitionradius} is significantly smaller than that
obtained in later work by Yuan et al. (2005; see
\S\ref{modelingresultsBHB} for details). The reason for the
discrepancy is explained in Yuan et al. (2005). This does not affect
the main conclusion.}

\subsubsection{Modeling observations of LLAGNs}
\label{modelingresultsllagn}

Following the initial application of the ADAF model to Sgr A* (Narayan
et al.~1995), a number of authors used similar ideas to explain a
variety of observations of other LLAGNs: nearby elliptical galaxies
(Fabian \& Rees 1995; Reynolds et al.~1996; Di Matteo et al.~2000,
2001; Loewenstein et al.~2001; Ho et al.~2003; Fabbiano et al.~2003),
LINERs (Lasota et al.~1996a, Quataert et al.~1999, Gammie et al.~1999,
Yuan et al.~2002b, Pellegrini et al.~2003, Ptak et al.~2004, Nemmen et
al.~2006, Xu \& Cao 2009, Liu \& Wu 2013, Nemmen et al.~2014),
BL Lac objects (Maraschi \& Tavecchio 2003), FR I sources (Reynolds
et al.~1996, Begelman \& Celotti 2004, Wu et al.~2007, Yuan et
al.~2009c, Yu et al.~2011), X-ray Bright Optically Normal Galaxies
(Yuan \& Narayan 2004), and even Seyferts (Chiang \& Blaes 2003;
Yuan \& Zdziarski 2004).

NGC~1097, a famous LINER, is an interesting case. It is the first and
best-studied LLAGN to display broad, double-peaked H$\alpha$ and
H$\beta$ emission lines (e.g., Storchi-Bergmann et al.~1997). Such
double-peaked lines are believed to be the result of irradiation of a
truncated thin disk, most likely by radiation from the inner hot
accretion flow (Chen \& Halpern 1989). In the case of NGC~1097, the
H$\alpha$ line profile requires the transition radius to be at $R_{\rm
  tr}\approx 225 R_S$ (Storchi-Bergmann et al.~1997). Using the
truncated thin disk plus hot accretion flow model, Nemmen et
al.~(2006) successfully modeled the optical-to-X-ray continuum
spectrum of the source. In their model, the truncated thin disk
dominates the optical-UV emission while the inner hot accretion flow
produces the X-ray emission.  Impressively, the $R_{\rm tr}$ they
require to fit the continuum spectrum agrees very well with the
$R_{\rm tr}$ estimated from fitting the double-peaked H$\alpha$ line.

\subsubsection{Modeling observations of BHBs}
\label{modelingresultsBHB}

\begin{figure}
\hskip 0.0truein
\psfig{figure=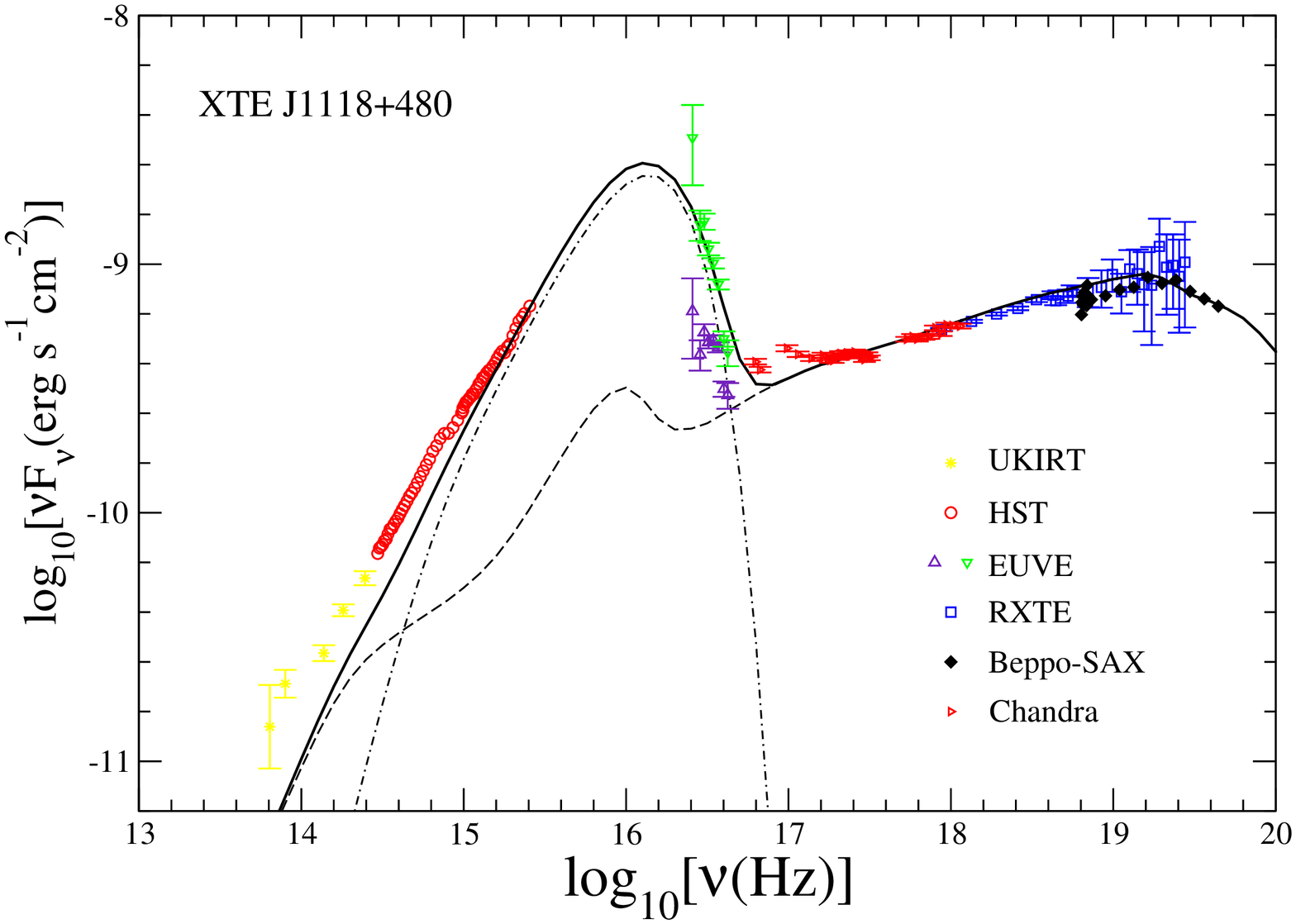,width=6.8cm,angle=270}
\hskip -0.1truein
\psfig{figure=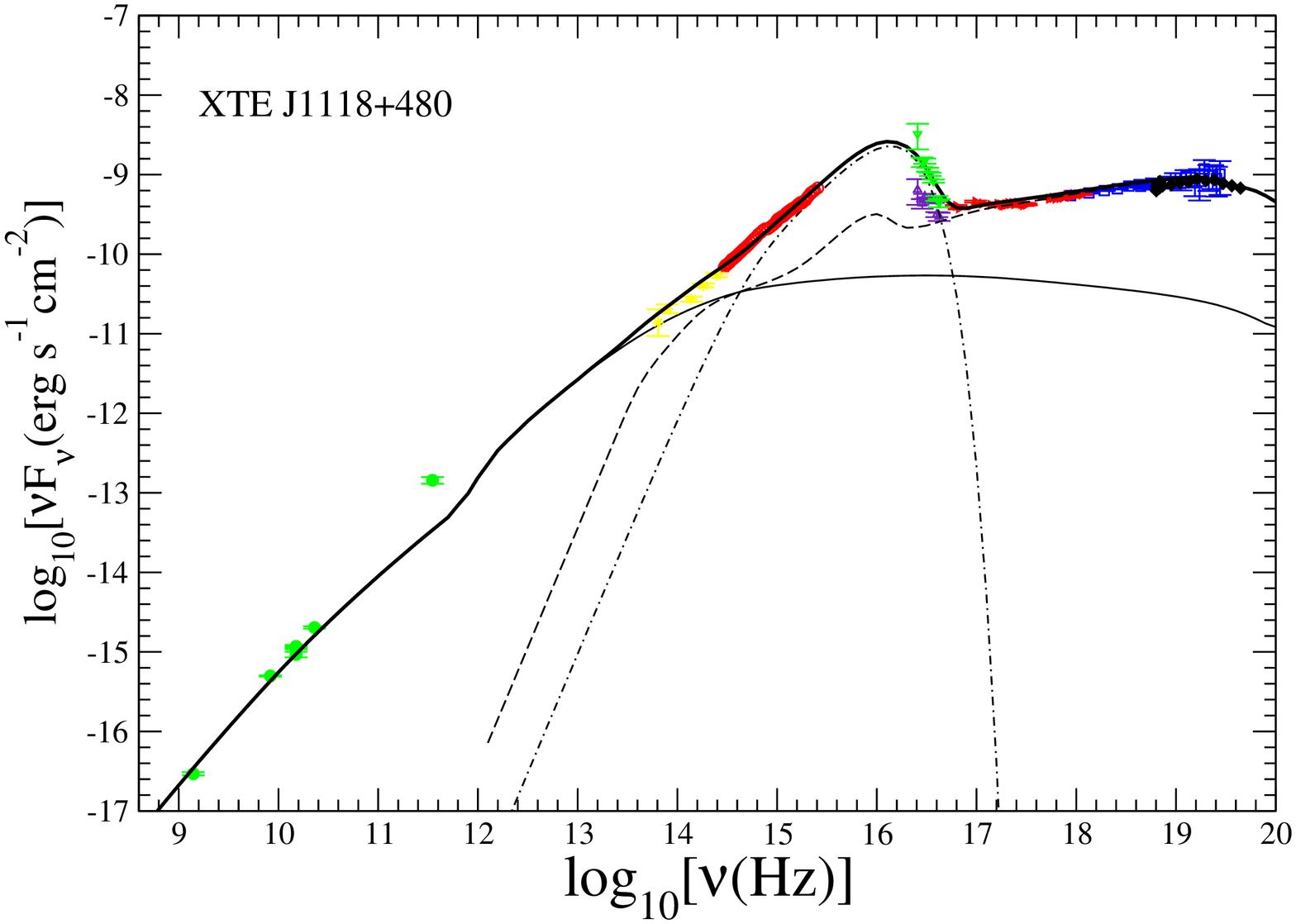,width=6.80cm,angle=270}
\caption{Modeling the spectrum of the BHB XTE J1118+480 in the hard
  state (from Yuan et al.~2005). {\em Left}: Calculated spectrum for a
  model consisting of an ADAF (dashed line) plus a truncated thin disk
  (dot-dashed line). The thin disk is essential to fit the
  optical--EUV part of the spectrum, while the hot ADAF is needed to
  fit the power-law X-ray emission. The model fails to explain the
  infrared and radio data. {\em Right}: A model that includes an
  additional jet component (thin solid line). This model explains all
  the observations.}
\label{Fig:1118}
\end{figure}

In the field of BHBs, the truncated thin disk plus hot accretion flow
scenario was first applied to the quiescent state (Narayan et
al.~1996, 1997a; Menou et al.~1999ab)\footnote{When the source is
  extremely dim, the X-ray radiation is likely to come primarily from
  the jet rather than the ADAF (see \S\ref{correlation}).} and soon
after to the hard state (Esin et al.~1997, 1998, 2001).  The source
XTE J1118+480 is a spectacular example where there is (i) good EUV
data which is crucial for constraining the radius of the inner edge of
the thin disk (McClintock et al.~2003), and (2) good timing
information, including a low-frequency QPO ($\sim 0.1$Hz) and
measurements of time lags between different wavebands (see review in
Yuan et al.~2005). Spectral fitting of the EUV data indicates that the
thin disk must be truncated (Esin et al.~2001, Chaty et
al.~2003). Figure \ref{Fig:1118} shows a comprehensive model (Yuan et
al.~2005) which uses a more modern hot accretion flow model, including
a jet.  In this model, the radio/infrared emission is dominated by the
jet, the optical--EUV by the truncated thin disk, and the X-ray region
of the spectrum is produced by the hot accretion flow.  The transition
radius is constrained to be $R_{\rm tr}\approx 300R_S$. The same
truncated thin disk scenario can also explain the low-frequency
QPO. In the model proposed by Giannios \& Spruit (2004; see also
Rezzolla et al.~2003 for a similar idea), the QPO arises from a global
p-mode oscillation of the hot ADAF.\footnote{As mentioned in
  \S\ref{globalsimulation}, another QPO model invokes the precession
  of the ADAF, with the frequency again determined by the size of the
  ADAF.}  The QPO frequency is roughly determined by the Keplerian
frequency at $R_{\rm tr}$ and agrees well with the observed frequency
for $R_{\rm tr}\approx 300R_s$. In addition, the model also
qualitatively explains other timing features such as time-lags between
different wavebands (Yuan et al.~2005, see also Malzac et al. 2004 for
a similar model with similar conclusions).

A number of observations have confirmed the basic features of the
truncated thin disk scenario in BHBs (see reviews by Zdziarski \&
Gierli\'nski 2004, McClintock \& Remillard 2006, Done et al.~2007,
Poutanen \& Veledina 2014): (1) A truncated thin disk is required to
model the thermal spectral component observed in BHBs in the quiescent
state (Narayan et al.~1996, 1997a; Yuan \& Cui 2005) and hard state
(Esin et al.~2001, Di Salvo et al.~2001, Chaty et al.~2003, Yuan et
al.~2005, Cabanac et al.~2009, Tomsick et al.~2009). (2) The observed
transient behavior of BHBs requires a truncated disk (Lasota et
al.~1996b, Menou et al.~2000, Dubus et al.~2001), as does the time
delay between the optical and X-ray outbursts (Hameury et
al.~1997). (3) A reflection component is seen in the X-ray continuum
spectrum in the hard state. When the X-ray spectrum steepens, both the
solid angle subtended by the reflection material and the amount of
relativistic smearing increase, consistent with the truncation radius
moving in and thereby increasing the flux of soft photons irradiating
the inner hot accretion flow (Gilfanov et al.~1999, Zdziarski et
al. 2003). (4) The truncated disk model explains the correlation
between the luminosity and the photon index of the X-ray spectrum in
the hard state (Qiao \& Liu 2013, Gardner \& Done 2013).  (5) The
model naturally explains why the QPO frequency increases with
increasing X-ray luminosity (Cui et al.~1999, Ingram \& Done
2011). (6) Across the soft-hard state transition, there is a sharp
change in observational features such as variability (Kalemci et
al.~2013, and references therein), photon index and X-ray flux (see
Fig.~6 in Zdziarski \& Gierli\'nski 2004), and high-energy cut-off of
the X-ray spectrum (Belloni et al. 2006), all of which suggest that
there must be a substantial qualitative change in the mode of
accretion during this transition.

Most investigations of hot accretion flows tend to focus on
  thermal electrons. However, it is quite plausible that some
  nonthermal electrons will also be present
  (\S\ref{heating}). Compared to a pure thermal model, a hybrid
  thermal-nonthermal model can explain a wider range of observations,
  with the nonthermal electrons playing a role similar to the jet
  component in disk-jet models.  For instance, the hybrid model
  explains the MeV tail in the hard state spectrum of some BHBs
  (Poutanen \& Vurm 2009), the power-law like optical/infrared
  spectrum, and the concave shape of the X-ray spectrum (Poutanen \&
  Veledina 2014).

The hard state often reaches luminosities $\sim10\%L_{\rm Edd}$ during
the hard-to-soft transition. This is a factor of several too high for
an ADAF (\S\ref{onedimension}). Yuan \& Zdziarski (2004) suggested
that such luminous hard state systems, and also some Seyfert galaxies,
might be explained by the LHAF model. This is confirmed in detailed
modeling of XTE J1550$-$564 (Yuan et al.~2007), where the X-ray
spectrum is naturally explained by the LHAF model. Features such as
the slope of the X-ray spectrum, the cutoff energy, and the
normalization are reproduced well. The agreement in the cutoff energy
indicates that the predicted electron temperature is consistent with
that required by observations (e.g., Fig. 3 in Yuan et al.~2007),
although in some cases it is claimed that the ADAF model is too hot
and that a hybrid thermal-nonthermal electron distribution is required
to reconcile the model with observations (e.g., Poutanen \& Veledina
2014).  Some BHBs achieve even higher luminosities of up to $\sim
30\%L_{\rm Edd}$ in the ``bright hard state'' (Gierli\'nski \& Newton
2006). Oda et al.~(2010) propose that a magnetically supported
accretion flow model might explain these systems.

\subsubsection{Assessment of arguments against disk truncation}
\label{notruncation}

The hard state sometimes shows a dim blackbody-like thermal component
and also a broad iron K$\alpha$ line at X-ray luminosities $L_{\rm
  0.5-10\,{\rm keV}} > 10^{-3}L_{\rm Edd}$. These observations have
been used to argue that the thin disk is not truncated but extends
down to the ISCO (e.g., Miller et al.~2006, Rykoff et al.~2007,
Ramadevi \& Seetha 2007, Reis et al.~2010). However, the soft
component in the spectrum typically has only 10\% of the total
observed luminosity. It is hard to understand how a radiatively
efficient thin disk can extend down to the ISCO and yet contribute so
little to the emitted luminosity. In fact, other groups have obtained
different estimates of the disk inner radius using the same data
(e.g., Done \& Gierli\'nski 2006, Done et al.~2007, Gierli\'nski et
al.~2008).  Cabanac et al.~(2009) carried out a detailed analysis of
systematic uncertainties in spectral fitting and concluded that, when
$L_{\rm 0.5-10\,{\rm keV}}\ \sgreat\ 0.01L_{\rm Edd}$ one has $R_{\rm
  tr}\ \sles\ 10R_g$; while, when $10^{-3}L_{\rm Edd}\ \sles\ L_{\rm
  0.5-10\,{\rm keV}}\ \sles\ 10^{-2}L_{\rm Edd}$, the disk inner edge
recedes well away from the ISCO.

Alternate explanations have been proposed for the weak thermal
spectral component in the hard state. These do not require the thin
disk to extend down to the ISCO.  One possibility is that the thermal
component might originate near the inner edge of the truncated thin
disk which is illuminated by hard X-rays from the hot flow (D'Angelo
et al.~2008). Another possibility is that the emission is from cold
clumps in the hot accretion flow (Chiang et al 2010). The same clumps
may also explain timing features in the X-ray emission such as the
power spectrum of rapid aperiodic variability (B\"ottcher \& Liang
1999). Note that, at accretion rates approaching the upper limit of
the hot accretion flow solution, cold clumps are expected to form
naturally as a result of thermal instability in the hot gas (Yuan
2003; see \S\ref{solutionsummary}) or condensation of the hot flow
(R\'o\.za\'nska \& Czerny 2000; Meyer et al.~2007; Liu et al.~2007,
2011; Mayer \& Pringle 2007; Meyer-Hofmeister et al.~2009).

Disk inner radii derived from iron line profile are even more
controversial. Hartnoll \& Blackman (2001) showed that iron lines can
be readily produced in a two-phase (hot gas plus cold clumps)
accretion flow. Provided the clumps are able to survive long enough,
the line profiles are similar to those produced by a thin disk
extending down to the ISCO.

Done \& Diaz Trigo (2010) re-analyzed MOS iron line data in GX 339$-$4
in the hard state. Miller et al.~2006 (see also Reis et al.~2008) had
previously claimed that the data indicated an extremely broad iron
line. However, Done \& Diaz Trigo (2010) showed that the line shape is
strongly affected by pile-up. Furthermore, using the simultaneous PN
timing-mode data, which should not be affected by pile-up, they
obtained a significantly narrower line, which is easily consistent
with a truncated disk. Recently, Plant et al.~(2013) have carried out
a systematic study of the iron line in GX 339$-$4 and have tracked the
evolution of the thin disk inner radius over a range of two orders of
magnitude in luminosity. They find that the data are consistent with
the thin disk being truncated throughout the hard state, with the
truncation radius moving closer to the black hole as the luminosity
increases. In another recent study, Kolehmainen et al.~(2013) used
data on both the weak thermal component and the iron line to constrain
the inner radius of the thin disk.  They find that the data are
consistent with a truncated disk.

The question of whether or not the disk is truncated in the hard state
is crucially important for efforts to measure black hole spin using
X-ray reflection spectroscopy (see Reynolds 2013 for a current
review).  A key assumption of this method is that reflection occurs
from a cool disk that extends down to the ISCO. Effectively, the
measured profile of the iron K$\alpha$ emission line is used to fit
for the radius of the inner edge of the disk. Then, assuming that the
disk edge is located at the ISCO, the black hole spin is
estimated. However, almost all applications of the reflection method
to stellar mass black holes have been carried out on X-ray data in the
hard state. If the cool disk in BHBs in the hard state is truncated
outside the ISCO, as the preponderence of evidence suggests (in the
authors' view), most spin measurements by the reflection method will
be affected. Measurements of the spins of supermassive black holes
(where the reflection method originated and where considerable work
has been done over many years, Reynolds 2013) are not affected by this
criticism unless those systems also have truncated disks.

\subsubsection{Radio/X-ray correlation and the role of jet radiation}
\label{correlation}

Corbel et al.~(2003) and Gallo et al.~(2003) discovered a remarkable
correlation between the radio luminosity $L_R$ and X-ray luminosity
$L_X$ of BHBs in the hard state. Soon after, Merloni et al.~(2003) and
Falcke et al.~(2004) considered the effect of black hole mass $M$ and
showed that the correlation extends also to LLAGNs (see Fig.~9; and
also K\"ording et al.~2006, Wang et al.~2006, Li et al.~2008,
G\"ultekin et al.~2009, Yuan et al.~2009c, de Gasperin et al.~2011,
Younes et al.~2012 for later work).  Their generalized correlation
takes the form (G\"ultekin et al.~2009),
\begin{equation}
\log \left(\frac{L_R}{\ergs}\right)=(0.7\pm0.1)
\,\log\left(\frac{L_X}{\ergs}\right)+(0.8\pm0.3)
\,\log\left(\frac{M}{M_\odot}\right)+(4.8\pm0.2),
\label{fundplane}
\end{equation}
and is referred to as the ``fundamental plane of black hole
activity''.

Using a simple model in which thermal gas in a hot accretion flow is
responsible for the X-ray emission and relativistic electrons in a jet
produce the radio emission, Heinz \& Sunyaev (2003) showed that the
fundamental plane can be naturally explained (see also Merloni et
al.~2003, Heinz 2004, Yuan et al.~2005, Yuan \& Cui 2005, Li et
al.~2008). An alternative explanation has also been advanced in which
the radio and X-ray emission are both produced by the jet (Markoff et
al.~2003)\footnote{Heinz (2004) pointed out that, if the ``cooling
  break'' is properly taken into account for the electron energy
  distribution, the predicted correlation is different from that
  obtained by Markoff et al.~(2003). This is confirmed by Yuan \& Cui
  (2005), who find a steeper correlation when radiation from the jet
  dominates both the radio and X-ray emission
  (eq.~\ref{steepercorrelation}). Plotkin et al.~(2012) therefore
  suggest that supermassive black holes should be excluded from the
  sample since the cooling break ``is a concern'' for these
  objects. However, Zdziarski et al.~(2012) find that the cooling
  break is independent of black hole mass, so the cooling break should
  be equally important for both LLAGNs and BHBs.}.

Most sources in the Merloni et al.~(2003) sample are relatively
luminous and correspond to the upper range of allowed $\dot{M}_{\rm BH}$ for
the hot accretion flow solution. Yuan \& Cui (2005) extrapolated the
coupled ADAF-jet model of Yuan et al.~(2005) to lower luminosities and
predicted that, below a critical luminosity $L_{\rm X,crit}$ given by
${\rm log}\left(L_{\rm X,crit}/{L_{\rm Edd}}\right) \approx
-5.36-0.17~{\rm log}({M}/{\msun})$, the correlation should steepen to
(see segment ``CD'' in the middle panel of Fig.~9),
\begin{equation}
\log\left(\frac{L_R}{\ergs}\right) =1.23
\,\log\left(\frac{L_X}{\ergs}\right) +0.25
\,\log\left(\frac{M}{M_\odot}\right)
-13.45.\label{steepercorrelation}
\end{equation}
This is because radiation from a hot accretion flow at low
$\dot{M}_{\rm BH}$ is roughly $\propto \dot{M}_{\rm BH}^2$
(\S\ref{radiation}) while that from the jet is $\propto \dot{M}_{\rm
  BH}$ (Heinz 2004, Yuan \& Cui 2005). Therefore, at a low enough
$\dot{M}_{\rm BH}$, the X-ray emission from the jet will
dominate. This can explain some otherwise puzzling observations of
quiescent black holes
(Yuan \& Cui 2005). Early models of M87, for example, assumed
that the X-ray emission is produced by the ADAF (e.g., Reynolds et
al.~1996, Di Matteo et al.~2003). However, {\it Chandra} observations
suggest that the emission is dominated by the jet (Wilson \& Yang
2002), which is consistent since M87 has $L_X < L_{\rm X,crit}$.

On the whole, there is not enough data on quiescent BHBs to verify
eq.~(\ref{steepercorrelation}). The two most promising sources are
V404 Cyg (Corbel et al.~2008) and A0620-00 (Gallo et al.~2006), but
the former is not dim enough to explore the low-luminosity end of the
correlation, while the latter has data only in the quiescent state,
not the hard state. The situation is much better in the case of
LLAGNs, where recent work has confirmed the change of slope of the
$L_R$-$L_X$ correlation at low luminosities (Pellegrini et al.~2007,
Wu et al.~2007, Wrobel et al.~2008, Yuan et al.~2009c, de Gasperin et
al.~2011, Younes et al.~2012). Yuan et al.~(2009c) considered 22
LLAGNs with $L_{\rm X} < L_{\rm X,crit}$ and found a correlation slope
$\sim 1.22$ (Fig.~9 bottom), in excellent agreement with
eq.~(\ref{steepercorrelation}).

\begin{figure}
\begin{center}
\hskip 0.2truein
\vspace{0.6cm}
\psfig{figure=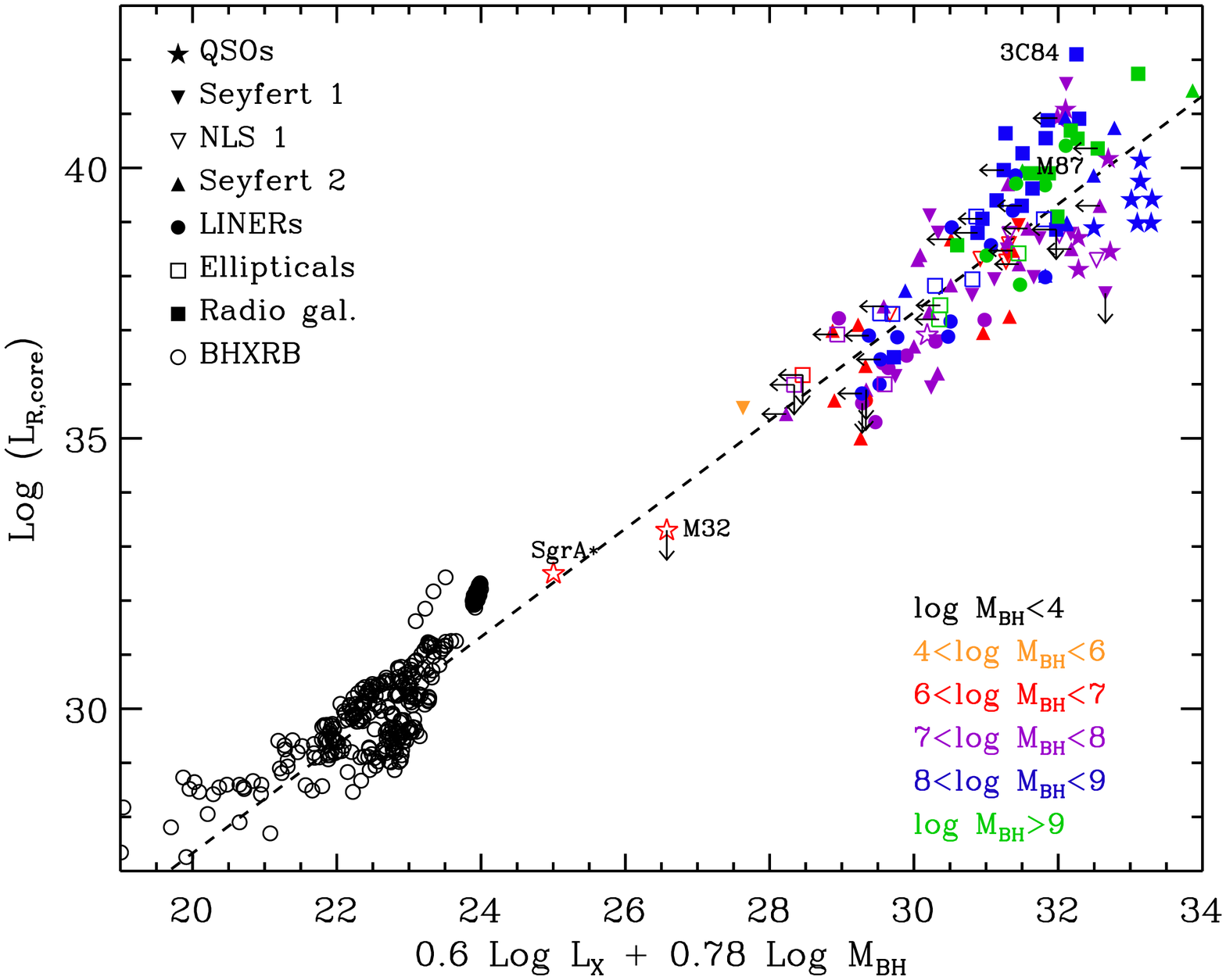,width=8.2cm,angle=0}
\vspace{-0.9cm}
\hskip 3.4truein
\psfig{figure=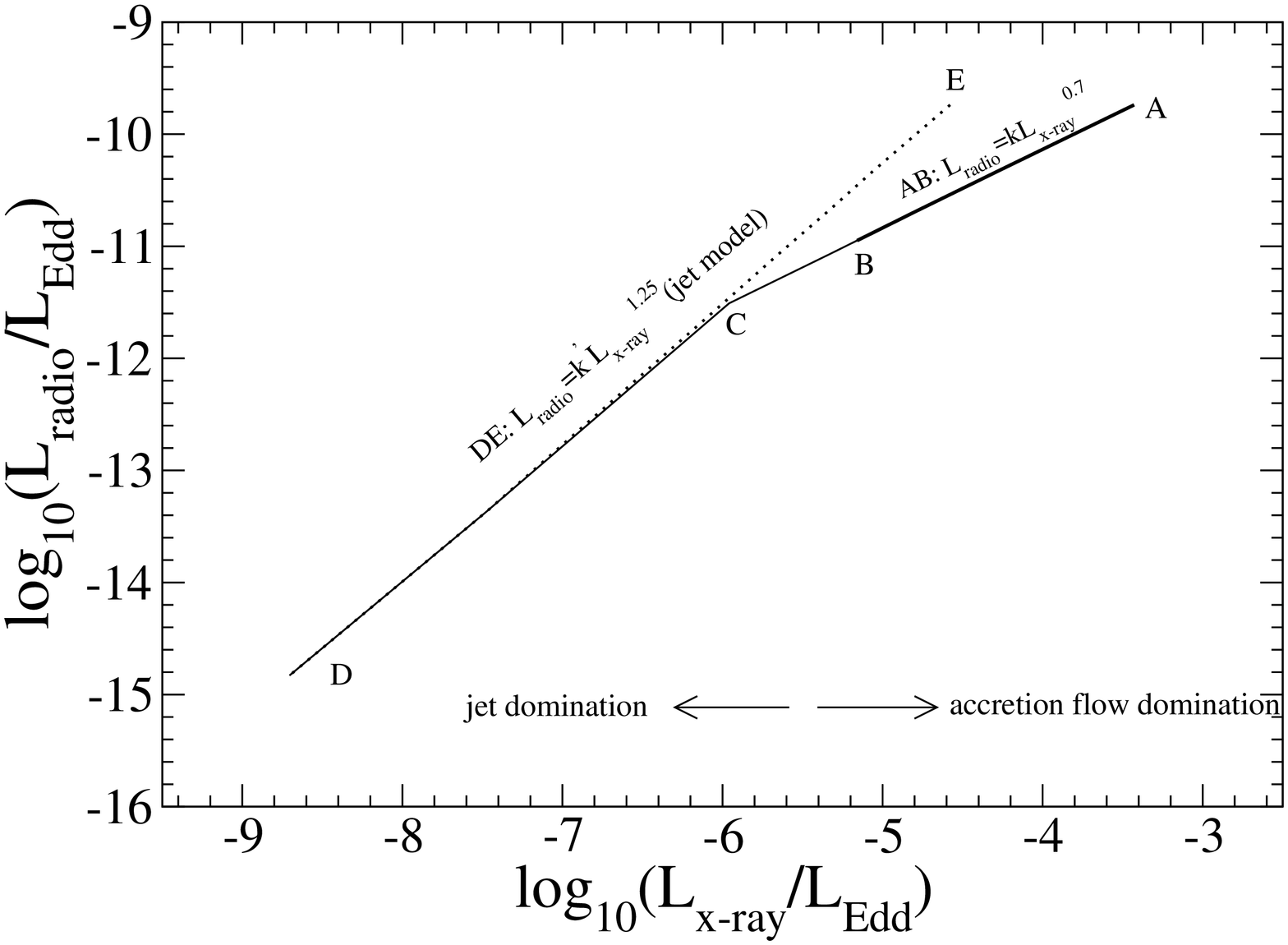,width=8.5cm,angle=270,rwidth=7.25cm}
\vspace{0.6cm}
\hskip 1.2truein
\psfig{figure=fig9c.eps,width=7.7cm,angle=0,rwidth=7.2cm}
\caption{{\em Top}: The ``fundamental plane of black hole activity''
  (from Merloni et al.~2003). Note that the correlation extends over
  many decades of black hole mass and accretion luminosity, and
  includes many different source types. (Figure courtesy of
  A.~Merloni.)  {\em Middle}: The predicted radio/X-ray correlation
  according to the ADAF-jet model (from Yuan \& Cui 2005). The segment
  AB, which represents luminous sources, has the same slope ($\sim
  0.6$) as the top panel. The segment CD corresponds to objects at
  lower luminosities, where the jet is expected to dominate both the
  radio and X-ray emission. {\em Bottom}: Observational data on 22
  LLAGNs with luminosities corresponding to CD in the middle panel
  (from Yuan et al.~2009c). The measured slope ($\sim 1.22$) agrees
  well with the theoretically predicted slope ($\sim 1.23$).}
\end{center}
\label{Fig:correlation}
\end{figure}

\subsubsection{Jets and black hole spin}

The fundamental plane of black hole activity (Fig.~9 top,
eq.~\ref{fundplane}) involves only the mass accretion rate $\dot{M}_{\rm BH}$
(through the luminosities $L_R$ and $L_X$) and the black hole mass
$M$; it does not involve the black hole spin. While there is a large
scatter in most of the data, van Velzen \& Falcke (2013) have recently
obtained a very homogeneous sample of radio quasars in which they
claim that most of the remaining scatter is due to environment,
leaving little room for additional scatter due to variations in the
black hole spin.  If jets are powered by black hole spin (BZ jet,
\S\ref{jet1}), we expect a strong dependence of jet power on the
angular velocity of the black hole $\Omega_H$ (eqs. \ref{BZpower},
\ref{Pjet}).  Why do the data not show this?

One possibility is that the range of black hole spins is not very
large. Since the data are usually shown in a log-log plot, any modest
variations due to spin could be hidden.  Even the van Velzen \& Falcke
(2013) data are potentially consistent with equation (\ref{Pjet}),
provided the black hole population does not span the full range of
spin values but is restricted to a smaller range, say from 0.3 to 1.

Alternatively, perhaps hot accretion flows in nature do not approach
anywhere near the MAD limit discussed in \S\ref{mad}. If the flux
$\Phi$ around the black hole is sufficiently below $\Phi_{\rm MAD}$,
the BZ jet mechanism will be sub-dominant. The quasi-relativistic disk
jet will then take over and we no longer expect a strong correlation
between jet power and spin (\S\ref{jet2}).  Note that jets in BHBs in
the hard state do not appear to be relativistic (Fender 2006) and may
well be disk jets rather than BZ jets. Even in the case of LLAGNs,
only a few sources (e.g., M87: Hada et al.~2011) are known to have
relativistic jets.

As discussed in \S\ref{jet3}, BHBs have two distinct kinds of jets:
(i) steady quasi-relativistic jets which are associated with the hard
state, and (ii) episodic jets which are most obviously seen when black
holes undergo state transitions from hot to cold mode accretion. So a
third possibility is that the van Velzen \& Falcke (2013) sample,
which consists of radio loud quasars, is dominated by episodic jets
(e.g., Merloni et al.~2003, Nipoti et al.~2005). If episodic jets are
powered by the disk mechanism proposed by Yuan et al.~(2009a; see
\S\ref{jet3} for details), no strong correlation between jet power and
black hole spin is expected.

In the case of BHBs, thanks to recent progress in measuring the spins
of black holes (McClintock et al.~2011, 2013), it is now possible to
check directly for a correlation between spin and jet power. A strong
correlation has been seen in the case of episodic jets (Narayan \&
McClintock 2012, Steiner et al.~2013, McClintock et al.~2013; but see
Fender et al.~2010, Russell et al.~2013a, who question the reality of
the correlation). However, the episodic jets in question are all
associated with the transition from a hard to a soft state (i.e., from
a hot accretion flow to a cold disk) rather than with a pure hard
state. Hence the simulation results discussed in \S\ref{jet1} may not
be relevant.
Also, since the jets were produced when the source luminosities were
close to Eddington (Steiner et al.~2013), it is possible that
additional physical effects, e.g., radiation, play an important
role. There have been no investigations of the effect of radiation on
the BZ mechanism.

\subsubsection{Alternative models for the hard state}

An alternative to the hot accretion flow model of hard state BHBs and
LLAGNs is the ``disk-corona'' model. Here the thin disk extends down
to the ISCO and the hard radiation is produced above the disk in a hot
corona which is heated perhaps by magnetic reconnection (Liang \&
Price 1977; Haardt \& Maraschi 1991, 1993). Because of the abundance
of soft photons from the disk, coupled with the strong reprocessing of
the coronal emission in the disk, the ``slab'' geometry of disk-corona
models generally gives relatively soft X-ray spectra (photon indices
$\Gamma\ \sgreat\ 2$). This is not consistent with observed spectra
(Stern et al. 1995, Zdziarski et al.~2003, Done et al.~2007). One way
out is to postulate that the corona forms the base of an
outward-moving jet (Beloborodov 1999, Malzac et al. 2001, Merloni \&
Fabian 2002). In this case, relativistic beaming reduces the amount of
reprocessing in the cold disk and it is possible to obtain hard
spectra.  A patchy corona is another possibility (e.g., Malzac et
al. 2001). In a recent study, Schnittman et al.~(2013) calculated
spectra for the disk-corona model using MHD simulation data.  They
claim to obtain relatively hard X-ray spectral slopes, although their
spectra are inconsistent with observations in certain other respects.
Apart from the spectral slope, the corona model also has trouble
explaining the observed correlations between the spectral slope, iron
line width and characteristic variability frequencies (Poutanen \&
Veledina 2014).

Another model for the hard state is the ``jet model''. As in the
standard hot accretion flow models, the thin disk here is truncated.
The main difference is that, not only the radio emission but also the
X-ray emission is produced by the jet. In early work, it was proposed
that the X-ray emission is due to synchrotron radiation (Markoff et
al.~1999, 2003). However, it has been pointed out that a pure
synchrotron model cannot reproduce the observed shape of the
high-energy cutoff of the X-ray spectrum (Zdziarski et
al.~2003). Later versions of the jet model invoke synchrotron
self-Compton radiation from the base of the jet to explain the X-ray
emission (Markoff et al.~2005). The model is then similar to the
standard hot accretion flow model, except that the required electron
temperature is much higher, $\sim$ several MeV.
Of course, a jet involves outflowing gas, so one way to distinguish
between the models is by measuring the velocity of the X-ray-emitting
gas. However, the outflowing gas at the base of the jet is only weakly
relativistic. Therefore, while there may be some modest beaming (as in
the Beloborodov 1999 model cited above), it is not expected to be a
dominant effect. Observationally, there is no evidence that X-ray
emission in the hard state has any dependence on the inclination of
the system (Fender et al.~2004, Narayan \& McClintock 2005).  Other
problems with the jet model are pointed out in Malzac et al. (2009)
and Poutanen \& Veledina (2014).

\section{Hot accretion and AGN feedback}
\label{agnfeedback}

There is considerable observational evidence that AGN feedback plays
an important role in the evolution of galaxies and galaxy clusters
(Fabian 2012, Kormendy \& Ho 2013). Arguments usually mentioned
include: (1) the famous correlation between the black hole mass and
the luminosity of the host galaxy or the velocity dispersion of the
galaxy bulge (Magorrian et al.~1998, Ferrarese \& Merritt 2000,
Gebhardt et al.~2000, Kormendy \& Ho 2013); (2) the observed
exponential cutoff in the number density of galaxies at the high
mass/luminosity end (Schechter 1976), even though there is no cutoff
at the same mass scale in the distribution of dark matter halos; (3)
the ``downsizing'' puzzle, where the most massive galaxies and black
holes are the oldest (Cowie et al.~1996, Kriek et al.~2007, Babi\'c et
al.~2007, Fanidakis et al.~2012); and (4) the ``cooling flow problem''
in galaxy clusters, where the lack of any significant cooling in
cluster cores (Peterson et al.~2001), despite a short cooling time,
suggests that a central AGN serves as an extra source of energy
(Pedlar et al.~1990; Churazov et al.~2000, 2002; Ciotti \& Ostriker
2001; Br\"uggen \& Kaiser 2002).

Two major modes of AGN feedback have been identified (Fabian 2012,
Kormendy \& Ho 2013): (1) ``Radiative'' mode, also known as ``quasar''
mode, which operates when the black hole accretes at a good fraction
($\sgreat\ 0.1$) of the Eddington rate. (2) ``Kinetic'' mode, also
known as ``radio'' mode or ``maintenance'' mode, which typically
operates when $\dot{M}_{\rm BH}$ is low, i.e., when the AGN is fed by
a hot accretion flow (the topic of this review). Maintenance mode
feedback has been considered in the context of the cooling flow
problem in galaxy clusters (Churazov 2002, Ruszkowski \& Begelman
2002), it is included in semi-empirical models of galaxy formation (Croton
et al.~2006, Hopkins et al.~2006, Somerville et al.~2008), and
incorporated in hydrodynamical simulations of galaxy formation and
evolution (Ciotti et al.~2010, Novak et al.~2011, Gaspari et
al.~2012). We review here some of the physics of maintenance mode
feedback, highlighting simplifications in current models that could be
improved using our current knowledge of hot accretion flows.

\subsection{Feedback from jets and outflows}
\label{jetfeedback}

Since the majority of supermassive black holes accrete at highly
sub-Eddington rates via hot accretion flows, maintenance mode feedback
is much more prevalent in the universe compared to quasar mode
feedback. In particular, most nearby galaxies contain LLAGNs
(\S\ref{llagn}), and any feedback activity in these systems occurs via
the maintenance mode.  Hot accretion flows tend to be radiatively
inefficient, so the maintenance mode is believed to be dominated by
mechanical feedback via jets and winds rather than radiation from the
accretion disk (but see \S\ref{radfeedback}).  Bubbles and cavities in
X-ray and radio images (e.g., Fabian 2012, Morganti et al.~2013)
provide direct observational evidence for interaction between jets and
the interstellar or intracluster medium. The two {\it Fermi} bubbles
detected above and below the Galactic center (Su et al.~2010) are also
thought to be inflated by a jet or wind from Sgr A* during a period of
activity in the last few million years (Zubovas et al.~2011, Guo \&
Mathews 2012).

Most studies of maintenance mode feedback in the literature focus on
the role of a collimated jet rather than a more isotropic
non-relativistic disk wind. In large part, this is driven by the fact
that jets are easily observed in the radio and their power can be
estimated directly from observations. In contrast, there is virtually
no direct evidence for uncollimated winds in LLAGNs (but see Crenshaw
\& Kraemer 2012). In addition, jets have been investigated via
simulations for a number of years (\S\ref{jetformation}), whereas the
study of winds from hot accretion flows has only just begun
(\S\ref{outflow}).

The relevant quantities that determine the effectiveness of feedback
are the rate of injection of energy and momentum into the external
medium and the degree of collimation of the jet or wind. Numerical
simulations of hot accretion flows can provide some useful
information. There is relatively better information in the case of
jets (\S\ref{jetformation}) compared to winds (\S\ref{outflow}), but
rapid progress is expected on both fronts.
One important parameter for feedback studies is ``feedback
efficiency'' $\epsilon$, defined as the ratio of the kinetic power in
the jet or wind to the accretion power $\dot{M}_{\rm BH}c^2$. In
almost all current cosmological models of feedback, $\epsilon$ is
regarded as a free parameter, but it could in principle be estimated
via simulations. Typically, one requires $\epsilon\sim10^{-3}-10^{-4}$
to explain observations (e.g., Di Matteo et al.~2005, Springel et
al.~2005, Ciotti et al.~2009, Ostriker et al.~2010, Gaspari et
al.~2012)\footnote{Note that many studies of AGN feedback on galaxies
  do not discriminate between quasar and maintenance mode feedback and
  adopt a single value of $\epsilon$ for both.}.

Both jets and winds carry with them substantial fluxes of energy and
momentum. Typically, the jet dominates the energy output, whereas the
wind dominates the momentum output (Yuan et al.~2012a, Sadowski et
al.~2013a). By and large, studies of AGN feedback have tended to focus
on energy feedback (e.g., Springel et al.~2005, Di Matteo et
al.~2005). However, momentum feedback can more effectively push the
surrounding gas and may be equally important for controlling the
growth of the black hole and switching off star formation (King 2003,
2005, 2010; Ostriker et al.~2010; Debuhr et al.~2010; Silk 2013).

The jet and wind differ significantly in their degree of
collimation. Even though the jet dominates the energy flux, it is not
clear that this energy couples very well to the interstellar medium of
the host galaxy.  The jet might simply drill through the surrounding
gas, depositing little energy within the galaxy. Indeed some
simulations indicate that the jet is ineffective even on the scale of
galaxy clusters (Vernaleo \& Reynolds 2006), though the problem is
much alleviated when shear, rotation and large-scale flows in the
intracluster medium are included (Heinz et al.~2006). Models that
invoke efficient jet heating generally require a relatively slow
(sub-relativistic) jet and a mass loss rate in the jet as large as the
Eddington rate, which seems unlikely (e.g., Omma et al.~2004).  In
contrast, the less collimated disk wind, despite its lower energy
budget, may actually be more important for galaxy scale feedback.

One other difference between jets and winds is the dependence of
feedback efficiency on the parameters of the system. Apart from the
obvious dependence on mass accretion rate $\dot{M}_{\rm BH}$, jet
power is strongly affected by the black hole spin and the magnetic
flux around the hole (at least in numerical simulations,
\S\ref{jetformation}). Thus, realistic incorporation of jet feedback
in cosmological simulations requires keeping track of the spins of
supermassive black holes. It is also necessary to decide whether or
not the MAD configuration is viable, which depends on whether
advection of magnetic field into the black hole is efficient
(\S\ref{mad}). In contrast, the energy and momentum flux in winds is
relatively insensitive to the black hole spin and magnetic flux, and
is primarily determined by $\dot{M}_{\rm BH}$. Thus, wind feedback
ought to be simpler to model. Unfortunately, estimating the mass
accretion rate itself involves large uncertainties (\S\ref{outflow}).

\subsection{Feedback from radiation}
\label{radfeedback}

The most obvious output of black hole accretion is radiation, which
can impart energy and momentum to the surrounding interstellar medium
via electron scattering, photoionization, atomic resonance scattering,
and absorption by dust grains. In some semi-analytical models and
numerical simulations (e.g., Wyithe \& Loeb 2003, Di Matteo et
al.~2005, Croton et al.~2006), it is assumed that a small and constant
fraction $\sim 0.05$ of the radiated luminosity from an AGN couples
thermodynamically to the surrounding gas. In other studies, a more
elaborate calculation is used to calculate the heating rate using the
Compton temperature $T_{\rm C}$ of the radiation field, which measures
the frequency-weighted average energy of the emitted photons. Usually
$T_{\rm C}\sim 10^7\,{\rm K}$ is adopted, as appropriate for typical
spectra of quasars (Sazonov et al.~2005, Ciotti et al.~2010, Novak et
al.~2011).

Whereas in quasar mode there is no doubt that feedback from radiation
and radiatively-driven winds (Proga 2007, Proga et al. 2008, Liu et
al. 2013) are very important, in the case of maintenance mode,
radiative feedback is usually assumed to be negligible compared to
mechanical feedback. Two reasons are invoked (e.g., Churazov et
al.~2005): (1) The kinetic power of the outflow is larger than the
radiative output of the disk. (2) The efficiency of radiative heating
is low. However, the radiative luminosity may actually be larger than
the kinetic outflow power for luminosities\footnote{This transition
  luminosity is different from the ``critical luminosity'' mentioned
  in \S\ref{correlation}.}  $\sgreat\ 10^{-4}L_{\rm Edd}$ (Fender et
al.~2003). In addition, the efficiency of radiative heating is not as
low as usually imagined, since the efficiency is proportional to
$T_c$. The spectrum of a hot accretion flow is much harder than that
of a quasar (\S\ref{llagn}). Hence the Compton temperature can be as
high as $T_c\sim 10^9$\,K (Yuan et al.~2009b)\footnote{The value of
  $T_c\sim 10^9{\rm K}$ is obtained for a pure hot accretion
  flow. When the contribution from a truncated thin disk is included
  (\S\ref{llagn}), the value will be somewhat lower.}, which means
that the radiative heating efficiency is much greater than in the
quasar mode.

Although the effect of a larger $T_c$ has not been included in studies
of galaxy-wide feedback, Ostriker and collaborators (e.g., Park \&
Ostriker 2001, 2007; Yuan et al.~2009b) have for many years
investigated its role on the scale of the accretion flow itself and
have demonstrated its importance in that context.
When the accretion rate is relatively high, non-local radiative
feedback via Compton heating is dynamically important and can change
the temperature profile of the accretion flow. Moreover, if
$L\ \sgreat\ 2\%L_{\rm Edd}$, radiative heating at and beyond the
Bondi radius ($R\ \sgreat\ 10^5R_S$) can be so strong that the gas is
heated above the virial temperature and wants to flow out rather than
in.  In this case, no steady accretion solution can be found and the
accretion flow oscillates between active and inactive phases (Cowie et
al.~1978; Ciotti \& Ostriker 1997, 2001, 2007; Yuan et
al.~2009b). This ``small-scale AGN radiative feedback'' effect has
also been invoked to explain the intermittent activity of compact
young radio sources (Yuan \& Li 2011).

\subsection{Estimating the mass accretion rate}

Both mechanical and radiative feedback depend strongly on the mass
accretion rate $\dot{M}_{\rm BH}$ of the black hole. Various
approaches have been adopted in the literature for estimating
$\dot{M}_{\rm BH}$. These include assuming that $\dot{M}_{\rm BH}$ is
equal to the Eddington rate, or the Bondi rate, or some variant of
these (e.g., Springel et al.~2005, Debuhr et al.~2010).
The Bondi accretion rate is more relevant for maintenance mode
feedback. However, whether or not the Bondi model is a reasonable
proxy for a hot accretion flow is still very much in debate (see
Narayan \& Fabian 2011 and references therein).

Given the density and temperature of the external gas at the Bondi
radius $R_{\rm B}\sim 10^5-10^6R_S$, it is straightforward to
calculate the Bondi accretion rate $\dot{M}_{\rm B}$ at that
radius. But how much of this gas actually reaches the black hole is
highly uncertain. If the accretion rate declines with decreasing
radius as $r^s$, with $s\sim0.5$ (\S\ref{outflow}), then as little as
$0.1-0.3\%$ of $\dot{M}_{\rm B}$ will make it to the black
hole. Although this is on the low side, it is not inconsistent with
some models of Sgr A* (\S\ref{sgra}). However, other systems with
powerful jets seem to require much more gas to reach the black hole to
power the observed jets (Allen et al.~2006, Rafferty et al.~2006,
McNamara et al.~2009, Russell et al. 2013b). A possible solution
is that the value of $s$ depends on boundary conditions. Perhaps $s$
is effectively lower whenever the accreting gas has very low angular
momentum (Narayan \& Fabian 2011, Bu et al.~2013) or when accretion
occurs via the MAD mode (Narayan et al.~2012b; Sadowski et
al.~2013a). Unfortunately, until one has a better understanding of the
mapping between $\dot{M}_{\rm B}$ and $\dot{M}_{\rm BH}$, it is hard
to imagine any kind of quantitative modeling of maintenance mode
feedback.

Another complication is that the mass accretion rate may be dominated
by cold gas from the external medium rather the hot gas usually
considered in hot accretion flow models (Pizzolato \& Soker 2005,
Rafferty et al. 2006). This could potentially boost the accretion rate
by up to two orders of magnitude compared to the Bondi rate calculated
based purely on hot gas (Gaspari et al. 2013).  The cold gas would
presumably first form a thin accretion disk and then evaporate to
become a hot accretion flow closer to the black hole. Modeling this
mode of accretion would require an understanding of the multi-phase
nature of the interstellar medium as well as the specific angular
momentum of the external cold clouds (Inogamov \& Sunyaev 2010).

\section{Prospects and remaining open questions}
\label{prospect}

The discovery of the self-similar ADAF solution 20 years ago (Narayan
\& Yi 1994), and the subsequent development of the ADAF model of hot
accretion flows (Abramowicz et al.~1995, Narayan \& Yi 1995b, Chen et
al.~1995), triggered a flurry of activity which has contributed
greatly to our understanding of the dynamics and thermodynamics of hot
accretion flows, as well as the recognition that these flows are
relevant for numerous astrophysical objects: Sgr A*, low luminosity
AGNs, BHBs.

Despite the impressive progress of the last two decades, there are
presently more questions than answers in this field. Below, in the
authors' view, are some of the more important questions:

\begin{itemize}

\item
How are electrons and ions heated in a hot accretion flow? What
particle energy distributions do these processes generate?  What role
do non-thermal particles play in the dynamical and radiative
properties of the system?

\item
Are there processes in addition to Coulomb collisions which transfer
energy from ions to electrons, and how do they influence the
temperatures of the two species?

\item
How strong are mass outflows from hot accretion flows, and how does
the mass accretion rate at the black hole ($\dot{M}_{\rm BH}$) depend
on boundary conditions at large radius? What if gas is supplied from
an external two-phase (or even multi-phase) medium?

\item
Why do hot accretion flows produce jets, whereas cool thin disks
apparently do not?  What role does the black hole, especially its
spin, play in determining the properties of the jet? What fraction of
the observed radiation comes from the jet versus the hot accretion
flow?

\item
How efficiently do hot accretion flows advect large-scale ordered
magnetic field towards the center, and how often do accreting black
holes approach the ``magnetically arrested disk'' limit?

\item
Why and how do state transitions in black hole binaries occur? What
are the physical processes responsible for converting cold optically
thick gas into hot optically thin gas in a ``truncated thin disk and
hot inner accretion flow'' configuration, and how do they relate to
the hysteresis phenomenon? How do the same processes behave in the
case of supermassive black holes?

\item
What is the thermal state of a hot accretion flow when $\dot{M}_{\rm
  BH}$ is close to the upper limit for a hot solution? Does the
accreting gas become a two-phase medium, and what observational
signatures do the hot and cold phases produce?

\item
What determines whether a hot accretion flow produces a steady jet or
an episodic jet, and why are the latter often associated with the hard
to soft state transition in black hole binaries? How does this map to
supermassive black holes, AGN jets and the radio loud/quiet dichotomy?

\item
What is the angular distribution of mass, momentum and energy outflow
from a hot accretion flow around a supermassive black hole, and how do
they determine the efficiency of feedback processes?

\item
How does the relative importance of mechanical versus radiative
feedback depend on $\dot{M}_{\rm BH}$ and other parameters of
the accretion flow?

\item
What are the properties of hot accretion flows around compact objects
with a surface?\footnote{An object with a surface introduces two
  important modifications compared to the black hole flows considered
  in this review. First, since gas comes to rest at the stellar
  surface, the inner boundary condition on the dynamical equations is
  very different and will result in vastly different density,
  velocity, pressure, etc. at small radii. Second, radiation from the
  surface will Compton-cool the hot accreting gas and modify its
  temperature. There have been only a few applications of the ADAF
  model to accreting neutron stars and white dwarfs. For lack of
  space, we have not reviewed this work here. Much more could be done
  in this area.}

\end{itemize}

\vskip 0.2cm {The authors are grateful to M.~C.~Begelman, C.~Done,
  J.-P.~Lasota, J.~E.~McClintock, J.~P.~Ostriker, J.~Poutanen,
  E.~Quataert, J.~Stone, A.~Tchekhovskoy and A.~Zdziarski for helpful
  comments on the manuscript. This work was supported by grants
  11133005 and 11121062 from the NSFC (FY) and AST1312651 from the NSF
  (RN).}

\vskip 0.5cm


\noindent {LITERATURE CITED}

\frenchspacing

\vskip 0.5cm

\nhi
Abarca D, Sadowski A, Sironi L. 2013. {\it MNRAS} submitted. arXiv:1309.2313

\nhi
Abramowicz MA, Blaes O, Hor\'ak J, Klu\'zniak W, Rebusco P. 2006. {\it Class. Quant. Grav.} 23:1689-96

\nhi
Abramowicz MA, Chen X, Granath M, Lasota JP. 1996. {\it Ap. J.} 471:762-73

\nhi
Abramowicz MA, Chen X, Kato S, Lasota, JP, Regev O. 1995. {\it Ap. J. Lett.} 438:L37-9

\nhi
Abramowicz MA, Czerny B, Lasota JP, Szuszkiewicz E. 1988. {\it Ap. J.} 332:646-58

\nhi
Abramowicz MA, Fragile PC. 2013. {\it Living Reviews in Relativity} 16:1-88

\nhi
Abramowicz MA, Igumenshchev IV, Lasota JP. 1998. {\it MNRAS} 293:443-6

\nhi
Abramowicz MA, Igumenshchev IV, Quataert E, Narayan R. 2002a. {\it Ap. J.} 565:1101-6

\nhi
Abramowicz MA, Kluzniak W, Lasota JP. 2002b. {\it Astron. Astrophys.} 396:L31-L34

\nhi
Agol E. 2000. {\it Ap. J. Lett.} 538:L121-4

\nhi
Aitken DK, Greaves J, Chrysostomou A, et al. 2000. {\it Ap. J. Lett.} 534:L173-6

\nhi
Allen SW, Dunn RJH, Fabian AC, Taylor GB, Reynolds CS. 2006. {\it MNRAS} 372:21-30

\nhi
Anninos P, Fragile PC, Salmonson JD. 2005. {\it Ap. J.} 635:723-40

\nhi
Anninos P, Fragile PC, Wilson J, Murray SD. 2012. {\it Ap. J.} 759:132

\nhi
Appl S, Camenzind M. 1992. {\it Astron. Astrophys.} 256:354-70

\nhi
Appl S, Camenzind M. 1993. {\it Astron. Astrophys.} 270:71-82

\nhi
Armitage P. 1998. {\it  Ap. J. Lett.} 501:L189-92

\nhi
Babi\'c A, Miller L, Jarvis MJ, et al. 2007. {\it Astron. Astrophys.} 474:755-62

\nhi
Baganoff FK, Bautz MW, Brandt WN, et al. 2001. {\it Nature} 413:45-48

\nhi
Baganoff FK, Maeda Y, Morris M,  et al. 2003. {\it Ap. J.} 591:891-915

\nhi
Bai X, Stone JM. 2013. {\it Ap. J.} 767:30

\nhi
Balbus SA. 2001. {\it Ap. J.} 562:909-17

\nhi
Balbus SA, Hawley JF. 1991. {\it Ap. J.} 376:214-33

\nhi
Balbus SA, Hawley JF. 1998. {\it Rev. Mod. Phys.} 70:1-53


\nhi
Bardeen JM, Petterson JA. 1975. {\it Ap. J. Lett.} 195:L65-L67

\nhi
Beckwith K, Hawley J, Krolik JH. 2008. {\it Ap. J.} 678:1180-99

\nhi
Beckwith K, Hawley J, Krolik JH. 2009. {\it Ap. J.} 707:428-45

\nhi
Begelman MC. 1979. {\it MNRAS} 187:237-51

\nhi
Begelman MC. 2012. {\it MNRAS} 420:2912-23

\nhi
Begelman MC, Celotti A. 2004. {\it MNRAS} 352:L45-8

\nhi
Begelman MC, Chiueh T. 1988. {\it MNRAS} 332:872-90

\nhi
Begelman MC, Meier DL. 1982. {\it Ap. J.} 253:873-96

\nhi
Belloni T, Parolin I, Del Santo M, et al. 2006. {\it MNRAS} 367:1113-20

\nhi
Beloborodov AM. 1999. {\it Ap. J. Lett.} 510:L123-6



\nhi
Beskin VS, Istomin YN, Parev VI. 1992. {\it Astronomicheskij Zhurnal} 69:1254-74

\nhi
Beskin VS, Malyshkin LM.  2000. {\it Astronomy Letters} 26:208-18

\nhi
Binder B, Markowitz A, Rothschild RE. 2009. {\it Ap. J.} 691:431-40

\nhi
Bisnovatyi-Kogan G, Lovelace RVE. 1997. {\it Ap. J.} 486:L43-46

\nhi
Bisnovatyi-Kogan G, Lovelace RVE. 2007. {\it Ap. J.} 667:L167-9

\nhi
Bisnovatyi-Kogan GS, Ruzmaikin AA. 1974. {\it Ap\&SS} 28:45-59

\nhi
Bj\"ornsson G, Abramowicz MA, Chen, X. Lasota JP. 1996. {\it
Ap. J.} 467:99-104

\nhi
Blackman EG. 1999. {\it MNRAS} 302:723-30

\nhi
Blackman EG, Penna RF,  Varni\'ere P. 2008. {\it New Astron.} 13:244-51

\nhi
Blaes O. 2013. {\it Space Sci. Rev.} (arXiv:1304.4879)

\nhi
Blaes OM, Arras P, Fragile PC. 2006. {\it Ap. J.} 369:1235-52

\nhi
Blandford RD, Begelman MC. 1999. {\it MNRAS} 303:L1-L5

\nhi
Blandford RD, Begelman MC. 2004. {\it MNRAS} 349:66-86

\nhi
Blandford RD, Payne DG. 1982. {\it MNRAS} 199:883-903

\nhi
Blandford RD, Znajek RL. 1977. {\it MNRAS} 179:433-56

\nhi
Bondi H. 1952. {\it MNRAS} 112:195-204

\nhi
B\"ottcher M, Liang EP. 1999. {\it Ap. J. Lett.} 511:L37-40

\nhi
Bower GC, Falcke H, Herrnstein RM, et al. 2004. {\it Science} 304:704-8

\nhi
Bower GC, Wright MCH, Falcke H, Backer DC. 2003. {\it Ap. J.} 588:331-7

\nhi
Brandenburg A, Nordlund A, Stein RF, Torkelsson U. 1995. {\it Ap. J.} 446:741-54

\nhi
Broderick AE, Fish VL, Doeleman SS, Loeb A. 2011a. {\it Ap. J.} 735:110

\nhi
Broderick AE, Fish VL, Doeleman SS, Loeb A. 2011b. {\it Ap. J.} 738:38

\nhi
Broderick AE, Loeb A, Narayan R. 2009. {\it Ap. J.} 701:1357-66

\nhi
Broderick AE, Narayan R. 2006. {\it Ap. J. Lett.} 638:L21-4

\nhi
Br\"uggen M, Kaiser CR. 2002. {\it Nature} 418:301-3

\nhi
Bu DF, Yuan F, Stone JM. 2011. {\it MNRAS} 413:2808-14

\nhi
Bu DF, Yuan F, Wu M, Cuadra J. 2013. {\it MNRAS} 434:1691-701

\nhi
Burkert A, Schartmann M, Alig C, et al. 2012. {\it Ap. J.} 750:58

\nhi
Bursa M, Abramowicz MA, Karas V, Klu\'zniak W. 2004. {\it Ap. J. Lett.}
617:L45-L48

\nhi
Cabanac C, Fender RP, Dunn RJH, K\"ording EG. 2009. {\it MNRAS} 396:1415-40

\nhi
Cao X. 2011. {\it Ap. J.} 737:94

\nhi
Chan CK, Liu S, Fryer CL,  et al. 2009. {\it Ap. J.} 701:521-34

\nhi
Chan CK, Psaltis D, \"Ozel F. 2013. {\it Ap. J.} 777:13

\nhi
Chaty S, Haswell CA, Malzac J, et al. 2003. {\it MNRAS} 346:689-703

\nhi
Chen K, Halpern JP. 1989. {\it Ap. J.} 344:115-24

\nhi
Chen X, Abramowicz MA, Lasota JP, Narayan R, Yi I. 1995. {\it Ap. J. Lett.} 443:L61-64

\nhi
Chen X, Abramowicz MA, Lasota JP. 1997. {\it Ap. J.} 476:61-69

\nhi
Chen X, Taam RE. 1993. {\it Ap. J.} 412:254-66

\nhi
Chiaberge M, Gilli R, Macchetto FD, Sparks WB. 2006. {\it Ap. J.} 651:728-34

\nhi
Chiang CY, Done C, Still M, Godet O. 2010. {\it MNRAS} 403:1102-12

\nhi
Chiang J, Blaes O. 2003. {\it Ap. J.} 586:97-111

\nhi
Churazov E, Forman W, Jones C, B\"ohringer H. 2000. {\it Astron. Astrophys.} 356:788-94

\nhi
Churazov E, Sazonov S, Sunyaev R, et al. 2005. {\it MNRAS} 363:L91-5

\nhi
Churazov E, Sunyaev R, Forman W, B\"ohringer H. 2002. {\it MNRAS} 332:729-34

\nhi
Ciotti L, Ostriker JP. 1997. {\it Ap. J. Lett.} 487:L105-8

\nhi
Ciotti L, Ostriker JP. 2001. {\it Ap. J.} 551:131-52

\nhi
Ciotti L, Ostriker JP. 2007. {\it Ap. J.} 665:1038-56

\nhi
Ciotti L, Ostriker JP, Proga D. 2009. {\it Ap. J.} 699:89-104

\nhi
Ciotti L, Ostriker JP, Proga D. 2010. {\it Ap. J.} 717:708-23


\nhi
Contopoulos J, Lovelace RVE. 1994. {\it Ap. J.} 429:139-52

\nhi
Corbel S, Fender RP, Tzioumis AK, et al. 2000. {\it Astron. Astrophys.} 359:251-68

\nhi
Corbel S, Koerding E, Kaaret P. 2008. {\it MNRAS} 389:1697-702

\nhi
Corbel S, Nowak MA, Fender RP, Tzioumis AK, Markoff S. 2003. {\it Astron. Astrophys.} 400:1007-12

\nhi
Cowie LL, Ostriker JP, Stark AA. 1978. {\it Ap. J.} 226:1041-62

\nhi
Cowie LL, Songaila A, Hu E, Cohen JG. 1996. {\it Astron. J.} 112:839-64

\nhi
Cowling TG. 1933. {\it MNRAS} 94:39-48

\nhi
Cox TJ, Jonsson P, Primack JR, Somerville RS. 2006. {\it MNRAS} 373:1013-38

\nhi
Crenshaw DM,  Kraemer SB. 2012. {\it Ap. J} 753:75

\nhi
Croton DJ, Springel V, White SD, et al. 2006. {\it MNRAS} 365:11-28

\nhi
Crumley P, Kumar P. 2013. {\it MNRAS} 436:1955-60

\nhi
Cuadra J, Nayakshin S, Martins F. 2008. {\it MNRAS} 383:458-66

\nhi
Cui W, Zhang SN, Chen W, Morgan EH. 1999. {\it Ap. J. Lett.} 512:L43-46

\nhi
D'Angelo C, Giannios D, Dullemond C, Spruit H. 2008. {\it Astron. Astrophys.} 488:441-50

\nhi
Davis SW, Stone JM, Pessah ME. 2010. {\it Ap. J.} 713:52-65

\nhi
de Gasperin F, Merloni A, Sell P, et al. 2011. {\it MMRAS} 415:2910-19

\nhi
De Villiers JP, Hawley JF. 2002. {\it Ap. J.} 577:866-79

\nhi
De Villiers JP, Hawley JF. 2003a. {\it Ap. J.} 589:458-80

\nhi
De Villiers JP, Hawley JF. 2003b. {\it Ap. J.} 592:1060-77

\nhi
De Villiers JP, Hawley JF, Krolik JH. 2003. {\it Ap. J.} 599:1238-53

\nhi
De Villiers JP, Hawley JF, Krolik JH, Hirose S. 2005. {\it Ap. J.} 620:878-88

\nhi
Debuhr J, Quataert E, Ma CP, Hopkins P. 2010, {\it MNRAS}, 406, L55-9

\nhi
Dexter J, Agol E, Fragile PC, McKinney JC. 2010. {\it Ap. J.} 717:1092-104

\nhi
Dexter J, Fragile PC. 2011. {\it Ap. J.} 730:36

\nhi
Dexter J, Fragile PC. 2013. {\it MNRAS} 432:2252-72

\nhi
Di Matteo T, Allen SW, Fabian AC, Wilson AS, Young AJ. 2003. {\it Ap. J.} 582:133-40

\nhi
Di Matteo T, Johnstone RM, Allen SW, Fabian AC. 2001. {\it Ap. J. Lett.} 550:L19-23

\nhi
Di Matteo T, Springel V, Hernquist L. 2005. {\it Nature} 433:604-7

\nhi
Di Matteo T, Quataert E, Allen SW, Narayan R, Fabian AC. 2000. {\it MNRAS} 311:507-21

\nhi
Di Salvo T, Done C, Zycki PT, Burderi L, Robba NR. 2001. {\it Ap. J.} 547:1024-33

\nhi
Diaz Trigo M, Miller-Jones JCA, Migliari S, Broderick JW, Tzioumis T. 2013. {\it Nature} in press (arXiv:1311.5080)

\nhi
Dibi S, Drappeau S, Fragile PC, Markoff S, Dexter J. 2012. {\it MNRAS} 426:1928-39

\nhi
Ding J, Yuan F,  Liang E. 2010. {\it Ap. J.} 708:1545-50

\nhi
Dodds-Eden K, Porquet D, Trap G,  et al. 2009. {\it Ap. J.} 698:676-92

\nhi
Dodds-Eden K, Sharma P, Quataert E, et al. 2010. {\it Ap. J.} 725:450-65

\nhi
Doeleman SS, Agol E, Backer D, et al. 2010. Astro2010: Science
White Paper 68. arXiv:0906.3899

\nhi
Doeleman SS, Fish VL, Schenck DE, et al. 2012. {\it Science} 338:355-8

\nhi
Doeleman SS, Weintroub J, Rogers AEE, et al. 2008. {\it Nature} 455:78-80

\nhi
Dolence JC, Gammie CF, Shiokawa H, Noble SC. 2012. {\it Ap. J.}
746:L10

\nhi
Done C,  Diaz Trigo M. 2010. {\it MNRAS} 407:2287-96

\nhi
Done C, Gierli\'nski M. 2006. {\it MNRAS} 367:659-68

\nhi
Done C, Gierli\'nski M, Kubota A. 2007. {\it Astron. Astrophys. Rev.} 15:1-66

\nhi
Dubus G, Hameury JM, Lasota JP. 2001. {\it Astron. Astrophys.} 373:251-71

\nhi
Dyda S, Lovelace RVE,  Ustyugova GV, et al. 2013. {\it MNRAS} 432:127-37


\nhi
Eckart A, Baganoff FK, Morris M, et al. 2004. {\it Astron. Astrophys.} 427:1-11

\nhi
Eckart A, Baganoff FK, Sch\"odel R, et al. 2006. {\it Astron. Astrophys.} 450:535-55

\nhi
Eracleous M, Hwang JA, Flohic, HMLG. 2010. {\it Ap. J. Suppl.} 187:135-48

\nhi
Esin AA. 1999. {\it Ap. J.} 517:381-95

\nhi
Esin AA, McClintock JE, Narayan R. 1997. {\it Ap. J.} 489:865-89 (erratum: 500:523)

\nhi
Esin AA, McClintock JE, Drake JJ, et al. 2001. {\it Ap. J.} 555:483-8

\nhi
Esin AA, Narayan R, Cui W, Grove JE, Zhang SN. 1998. {\it Ap. J.} 505:854-68

\nhi
Fabbiano G, Elvis M, Markoff S, et al. 2003. {\it Ap. J.} 588:175-85

\nhi
Fabian AC. 2012. {\it Annu. Rev. Astron. Astrophys.} 50:455-89

\nhi
Fabian AC, Canizares CR. 1988. {\it Nature} 333:829-31

\nhi
Fabian AC, Rees MJ. 1995. {\it MNRAS} 277:L55-8

\nhi
Fabrika S. 2004. {\it Astrophy. and Space Phys. Rev.} 12:1-152

\nhi
Falcke H, K\"ording E, Markoff S. 2004. {\it Astron. Astrophys.} 414:895-903

\nhi
Falcke H, Markoff S. 2000. {\it  Astron. Astrophys.} 362:113-8

\nhi
Falcke H, Melia F. 1997. {\it Ap. J.} 479:740-51

\nhi
Falcke H, Nagar NM, Wilson AS, Ulvestad JS. 2000. {\it Ap. J.} 542:197-200

\nhi
Fanidakis N, Baugh CM, Benson AJ, et al. 2012. {\it MNRAS} 419:2797-820


\nhi
Fender RP. 2001. {\it MNRAS} 322:31-42

\nhi
Fender RP. 2006. In {\it Compact stellar X-ray sources.} Edited by W Lewin \& M van der Klis. Cambridge Astrophysics Series, 39:381-419

\nhi
Fender RP, Belloni TM, Gallo E. 2004. {\it MNRAS} 355:1105-18

\nhi
Fender RP, Gallo E, Jonker PG.  2003. {\it MNRAS} 343:L99-103

\nhi
Fender RP, Gallo E, Russell D. 2010. {\it MNRAS} 406:1425-34

\nhi
Fender R, Stirling AM, Spencer RE, et al. 2006. {\it MNRAS} 369:603-7

\nhi
Ferrarese L, Merritt D. 2000. {\it Ap. J. Lett.} 539:L9-12

\nhi
Fish VL, Broderick AE, Doeleman SS, Loeb A. 2009. {\it Ap. J. Lett.} 692:L14-L18

\nhi
Fish VL, Doeleman SS, Beaudoin C, et al. 2011. {\it Ap. J. Lett.}
727:L36

\nhi
Fragile PC. 2009. {\it Ap. J. Lett.} 706:L246-50

\nhi
Fragile PC, Anninos P. 2005. {\it Ap. J.} 623:347-61

\nhi
Fragile PC, Blaes OM, Anninos P, Salmonson JD. 2007. {\it Ap. J.} 668:417-29

\nhi
Fragile PC, Lindner CC, Anninos P, Salmonson JD. 2009. {\it
Ap. J.} 691:482-94

\nhi
Fragile PC, Meier DL. 2009. {\it Ap. J.} 693:771-83

\nhi
Frank J, King A, Raine DJ. 2002. {\it Accretion Power in Astrophysics}, Cambridge, UK: Cambridge University Press

\nhi
Fromang S, Papaloizou J. 2007. {\it Astron. Astrophys.} 476:1113-22

\nhi
Fromang S, Papaloizou J, Lesur G, Heinemann T. 2007. {\it Astron. Astrophys.} 476:1123-32

\nhi
Gallo E, Fender RP, Miller-Jones JCA, et al. 2006. {\it MNRAS} 370:1351-60

\nhi
Gallo E, Fender RP, Pooley GG. 2003. {\it MNRAS} 344:60-72

\nhi
Gammie CF, McKinney JC, T\'oth G. 2003. {\it Ap. J.} 589:444-57

\nhi
Gammie CF, Narayan R, Blandford R. 1999. {\it Ap. J.} 516:177-86

\nhi
Gammie CF, Popham R. 1998. {\it Ap. J.} 498:313-26

\nhi
Gammie CF, Shapiro SL, McKinney JC. 2004. {\it Ap. J.} 602:312-9

\nhi
Garcia MR, McClintock JE, Narayan R, et al. 2001. {\it Ap. J. Lett.} 553:L47-50

\nhi
Gardner E, Done C. 2013. {\it MNRAS} 434:3454-62

\nhi
Gaspari M, Brighenti F, Temi P. 2012. {\it MNRAS} 424:190-209

\nhi
Gaspari M, Ruszkowski M, Peng Oh S. 2013. {\it MNRAS} 432:3401-22

\nhi
Gebhardt K, Bender R, Bower G, et al. 2000. {\it Ap. J. Lett.} 539:L13-6

\nhi
Genzel R, Eisenhauer F, Gillessen S. 2010. {\it Rev. Mod. Phys.} 82:3121-95

\nhi
Genzel R, Sch\"odel R, Ott T, et al. 2003. {\it Nature} 425:934-7

\nhi
Ghez AM, Duchene G, Matthews K, et al. 2003. {\it Ap. J. Lett.} 586:L127-31

\nhi
Ghez AM, Salim S, Weinberg NN, et al. 2008. {\it Ap. J.} 689:1044-62

\nhi
Ghez AM, Wright SA, Mattews K, et al. 2004. {\it Ap. J. Lett.} 601:L159-62

\nhi
Giannios D, Spruit HC. 2004. {\it Astron. Astrophys.} 427:251-61

\nhi
Gierli\'nski M, Done C, Page K. 2008. {\it MNRAS} 388:753-60

\nhi
Gierli\'nski M, Newton J. 2006. {\it MNRAS} 370:837-44

\nhi
Gilfanov M, Churazov E, Revnivtsev M. 1999. {\it Astron. Astrophys.} 352:182-8

\nhi
Gillessen S, Eisenhauer F, Quataert E, et al. 2006. {\it Ap. J. Lett.} 640:L163-6

\nhi
Gillessen S, Eisenhauer F, Fritz TK,  et al. 2009a. {\it Ap. J. Lett.} 707:L114-7

\nhi
Gillessen S, Eisenhauer F, Trippe S, et al. 2009b. {\it Ap. J.} 692:1075-109

\nhi
Gillessen S, Genzel R, Fritz TK, et al. 2012. {\it Nature} 481:51-4

\nhi
Gillessen S, Genzel R, Fritz TK, et al. 2013. {\it Ap. J.} 774:44

\nhi
Goldston JE, Quataert E, Igumenshchev IV. 2005. {\it Ap. J.} 621:785-92

\nhi
Goodson AP, Bohm KH, Winglee RM. 1999. {\it Ap. J.} 524:142-58

\nhi
Goodson AP, Winglee RM. 1999. {\it Ap. J.} 524:159-68

\nhi
Gu WM, Lu JF. 2000. {\it Ap. J.} 540:L33-6

\nhi
Guan X, Gammie CF, Simon JB, Hohnson BM. 2009. {\it Ap. J.} 694:1010-18

\nhi
Guilet J, Ogilvie GI. 2012. {\it MNRAS} 424:2097-117

\nhi
Guilet J, Ogilvie GI. 2013. {\it MNRAS} 430:822-35

\nhi
Guo F, Mathews WG. 2012. {\it Ap. J.} 756:181

\nhi
G\"ultekin K, Cackett EM, Miller JM, et al. 2009. {\it Ap. J.} 706:404-16

\nhi
Haardt F,  Maraschi L. 1993. {\it Ap. J.} 413:507-17

\nhi
Hada K, Doi A, Kino M, et al. 2011. {\it Nature} 477:185-7

\nhi
Hameury JM, Lasota JP, McClintock JE, Narayan R. 1997. {\it Ap. J.} 489:234-43

\nhi
Hartnoll SA,  Blackman EG. 2001. {\it MNRAS} 324:257-66

\nhi
Hawley JF. 2000. {\it Ap. J.} 528:462-79


\nhi
Hawley JF, Balbus SA. 1991. {\it Ap. J.} 376:223-33

\nhi
Hawley JF, Balbus SA. 1996. in Physics of Accretion Disks, eds.
S. Kato et al. New York: Gordon \& Breach

\nhi
Hawley JF, Balbus SA. 2002. {\it Ap. J.} 573:738-48

\nhi
Hawley JF, Balbus S, Stone JM. 2001. {\it Ap. J.} 554:L49-L52

\nhi
Hawley JF, Gammie CF, Balbus SA. 1995. {\it Ap. J.} 440:742

\nhi
Hawley JF, Gammie CF, Balbus SA. 1996. {\it Ap. J.} 464:690

\nhi
Hawley JF, Guan X,  Krolik JH. 2011. {\it Ap. J.} 738:84

\nhi
Hawley JF, Krolik JH. 2001. {\it Ap. J.} 548:348-67

\nhi
Hawley JF, Krolik JH. 2002. {\it Ap. J.} 566:164-80

\nhi
Hawley JF, Krolik JH. 2006. {\it Ap. J.} 641:103-16

\nhi
Hawley JF, Richers SA, Guan X, Krolik JH. 2013. {\it Ap. J.} 722:102

\nhi
Hayashi MR, Shibata K, Matsumoto R. 1996. {\it Ap. J. Lett.} 468:L37-40

\nhi
Heinz S. 2004. {\it MNRAS} 355:835-44

\nhi
Heinz S, Br\"uggen M, Young A, Levesque E. 2006. {\it MNRAS} 373:L65-9

\nhi
Heinz S, Sunyaev RA. 2003. {\it MNRAS} 343:L59-64

\nhi
Heyvaerts J, Norman C. 1989. {\it Ap. J.} 347:1055-81

\nhi
Hirose S, Krolik JH, Blaes O. 2009. {\it Ap. J.} 691:16-31

\nhi
Hirose S, Krolik JH, De Villiers JP, Hawley JH. 2004. {\it Ap. J.} 606:1083-97

\nhi
Hirotani K, Okamoto I. 1998. {\it Ap. J.} 497:563

\nhi
Ho LC. 1999. {\it Ap. J.} 516:672-82

\nhi
Ho LC. 2002. {\it Ap. J.} 564:120-32

\nhi
Ho LC. 2008. {\it Annu. Rev. Astron. Astrophys.} 46:475-539

\nhi
Ho LC. 2009. {\it Ap. J.} 699:626-37

\nhi
Ho LC, Terashima Y, Ulvestad JS. 2003. {\it Ap. J.} 589:783-9

\nhi
Honma F. 1996. {\it Publ. Astron. Soc. Jap.} 48:77-87

\nhi
Hoshino M. 2012. {\it Phys. Rev. Lett.} 108:135003

\nhi
Hoshino M. 2013. {\it Ap. J.} 773:118

\nhi
Hopkins PF, Hernquist L, Cox TJ, et al. 2006. {\it Ap. J.S.} 163:1-49

\nhi
Huang L, Cai M, Shen ZQ, Yuan F. 2007. {\it MNRAS} 379:833-40

\nhi
Ichimaru S. 1977. {\it Ap. J.} 214:840-55

\nhi
Igumenshchev IV. 2002. {\it Ap. J. Lett.} 577:L31-L34

\nhi
Igumenshchev IV. 2006. {\it Ap. J.} 649:361-72

\nhi
Igumenshchev IV. 2008. {\it Ap. J.} 677:317-26

\nhi
Igumenshchev IV, Aramowicz MA. 1999. {\it Ap. J. Lett.} 537:L27

\nhi
Igumenshchev IV, Aramowicz MA. 2000. {\it Ap. J. Supp.} 130:463-84

\nhi
Igumenshchev IV, Aramowicz MA, Narayan R. 2000. {\it Ap. J. Lett.} 537:L27-30

\nhi
Igumenshchev IV, Chen XM,  Abramowicz M. 1996. {\it MNRAS} 278:236-250

\nhi
Igumenshchev IV, Narayan R. 2002. {\it Ap. J.} 566:137-47

\nhi
Igumenshchev IV, Narayan R,  Abramowicz, MA. 2003. {\it Ap. J.} 592:1042-59

\nhi
Ingram A, Done C, Fragile PC. 2009. {\it MNRAS} 397:L101-05

\nhi
Ingram A, Done C. 2011. {\it MNRAS} 415:2323-35

\nhi
Inogamov NA, Sunyaev RA. 2010. {\it Astron. Lett.} 36:835-47

\nhi
Janiuk A, Sznajder M, Mo\'scibrodzka M, Proga D. 2009. {\it Ap. J.} 705:1503-21

\nhi
Jaroszynski M, Kurpiewski A. 1997. {\it  Astron. Astrophys.} 326:419-26

\nhi
Jiang YF, Stone JM, Davis SW. 2013. {\it Ap. J.} 778:65

\nhi
Jiao CL, Wu XB. 2011. {\it Ap. J.} 733:112


\nhi
Johnson BM, Quataert E. 2007. {\it Ap. J.} 660:1273-81

\nhi
Kalemci E, Dincer T, Tomsick JA, et al. 2013. {\it Ap. J.} 779:95

\nhi
Kato S, Abramowicz MA, Chen X. 1996. {\it Publ. Astron. Soc. Jap.} 48:67-75

\nhi
Kato S, Fukue J, Mineshige S. 2008. {\it Black-Hole Accretion Disks - Towards
a New Paradigm} (Kyoto, Japan: Kyoto Univ. Press)

\nhi
Kato S, Yamasaki T, Abramowicz MA, Chen X. 1997. {\it Publ. Astron. Soc. Jap.} 49:221-5

\nhi
Kato Y, Hayashi MR, Matsumoto R. 2004a. {\it Ap. J.} 600:338-42

\nhi
Kato Y, Mineshige S,  Shibata K. 2004b. {\it Ap. J.} 605:307-20

\nhi
Kato Y, Umemura M, Ohsuga K. 2009. {\it MNRAS} 400:1742-8

\nhi
Katz J. 1977. {\it Ap. J.} 215:265-75

\nhi
Kawamuro T, Ueda Y, Tazaki F, Terashima Y. 2013. {\it Ap. J.} 770:157

\nhi
King A. 2003. {\it Ap. J. Lett.} 596:L27-9

\nhi
King A. 2005. {\it Ap. J. Lett.} 635:L121-3

\nhi
King AR. 2010. {\it MNRAS} 402:1516-22

\nhi
Krolik JH, Hawley JF, Hirose S. 2005. {\it Ap. J.} 622:1008-23

\nhi
Koide S. 2003. {\it Phys. Rev. D} 67:104010

\nhi
Koide S, Shibata K, Kudoh T. 1999. {\it Ap. J.} 522:727-52

\nhi
Koide S, Meier DL, Shibata K, Kudoh T. 2000. {\it Ap. J.} 536:668-74

\nhi
Kolehmainen M, Done C, Diaz Trigo M. 2013. {\it MNRAS} submitted (arXiv:1310.1219)


\nhi
Komissarov SS. 1999. {\it MNRAS} 303:343-66

\nhi
Komissarov SS. 2001. {\it MNRAS} 326:L41-4

\nhi
Komissarov S, Barkov MV, Vlahakis N, Konigl A. 2007. {\it MNRAS} 380:51-70

\nhi
Komissarov SS, McKinney JC. 2007. {\it MNRAS} 377:L49-L53


\nhi
Koratkar A, Blaes O. 1999. {\it PASJ} 755:1-30

\nhi
K\"ording E, Falcke H, Corbel S. 2006. {\it Astron. Astrophys.} 456:439-50

\nhi
Kormendy J, Ho LC. 2013. {\it Annu. Rev. Astron. Astrophys.} 51:511-653

\nhi
Kriek M, van Dokkum PG, Franx M, et al. 2007. {\it Ap. J.} 669:776-90

\nhi
Krolik JH, Hawley JF, Hirose S. 2005. {\it Ap. J.} 622:1008-23

\nhi
Kudoh T, Matsumoto R, Shibata K 1998. {\it Ap. J.} 508:186-99

\nhi
Kudoh T, Matsumoto R, Shibata K 2002. {\it PASJ} 54:267-74

\nhi
Kusunose M, Mineshige S. 1996. {\it Ap. J.} 468:330-37

\nhi
Kuwabara T, Shibata K, Kudoh T, Matsumoto R. 2000. {\it Publ. Astron. Soc. Jap.} 52:1109-24

\nhi
Lasota JP. 1999. {\it Phys. Rep.} 311, 247-58

\nhi
Lasota LP, Abramowicz MA, Chen X, et al. 1996a. {\it Ap. J.} 462:142-6

\nhi
Lasota, JP, Gourgoulhon E, Abramowicz M, Tchekhovskoy A, Narayan
R.  2014. {\it Phys. Rev. D}, in press (arXiv:1310.7499)

\nhi
Lasota JP, Narayan R, Yi I. 1996b. {\it Astron. Astrophys.} 314:813-20

\nhi
Lehe R, Parrish IJ, Quataert E. 2009. {\it Ap. J.} 707:404-19

\nhi
Li J, Ostriker J, Sunyaev R. 2013a. {\it Ap. J.} 767:105

\nhi
Li Z, Morris MR, Baganoff FK. 2013b. {\it Ap. J.} 779:154

\nhi
Li Z, Wu XB, Wang R. 2008. {\it Ap. J.} 688:826-36

\nhi
Li ZY, Chiueh T, Begelman MC. 1992. {\it Ap. J.} 394:459-71

\nhi
Liang EPT, Price RH. 1977. {\it Ap. J.} 218:247-52

\nhi
Lightman AP, Eardley DM.  1974. {\it Ap. J.} 187:L1-4

\nhi
Liu BF, Done C, Taam RE. 2011. {\it Ap. J.} 726:10

\nhi
Liu BF, Meyer F, Meyer-Hofmeister E. 2005. {\it Astron. Astrophys.} 442:555-62

\nhi
Liu BF, Taam RE, Meyer-Hofmeister E, Meyer F. 2007. {\it Ap. J.} 671:695-705

\nhi
Liu BF, Yuan W, Meyer F, Meyer-Hofmeister E. 1999. {\it Ap. J.} 527:L17-20

\nhi
Liu C, Yuan F, Ostriker JP, Gan Z, Yang X. 2013. {\it MNRAS} 434:1721-35

\nhi
Liu H, Wu Q. 2013. {\it Ap. J.} 764:17


\nhi
Livio M, Ogilvie GI, Pringle JE. 1999. {\it Ap. J.} 512:100-04

\nhi
Lu JF, Lin YQ, Gu WM 2004. {\it Ap. J. Lett.} 602:L37

\nhi
Lodato G, Pringle JE. 2006. {\it MNRAS} 368:1196-1208

\nhi
Loewenstein M, Mushotzky RF, Angelini L, Arnaud KA, Quataert E. 2001. {\it Ap. J. Lett.} 555:L21-4

\nhi
Lovelace RVE. 1976. {\it Nature} 262:649-52

\nhi
Lovelace RVE, Mehanian CM, Mobarry CM, Sulkanen ME. 1986.
{\it Ap. J. Suppl.} 62:1-37

\nhi
Lynden-Bell D, Pringle JE. 1974. {\it MNRAS} 168:603-37

\nhi
Lynden-Bell D. 2003. {\it MNRAS} 341:1360-72

\nhi
MacDonald D, Thorne KS.  1982. {\it MNRAS} 198:345-82

\nhi
Machida M, Hayashi MR, Matsumoto R. 2000. {\it Ap. J. Lett.} 532:L67-70

\nhi
Machida M, Matsumoto R. 2003. {\it Ap. J.} 585:429-42

\nhi
Machida M, Matsumoto R, Mineshige S. 2001. {\it Publ. Astron. Soc. Jap.} 53:L1-L4

\nhi
Machida M, Nakamura KE, Matsumoto R. 2004. {\it Publ. Astron. Soc. Jap.} 56:671-9

\nhi
Machida M, Nakamura KE, Matsumoto R. 2006. {\it Publ. Astron. Soc. Jap.} 58:193-202

\nhi
Magorrian J, Tremaine S, Richstone D, et al. 1998. {\it Astron. J.} 115:2285-305

\nhi
Mahadevan R. 1997. {\it Ap. J.} 477:585-601

\nhi
Mahadevan R. 1998. {\it Nature} 394:651-3

\nhi
Mahadevan R. 1999. {\it MNRAS} 304:501-11

\nhi
Mahadevan R, Narayan R, Krolik J. 1997. {\it Ap. J.} 486:268-75

\nhi
Mahadevan R, Quataert E. 1997. {\it Ap. J.} 490:605-18

\nhi
Malzac J, Belmont R, Fabian AC. 2009. {\it MNRAS} 400:1512-20

\nhi
Malzac J,  Beloborodov AM, Poutanen J. 2001. {\it MNRAS} 326:417-27

\nhi
Malzac, J, Merloni, A, Fabian, AC. 2004. {\it MNRAS} 351:253-64

\nhi
Manmoto T. 2000. {\it Ap. J.} 534:734-46

\nhi
Manmoto T, Kato S. 2000. {\it Ap. J.} 538:295-306

\nhi
Manmoto T, Kato S, Nakamura KE, Narayan R. 2000. {\it Ap. J.} 529:127-37

\nhi
Manmoto T, Mineshige S, Kusunose M. 1997. {\it Ap. J.} 489:791-803

\nhi
Manmoto T, Takeuchi M, Mineshige S, Matsumoto R, Negoro H. 1996. {\it Ap. J. Lett.} 464:L135-8

\nhi
Maoz D. 2007. {\it MNRAS} 377:1696-710

\nhi
Maraschi L, Tavecchio F. 2003. {\it Ap. J.} 593:667-75

\nhi
Markoff S, Bower GC, Falcke H. 2007. {\it MNRAS} 379:1519-32

\nhi
Markoff S, Falcke H, Fender R. 1999. {\it Astron. Astrophys.} 372:L25-28

\nhi
Markoff S, Falcke H, Yuan F, Biermann P. 2001. {\it Astron. Astrophys.} 379:L13-16

\nhi
Markoff S, Nowak M, Corbel S, Fender R. et al. 2003. {\it Astron. Astrophys.} 397:645

\nhi
Markoff S, Nowak MA, Wilms J. 2005. {\it Ap. J.} 635:1203-16

\nhi
Marrone DP, Moran JM, Zhao JH, Rao R. 2006. {\it Ap. J.} 640:308-18

\nhi
Marrone DP, Moran JM, Zhao JH, Rao R. 2007. {\it Ap. J. Lett.} 654:L57-L60

\nhi
Marsch E. 2012. {\it Space Sci. Rev.} 172:23-39

\nhi
Massi M, Poletto G. 2011. {\it Memorie della Societa Astronomica Italiana} 82:145

\nhi
Matsumoto R, Tajima T. 1995. {\it Ap. J.} 445:767-79

\nhi
Matsumoto R, Shibata K. 1997. in IAU Colloq. 163, Accretion Phenomena
and Related Outflows, ed. D. Wickramsinghe, G. Bicknell \& L.
Ferrario, (ASP Conf. Ser. 121; San Francisco: ASP), 443

\nhi
Mayer M, Pringle JE. 2007. {\it MNRAS} 376:435-56

\nhi
McClintock JE, Remillard RA. 2006. In {\it Compact stellar X-ray sources.} Edited by W Lewin \& M van der Klis. Cambridge Astrophysics Series, 39:157-213

\nhi
McClintock JE, Narayan R, Davis SW, et al. 2011. {\it Classical and Quantum Gravity} 28:114009

\nhi
McClintock JE, Narayan R, Garcia MR, et al. 2003. {\it Ap. J.} 593:435-51


\nhi
McClintock JE, Narayan R, Steiner JF. 2013. {\it Space Sci. Rev.}, in
press (arXiv:1303.1583)

\nhi
McKinney JC. 2005. {\it Ap. J. Lett.} 630:L5-L8

\nhi
McKinney JC. 2006. {\it MNRAS} 368:1561-82

\nhi
McKinney JC, Blandford RD. 2009. {\it MNRAS} 394:L126-30

\nhi
McKinney JC, Gammie CF. 2004. {\it Ap. J.} 611:977-95



\nhi
McKinney JC,  Tchekhovskoy A,  Blandford RD. 2012. {\it MNRAS} 423:3083-117

\nhi
McKinney JC, Tchekhovskoy A, Blandford RD. 2013. {\it Science} 339:49-52

\nhi
McNamara BR, Kazemzadeh F, Rafferty DA, et al. 2009. {\it Ap. J.} 698:594-605

\nhi
Medvedev MV. 2000. {\it Ap. J.} 541:811-20

\nhi
Meier DL. 2005. {\it Ap. Space. Sci.} 300:55-65

\nhi
Melia F. 1992. {\it Ap. J. Lett.} 387:L25-8

\nhi
Melia F, Liu S, Coker R. 2001. {\it Ap. J.} 553:146-57

\nhi
Menou K, Esin AA, Narayan R, et al. 1999a. {\it Ap. J.} 520:276-91

\nhi
Menou K, Hameury JM, Lasota JP, Narayan R. 2000. {\it MNRAS}
314:498-510

\nhi
Menou K, Narayan R, Lasota JP. 1999b. {\it Ap. J.} 513:811-26

\nhi
Merloni A, Fabian AC. 2002. {\it MNRAS} 332:165-75

\nhi
Merloni A, Heinz S, Di Matteo T. 2003. {\it MNRAS} 345:1057-76

\nhi
Meyer F, Meyer-Hofmeister E. 1994. {\it Astron. Astrophys.} 288:175-82

\nhi
Meyer F, Liu BF, Meyer-Hofmeister E. 2007. {\it Astron. Astrophys.} 463:1-9

\nhi
Meyer F, Liu BF, Meyer-Hofmeister E. 2000. {\it Astron. Astrophys.} 361:175-88

\nhi
Meyer L, Ghez AM, Sch\"odel R, et al. 2012. {\it Sci.} 338:84-87

\nhi
Meyer-Hofmeister E, Liu BF, Meyer F. 2005. {\it Astron. Astrophys.} 432:181-7

\nhi
Meyer-Hofmeister E, Liu BF, Meyer F. 2009. {\it Astron. Astrophys.} 508:329-37

\nhi
Miller JM, Homan J, Steeghs D, et al. 2006. {\it Ap. J.} 653:525-35

\nhi
Mineshige S, Kawaguchi T, Takeuchi M, Hayashida K. 2000. {\it Publ. Astron. Soc. Jap.} 52:499-508


\nhi
Morganti R, Fogasy J, Paragi Z, Oosterloo T, Orienti M. 2013. {\it Science} 341:1082-5

\nhi
Mo\'scibrodzka M, Proga D, Czerny B, Siemiginowska A. 2007.  {\it Astron. Astrophys.} 474:1-13

\nhi
Mo\'scibrodzka M, Gammie CF, Dolence JC, Shiokawa H. 2011. {\it
Ap. J.} 735:9

\nhi
Mo\'scibrodzka M, Gammie CF, Dolence JC, Shiokawa H, Leung PK. 2009. {\it Ap. J.} 706:497-507


\nhi
Mo\'scibrodzka M, Shiokawa H, Gammie CF, Dolence JC. 2012. {\it Ap. J. Lett.} 752:L1


\nhi
Nagar NM, Falcke H, Wilson AS, Ho LC. 2000. {\it Ap. J.} 542:186-96

\nhi
Nakamura KE. 1998. {\it Publ. Astron. Soc. Jap.} 50:L11-L14

\nhi
Nakamura KE, Kusunose M, Matsumoto R, Kato S. 1997. {\it Publ. Astron. Soc. Jap.} 49:503-12

\nhi
Nakamura KE, Matsumoto R, Kusunose, M, Kato S. 1996. {\it Publ. Astron. Soc. Jap.} 48:761-69

\nhi
Narayan R. 1996. {\it Ap. J.} 462:136-41



\nhi
Narayan R, Barret D, McClintock JE. 1997a. {\it Ap. J.} 482:448-64

\nhi
Narayan R, Fabian A. 2011. {\it MNRAS} 415:3721-30

\nhi
Narayan R, Garcia MR,  McClintock JE. 1997b, {\it Ap. J. Lett.} 478:L79

\nhi
Narayan R, Igumenshchev IV, Abramowicz MA. 2000. {\it Ap. J.} 539:798-808

\nhi
Narayan R, Igumenshchev IV, Abramowicz MA. 2003. {\it Publ. Astron. Soc. Jap.} 55:L69-72

\nhi
Narayan R, Kato S, Honma F. 1997c. {\it Ap. J.} 476:49-60

\nhi
Narayan R, Mahadevan R, Grindlay JE, Popham RG, Gammie C. 1998a. {\it Ap. J.} 492:554-68

\nhi
Narayan R, Mahadevan R, Quataert E. 1998b. in Theory of Black Hole Accretion Disks, ed. M. A. Abramowicz, G. Bjornsson, \& J. E. Pringle (Cambridge Univ. Press) p148

\nhi
Narayan R, McClintock JE. 2005. {\it Ap. J.} 623:1017-25

\nhi
Narayan R, McClintock JE.  2008. {\it New Astro. Rev.} 51:733-51

\nhi
Narayan R, McClintock JE.  2012. {\it MNRAS} 419:L69-73


\nhi
Narayan R, McClintock JE, Yi I. 1996. {\it Ap. J.} 457:821-33


\nhi
Narayan R, Medvedev MV. 2001. {\it Ap. J.} 562:L129-32

\nhi
Narayan R, \"Ozel F, Sironi L. 2012a. {\it Ap. J. Lett.} 757:L20

\nhi
Narayan R, Quataert E, Igumenshchev IV, Abramowicz MA. 2002. {\it Ap. J.} 577:295-301

\nhi
Narayan R, Raymond J. 1999. {\it Ap. J. Lett.} 515:L69-72

\nhi
Narayan R, S\"adowski A, Penna RF,  Kulkarni AK. 2012b. {\it MNRAS} 426:3241-59


\nhi
Narayan R, Yi I. 1994. {\it Ap. J. Lett.} 428:L13-6

\nhi
Narayan R, Yi I. 1995a. {\it Ap. J.} 444:231-243

\nhi
Narayan R, Yi I. 1995b. {\it Ap. J.} 452:710-735

\nhi
Narayan R, Yi I, Mahadevan R. 1995. {\it Nature} 374:623-5

\nhi
Nemmen RS, Storchi-Bergmann T, Eracleous M. 2014. {\it MNRAS} in press (arXiv:1312.1982)

\nhi
Nemmen RS, Storchi-Bergmann T, Yuan F, et al. 2006. {\it Ap. J.} 643:652-9

\nhi
Nipoti C, Blundell KM, Binney J. 2005. {\it MNRAS} 361:633-7

\nhi
Noble SC, Krolik JH, Hawley JF. 2010. {\it Ap. J.} 711:959-73

\nhi
Noble SC, Leung PK, Gammie CF, Book LG. 2007. {\it Class. Quantum
Grav.} 24:S259-74

\nhi
Novak GS, Ostriker JP,  Ciotti L. 2011. {\it Ap. J.} 737:26

\nhi
Novikov ID, Thorne KS. 1973. {\it Black holes (Les astres occlus)} 343-450

\nhi
Oda H, Machida M, Nakamura KE, Matsumoto R. 2010. {\it Ap. J.} 712:639-52

\nhi
Ogilvie GI. 1999. {\it MNRAS} 306:L9-13

\nhi
Ohsuga K, Mineshige S. 2011. {\it Ap. J.} 736:2

\nhi
Ohsuga K, Mineshige S, Mori M, Kato Y. 2009. {\it Publ. Astron. Soc. Jap.} 61:L7-11

\nhi
Ohsuga K, Mori M, Nakamoto T, Mineshige S. 2005. {\it Ap. J.} 628:368-81

\nhi
Omma H, Binney J, Bryan G, Slyz A. 2004. {\it MNRAS} 348:1105-19

\nhi
Ostriker E. 1997. {\it Ap. J.} 486:291-306

\nhi
Ostriker JP, Choi E, Ciotti L, Novak GS, Proga D. 2010. {\it Ap. J.} 722:642-52

\nhi
\"Ozel F, Psaltis D, Narayan R. 2000. {\it Ap. J.} 541:234-49

\nhi
Paczy\'nski B,  Wiita PJ. 1980. {\it Astron. Astrophys.} 88:23-31

\nhi
Pang B, Pen UL, Matzner CD, Green SR, Liebend\"orfer M. 2011. {\it MNRAS} 415:1228-39

\nhi
Papaloizou JCB, Lin DNC. 1995. {\it Ap. J.} 438:841-51

\nhi
Park M, Ostriker JP. 2001. {\it Ap. J.} 549:100-17

\nhi
Park M, Ostriker JP. 2007. {\it Ap. J.} 655:88-97

\nhi
Parrish IJ, Stone JM. 2007. {\it Ap. J.} 664:135-48

\nhi
Pedlar A, Ghataure HS, Davies RD, et al., 1990. {\it MNRAS} 246:477-489

\nhi
Peitz J, Appl S. 1997. {\it MNRAS} 286:681-95

\nhi
Pellegrini S, Baldi A, Fabbiano G, Kim DM. 2003. {\it Ap. J.} 597:175-85

\nhi
Pellegrini S, Siemiginowska A, Fabbiano G,  et al. 2007. {\it Ap. J.} 667:749-59

\nhi
Pen UL, Matzner CD, Wong S. 2003. {\it Ap. J. Lett.} 596:L207-L210

\nhi
Penna RF, McKinney JC, Narayan R, et al. 2010. {\it MNRAS} 408:752-82

\nhi
Penna RF, Narayan R, Sadowsli A. 2013a. {\it MNRAS}, 436:3741-58

\nhi
Penna RF, Sadowski A, Kulkarni AK, Narayan R. 2013b. {\it MNRAS} 428:2255-74

\nhi
Penrose R. 1969. {\it Rivista del Nuovo Cimento, Numero Speziale I} 252

\nhi
Perna R, Raymond J, Narayan R. 2000. {\it Ap. J.} 541:898-907

\nhi
Pessah ME, Chan CK, Psaltis D. 2007. {\it Ap. J. Lett.} 668:L51-4

\nhi
Peterson JR, Paerels FBS, Kaastra JS, et al. 2001. {\it Astron. Astrophys.} 365:L104-9

\nhi
Phifer K, Do T, Meyer L, et al. 2013. {\it Ap. J. Lett.} 773:L13

\nhi
Phinney ES. 1983. in Astrophysical Jets, eds. A Ferrari, AG
Pacholczyk.  {\it Ap. Sp. Sci. Library} 103:201-12

\nhi
Piran T. 1978. {\it Ap. J.} 221:652-60

\nhi
Pizzolato F, Soker N. 2005. {\it ApJ} 632:821-30

\nhi
Plant DS, Fender RP, Ponti G, Munoz-Darias T, Coriat M. 2013. {\it MNRAS} submitted (arXiv:1309.4781)


\nhi
Plotkin RM, Markoff S, Kelly BC, K\"ording E, Anderson SF. 2012. {\it MNRAS} 419:267-86

\nhi
Popham R, Gammie CF. 1998. {\it Ap. J.} 504:419-30


\nhi
Poutanen J, Krolik JH, Ryde F. 1997. {\it MNRAS} 292:L21-5

\nhi
Poutanen J, Lipunova G, Fabrika S, Butkevich AG, Abolmasov
P. 2007. {\it MNRAS} 377:1187-94

\nhi
Poutanen J, Veledina A. 2014. {\it Space Sci. Rev.} in press (arXiv:1312.2761)

\nhi
Poutanen J, Vurm I. 2009. {\it Ap. J. Lett.} 690:L97-100

\nhi
Press WH, Teukolsky SA, Vetterling WT, Flannery BP. 1992. {\it
Numerical Recipes in FORTRAN. The Art of Scientific Computing},
Cambridge, UK: Cambridge University Press

\nhi
Press WH, Teukolsky SA, Vetterling WT, Flannery BP. 2002. {\it
Numerical Recipes in C++: The Art of Scientific Computing},
William H. Press

\nhi
Pringle JE. 1981. {\it Annu. Rev. Astron. Astrophys.} 19:137-162

\nhi
Proga D. 2007. {\it Ap. J.} 661:693-701


\nhi
Proga D, Begelman MC. 2003. {\it Ap. J.} 582:69-81

\nhi
Proga D, Ostriker JP, Kurosawa R. 2008. {\it Ap. J.} 676:101-12


\nhi
Ptak A, Terashima Y, Ho LC, Quataert E. 2004. {\it Ap. J.} 606:173-84

\nhi
Pudritz RE, Norman CA. 1983. {\it Ap. J.} 274:677-97

\nhi
Punsly B, Coroniti FV. 1989. {\it Phys. Rev. D.} 40:3834-57


\nhi
Qiao E, Liu BF. 2013. {\it Ap. J.} 764:2

\nhi
Quataert E. 1998. {\it Ap. J.} 500:978-91

\nhi
Quataert E. 2001. In {\it Probing the Physics of Active Galactic Nuclei}, ed. BM Peterson, RW Pogge, RS Polidan, p. 71. San Francisco: ASP

\nhi
Quataert E. 2002. {\it Ap. J.} 575:855-9

\nhi
Quataert E. 2004. {\it Ap. J.} 613:322-5

\nhi
Quataert E, Di Matteo T, Narayan R,  Ho LC. 1999. {\it Ap. J. Lett.} 525:L89-92


\nhi
Quataert E, Dorland W, Hammett GW. 2002. {\it Ap. J.} 577:524-33

\nhi
Quataert E, Gruzinov A. 1999. {\it Ap. J.} 520:248-255

\nhi
Quataert E, Gruzinov A. 2000. {\it Ap. J.} 539:809-814

\nhi
Quataert E,  Narayan R. 1999a. {\it Ap. J.} 516:399-410

\nhi
Quataert E,  Narayan R. 1999b. {\it Ap. J.} 520:298-315

\nhi
Rafferty DA, McNamara BR, Nulsen PEJ, Wise MW. 2006. {\it Ap. J.} 652:216-31

\nhi
Rakowski CE. 2005. {\it Advances in Space Research} 35:1017-26

\nhi
Ramadevi MC, Seetha S. 2007. {\it MNRAS} 378:182-8

\nhi
Rees MJ, Begelman MC, Blandford RD, Phinney ES. 1982. {\it Nature} 295:17-21

\nhi
Reis RC, Fabian AC, Miller JM. 2010. {\it MNRAS} 402:836-54

\nhi
Reis RC, Fabian AC, Ross RR, et al. 2008. {\it MNRAS} 387:1489-98

\nhi
Remillard RA, McClintock JE. 2006. {\it Annu. Rev. Astron. Astrophys.} 44:49-92

\nhi
Reynolds CS. 2013. {\it Space Sci. Rev.} in press (arXiv:1302.3260)

\nhi
Reynolds CS,  Di Matteo T, Fabian AC, Hwang U, Canizares CR. 1996. {\it MNRAS} 283:L111-6

\nhi
Reynolds CS, Fabian AC. 2008. {\it Ap. J.} 675:1048-56

\nhi
Rezzolla L, Yoshida S, Maccarone TJ, Zanotti O. 2003. {\it MNRAS} 344:L37-41

\nhi
Riquelme MA, Quataert E, Sharma P, Spitkovsky A. 2012. {\it Ap. J.} 755:50

\nhi
Romanova MM, Ustyugova GV, Koldoba AV, Chechetkin VM, Lovelace RVE. 1998. {\it Ap. J.} 500:703-13

\nhi
R\'o\.za\'nska A, Czerny B. 2000. {\it Astron. Astrophys.} 360:1170-86

\nhi
Rothstein DM, Lovelace RVE. 2008. {\it Ap. J.} 677:1221-32

\nhi
Ruffini R, Wilson JR. 1975. {\it Phys. Rev. D} 12, 2959-62

\nhi
Russell DM, Gallo E, Fender RP. 2013a. {\it MNRAS} 431:405-14

\nhi
Russell HR, McNamara BR, Edge AC, et al. 2013b. {\it MNRAS} 432:530-53

\nhi
Ruszkowski M, Begelman MC. 2002. {\it Ap. J.} 581:223-8

\nhi
Rykoff E, Miller JM, Steeghs D, Torres MAP. 2007. {\it Ap. J.} 666:1129-39

\nhi
Sadowski A, Narayan R, Penna R, Zhu Y. 2013a. {\it MNRAS}, 436:3856-74

\nhi
Sadowski A, Sironi L, Abarca D, et al. 2013b. {\it MNRAS} 432:478-91

\nhi
Saitoh TR, Makino J, Asaki Y, et al. 2013. {\it PASJ} in
press. arXiv:1212.0349

\nhi
Sazonov SY, Ostriker JP, Ciotti L,  Sunyaev RA. 2005. {\it MNRAS} 358:168-80

\nhi
Sazonov S, Sunyaev R, Revnivtsev M. 2012. {\it MNRAS} 420:388-404

\nhi
Schartmann M, Burkert A, Alig C, et al. 2012. {\it Ap. J.} 755:155

\nhi
Scheuer PAG, Feiler R. 1996. {\it MNRAS} 282:291-4

\nhi
Schechter P. 1976. {\it Ap. J.} 203:297-306

\nhi
Schnittman JD, Krolik JH,  Noblem SC. 2013. {\it Ap. J.} 769:156

\nhi
Sch\"odel R, Morris MR, Muzic K, et al. 2011. {\it  Astron. Astrophys.} 532:83

\nhi
Sch\"odel R, Ott T, Genzel R, et al. 2002. {\it Nature} 419:694

\nhi
Shafee R, McKinney JC, Narayan R, et al. 2008.{\it Ap. J. Lett.} 687:L25-8

\nhi
Shakura NI, Sunyaev RA. 1973. {\it Astron. Astrophys.} 24:337

\nhi
Shapiro SL, Lightman AP, Eardley DM. 1976. {\it Ap. J.} 204:187-99

\nhi
Sharma P, Hammett GW, Quataert E, Stone JM. 2006. {\it Ap. J.} 637:952-67

\nhi
Sharma P, Quataert E, Hammett GW, Stone JM. 2007a. {\it Ap. J.} 667:714-23

\nhi
Sharma P, Quataert E, Stone JM. 2007b. {\it Ap. J.} 671:1696-1707

\nhi
Sharma P, Quataert E, Stone JM. 2008. {\it MNRAS} 389:1815-27

\nhi
Shen ZQ, Lo KY, Liang MC, Ho PTP, Zhao JH. 2005. {\it Nature} 438:62-4

\nhi
Shcherbakov RV. 2013. {\it Ap. J.} submitted. arXiv:1309.2282

\nhi
Shcherbakov RV, Baganoff FK. 2010. {\it Ap. J.} 716:504-9

\nhi
Shcherbakov RV, McKinney JC. 2013. {\it Ap. J. Lett.} 774:L22

\nhi
Shcherbakov RV, Penna RF,  McKinney JC. 2012. {\it Ap. J.} 755:133

\nhi
Shibata K,  Uchida Y. 1985. {\it Publ. Astron. Soc. Jap.} 37:31-46

\nhi
Shibata K,  Uchida Y. 1986. {\it Publ. Astron. Soc. Jap.} 38:631-60

\nhi
Shiokawa H, Dolence JC, Gammie CF, Noble SC. 2012. {\it Ap. J.} 744:187

\nhi
Silk J. 2013. {\it Ap. J.} 772:112

\nhi
Somerville RS, Hopkins PF, Cox TJ, Robertson BE, Hernquist L. 2008. {\it MNRAS} 391:481-506

\nhi
Sorathia KA, Reynolds CS, Armitage PJ. 2010. {\it Ap. J.} 712:1241-7

\nhi
Sorathia KA, Reynolds CS, Stone, JM, Beckwith K. 2012. {\it
Ap. J.} 749:189

\nhi
Spruit HC, Deufel B. 2002. {\it Astron. Astrophys.} 387:918-30

\nhi
Spruit HC,  Matsuda T, Inoue M, Sawada K. 1987. {\it MNRAS} 229:517-27

\nhi
Spruit HC, Uzdensky DA. 2005. {\it Ap. J.} 629:960-8

\nhi
Springel V, Di Matteo T, Hernquist L. 2005. {\it MNRAS} 361:776-94

\nhi
Steiner JF, McClintock JE, Narayan R. 2013. {\it Ap. J.} 762:104

\nhi
Stern BE, Poutanen J, Svensson R, Sikora M, Begelman MC. 1995. {\it
Ap. J. Lett.} 449:L13-6

\nhi
Stone JM, Gardiner TA, Teuben P, Hawley JF, Simon JB. 2008. {\it Ap. J.} 178:137-77

\nhi
Stone JM, Hawley JF, Gammie CF, Balbus SA. 1996. {\it Ap. J.} 463:656-73

\nhi
Stone JM, Norman ML. 1992a. {\it Ap. JS} 80:753-90

\nhi
Stone JM, Norman ML. 1992b. {\it Ap. JS} 80:791-818

\nhi
Stone JM, Pringle JE. 2001. {\it MNRAS} 322:461-72

\nhi
Stone JM, Pringle JE, Begelman MC. 1999. {\it MNRAS} 310:1002-16

\nhi
Storchi-Bergmann T, Eracleous M, Ruiz MT, et al. 1997. {\it Ap. J.} 489:87-93

\nhi
Su M, Slatyer TR, Finkbeiner DP. 2010. {\it Ap. J.} 724:1044-82

\nhi
Taam RE, Liu B, Yuan W, Qiao E. 2012. {\it Ap. J.} 759:65

\nhi
Tanaka T, Menou K. 2006. {\it Ap. J.} 649:345-360

\nhi
Tchekhovskoy A, McKinney JC. 2012. {\it MNRAS} 423:L55-9

\nhi
Tchekhovskoy A, McKinney JC, Narayan R. 2012. {\it JPhCS} 372:012040

\nhi
Tchekhovskoy A, Metzger BD, Giannios D, Kelley LZ. 2013. {\it
MNRAS} in press (arXiv:1301.1982)

\nhi
Tchekhovskoy A, Narayan R,  McKinney JC. 2010. {\it Ap. J.} 711:50-63

\nhi
Tchekhovskoy A, Narayan R,  McKinney JC. 2011. {\it MNRAS} 418:L79-83

\nhi
Thorne KS, Price RH, MacDonald DA. 1986. Black holes: The
membrane paradigm. Yale Univ. Press

\nhi
Tomsick JA, Yamaoka K, Corbel S. 2009. {\it Ap. J. Lett.} 707:L87-91

\nhi
Trap G,  Goldwurm A, Dodds-Eden K, et al. 2011. {\it Astron. Astrophys.} 528:140

\nhi
van der Laan H. 1966. {\it Nature} 211:1131-3

\nhi
van Velzen S, Falcke H. 2013. {\it Astron. Astrophys.} 557:L7-10

\nhi
Veledina A, Poutanen J, Ingram A. 2013. {\it Ap. J.} 773:165

\nhi
Vernaleo JC, Reynolds C. 2006. {\it Ap. J.} 645:83-94

\nhi
Vlahakis N, Konigl A. 2003. {\it Ap. J.} 596:1080-103

\nhi
Wang QD, Nowak MA, Markoff SB, et al. 2013. {\it Science} 341:981-3

\nhi
Wang R, Wu XB, Kong MZ. 2006. {\it Ap. J.} 645:890-9

\nhi
Watarai K, Mizuno T,  Mineshige S. 2001. {\it Ap. J. Lett.} 549:L77-88

\nhi
Wilson AS, Yang Y. 2002. {\it Ap. J.} 568:133-40

\nhi
Wrobel JM, Terashima Y, Ho L. 2008. {\it Ap. J.} 675:1041-7

\nhi
Wu Q, Yuan F, Cao X. 2007. {\it Ap. J.} 669:96-105

\nhi
Wu XB. 1997. {\it MNRAS} 292:113-9

\nhi
Wu XB, Li Q. 1996. {\it Ap. J.} 469:776-83

\nhi
Wyithe JSB, Loeb A. 2003. {\it Ap. J.} 595:614-23

\nhi
Xie FG, Yuan F. 2008. {\it Ap. J.} 681:499-505

\nhi
Xie FG, Yuan F. 2012. {\it MNRAS} 427:1580-6

\nhi
Xu G, Chen X. 1997. {\it Ap. J.} 489:L29-32

\nhi
Xu YD, Cao XW. 2009. {\it Res. Astron. Astrophys.} 9:401-8

\nhi
Xu YD, Narayan R, Quataert E, Yuan F, Baganoff FK. 2006. {\it Ap. J.} 640:319-26

\nhi
Xue L,  Wang JC. 2005. {\it Ap. J.} 623:372-82

\nhi
Younes G, Porquet D, Sabra B, Reeves JN. 2011. {\it Astron. Astrophys.} 530:149

\nhi
Younes G, Porquet D, Sabra B, Reeves JN, Grosso N. 2012. {\it Astron. Astrophys.} 539:104

\nhi
Yu W, Yan Z. 2009. {\it Ap. J.} 701:1940-57

\nhi
Yu Z, Yuan F, Ho LC. 2011. {\it Ap. J.} 726:87

\nhi
Yuan F. 1999. {\it Ap. J. Lett.} 521:L55-58

\nhi
Yuan F. 2001. {\it MNRAS} 324:119-27

\nhi
Yuan F. 2003. {\it Ap. J. Lett.} 594:L99-102



\nhi
Yuan F, Bu D. 2010. {\it MNRAS} 408:1051-60

\nhi
Yuan F, Bu D, Wu M. 2012a. {\it Ap. J.} 761:130

\nhi
Yuan F, Cui W. 2005. {\it Ap. J.} 629:408-13

\nhi
Yuan F, Cui W, Narayan R. 2005. {\it Ap. J.} 620:905-14

\nhi
Yuan F, Li M. 2011. {\it Ap. J.} 737:23

\nhi
Yuan F, Lin J, Wu K, Ho LC. 2009a. {\it MNRAS} 395:2183-8

\nhi
Yuan F, Markoff S, Falcke H. 2002a. {\it Astron. Astrophys.} 383:854-63

\nhi
Yuan F, Markoff S, Falcke H, Biermann PL. 2002b. {\it Astron. Astrophys.}
391:139-48

\nhi
Yuan F, Narayan R. 2004. {\it Ap. J.} 612:724-28


\nhi
Yuan F, Quataert E, Narayan R. 2003, {\it Ap. J.} 598:301-12

\nhi
Yuan F, Quataert E, Narayan R. 2004, {\it Ap. J.} 606:894-99


\nhi
Yuan F, Taam RE, Xue R, Cui W. 2006. {\it Ap. J.} 636:46-55

\nhi
Yuan F, Wu M, Bu D. 2012b. {\it Ap. J.} 761:129

\nhi
Yuan F, Xie F, Ostriker JP. 2009b. {\it Ap. J.} 691:98-104

\nhi
Yuan F, Yu Z, Ho L. 2009c. {\it Ap. J.} 703:1034-43

\nhi
Yuan F, Zdziarski A. 2004. {\it MNRAS} 354:953-60

\nhi
Yuan F, Zdziarski A, Xue Y, Wu XB. 2007. {\it Ap. J.} 659:541-48

\nhi
Yusef-Zadeh F, Roberts D, Wardle M, Heinke CO, Bower GC. 2006. {\it Ap. J.} 650:189-94

\nhi
Yusef-Zadeh F, Bushouse H, Wardle M, et al. 2009. {\it Ap. J.} 706:348-75

\nhi
Yusef-Zadeh F, Wardle M. 2013. {\it Ap. J. Lett.} 770:L21

\nhi
Zdziarski AA, Lubi\'nski P, Gilfanov M, Revnivtsev M. 2003. {\it MNRAS} 342:355-72

\nhi
Zdziarski AA, Lubi\'nski P, Sikora M. 2012. {\it MNRAS} 423:663-75

\nhi
Zdziarski AA,  Gierli\'nski M. 2004. {\it Prog. Theor. Phys. Supp.}  155:99-119

\nhi
Zdziarski AA, Poutanen J,  Mikolajewska J, et al. 1998. {\it MNRAS} 301:435-50


\nhi
Zhang SN. 2013. {\it Frontiers of Phys.}  Eds. Bing Zhang \& P\'eter M\'esz\'aros (arXiv:1302.5485)

\nhi
Zubovas K, King AR, Nayakshin S. 2011. {\it MNRAS} 415:L21-L25

\end{document}